\documentclass[iop]{emulateapj}

\usepackage{amsmath,amssymb,latexsym, graphicx}
\usepackage{epsfig}
\usepackage{booktabs}
\usepackage{afterpage,lscape}
\usepackage[dvips]{color}

\slugcomment{Submitted to ApJ}

\shorttitle{A radio-selected sample of GRB afterglows}
\shortauthors{Chandra \& Frail}

\begin{document}

\title{A Radio-Selected Sample of Gamma Ray Burst Afterglows}
\author{
Poonam\,Chandra,\altaffilmark{1}
\&
Dale\,A.\,Frail,\altaffilmark{2}
}

\altaffiltext{1}{Department of Physics, Royal Military College of
  Canada, Kingston, ON K7M3C9, Canada; {\tt Poonam.Chandra@rmc.ca}}
\altaffiltext{2}{National Radio Astronomy Observatory, 1003 Lopezville
  Road, Socorro, NM 87801.}

\begin{abstract}

  We present a catalog of radio afterglow observations of gamma-ray
  bursts (GRBs) over a 14 year period from 1997 to 2011. Our sample of
  304 afterglows consists of 2995 flux density measurements (including
  upper limits) at frequencies between 0.6 GHz and 660 GHz, with the
  majority of data taken at 8.5 GHz frequency band (1539
  measurements).  We use this dataset to carry out a statistical
  analysis of the radio-selected sample. The detection rate of radio
  afterglows has stayed unchanged almost at 31\% before and after the
  launch of the {\em Swift} satellite. The canonical long-duration GRB
  radio light curve at 8.5 GHz peaks at 3-6 days in the source rest
  frame, with a median peak luminosity of $10^{31}$ erg s$^{-1}$
  Hz$^{-1}$. The peak radio luminosities for short-hard bursts, X-ray
  flashes and the supernova-GRB classes are an order of magnitude or
  more fainter than this value. There are clear relationships between
  the detectability of a radio afterglow and the fluence or energy of
  a GRB, and the X-ray or optical brightness of the afterglow.
  However, we find few significant correlations between these same
  GRB and afterglow properties and the peak radio flux density.  
  We also
  produce synthetic light curves at centimeter and millimeter bands
  using a range of blastwave and microphysics parameters derived from
  multiwavelength afterglow modeling, and we use them to compare to
  the radio sample.  Finding agreement, we extrapolate this behavior
  to predict the centimeter and millimeter behavior of GRBs observed
  by the Expanded Very Large Array (EVLA) and the Atacama Large
  Millimeter Array (ALMA).

\end{abstract}


\section{Introduction}
\label{intro}

Our understanding of gamma-ray bursts (GRBs) has advanced rapidly
since the discovery of the long-lived ``afterglows'' at X-ray, optical
and radio wavelengths \citep{cfh+97, vgg+97, fkn+97}. Since that time
many more $\gamma-$ray and hard X-ray satellites have monitored the
sky for GRBs providing accurate localizations and enabling detailed
followed up by the space and ground-based facilities at longer
wavelengths.  As a result of this large and sustained effort, several
hundred afterglows have been detected over the last 15 years. The
afterglow study has exposed a rich diversity of GRB behaviors and
several distinct progenitor classes \citep{grf09}.  While many of the
early advances in the field were the result of studying individual
GRBs and their afterglows, the field has now matured to the point
where much more can be learned from the study of large samples.

Statistical analysis have long been carried out on GRB-only samples
\citep[e.g.][]{kmf+93,sbb+11} but including afterglow data has only
been done recently. Compilations of X-ray and optical light curves
from the {\em Swift} satellite have revealed complex but canonical
light curve behaviors \citep{nkg+06,zfd06,mmk+08,rko+09,ebp+09}.  Optical
catalogs have derived mean dust extinction laws of GRB host galaxies
\citep{kkz06,sww+07}, and have lead to claims of clustering of the
optical afterglow luminosities \citep{lz06,kkz06,Nardini06}.  

Comparative studies of GRBs and their afterglows have proven even more
useful. The standard fireball model has been tested through spectral
and temporal comparisons of X-ray and optical light curves
\citep[e.g.][]{ops+09, ops+11,Schulze11}. Significant differences have been
found between the mean brightness and the redshift distribution of the {\em
  Swift} and the pre-{\em Swift} GRBs \citep{bkf+05,jlf+06,kkz+10}.
Correlations have been found for both short and long-duration bursts
between the gamma-ray fluence and the X-ray and optical afterglow
brightness \citep{gbb+08,nf09,kkz+08}. A population of dark bursts has been
identified \citep{rwk+05}, as has a population of nearby, low
luminosity events \citep{sazonov04,skn+06,pian06,lzz07,kkz+10}.

In comparison, very little effort has gone into compiling radio
afterglow data and carrying out correlative studies \citep{berger04}.
A catalog of the first five years of radio afterglow data was produced
by \citet{fkbw03}, but comparisons were limited just between detection
rates of X-ray, optical and radio afterglows. The mean flux densities
and luminosities of radio afterglows were given in \citet{frail05},
while \citet{skn+06} compiled radio light curves to compare GRB and
supernova luminosities. Recently \citet{alma} has compiled the
mm/submm data of GRBs and carried out the cumulative analysis in
view of upcoming Atacama Large Millimeter Array (ALMA). 
Since the launch of the {\em Swift} satellite
the number of GRBs with radio data has doubled. Moreover, in the near
future the continuum sensitivity of the Very Large Array\footnote{The
  Very Large Array is operated by the National Radio Astronomy
  Observatory, a facility of the National Science Foundation operated
  under cooperative agreement by Associated Universities, Inc.} (VLA; 
now Expanded Vary Large Array)
- the primary telescope for radio afterglow follow-up -- will increase
by a factor of 5 to 20 (depending on wavelength). It is therefore
timely to pull together all the past radio data in one place and to
use the catalog to define the average properties of radio afterglows,
search for trends in different sub-classes of GRBs (e.g. short vs long),
search for differences in the {\em Swift} and pre-{\em Swift} sample,
and to investigate if there are any correlations between the radio
afterglows and the prompt or afterglow emission at shorter
wavelengths.

In \S\ref{radiosample}, we present a summary of our radio sample and
its properties.  We also compare the detection statistics with respect
to other wave bands.  We discuss the details and statistics of
detection versus upper limits of our sample in \S \ref{sec:det}. In
\S\ref{sec:multi} and \S\ref{corr}, we investigate the distribution
and correlative properties of our radio sample with respect to the GRB
prompt emission properties as well as the X-ray and optical afterglow
properties.  We plot the synthetic radio light curves and their
dependence on various parameters in \S\ref{sec:synthetic}. We use
these light curves to make some predictions for the Atacama Large
Millimeter Array (ALMA) in \S\ref{sec:alma}, while in
\S\ref{sec:discussion} we bring together to summarize the various
results of the paper.

We adopt a $\Lambda$-CDM cosmology throughout this paper with $H_0=71$
km s$^{-1}$ Mpc$^{-1}$, $\Omega_m=0.27$ and $\Omega_\Lambda=0.73$
\citep{sbd+07}.

\section{The radio sample and statistics} 
\label{radiosample}

\subsection{Radio Sample}
\label{sample}

\begin{figure}
\centering
\includegraphics[width=0.48\textwidth]{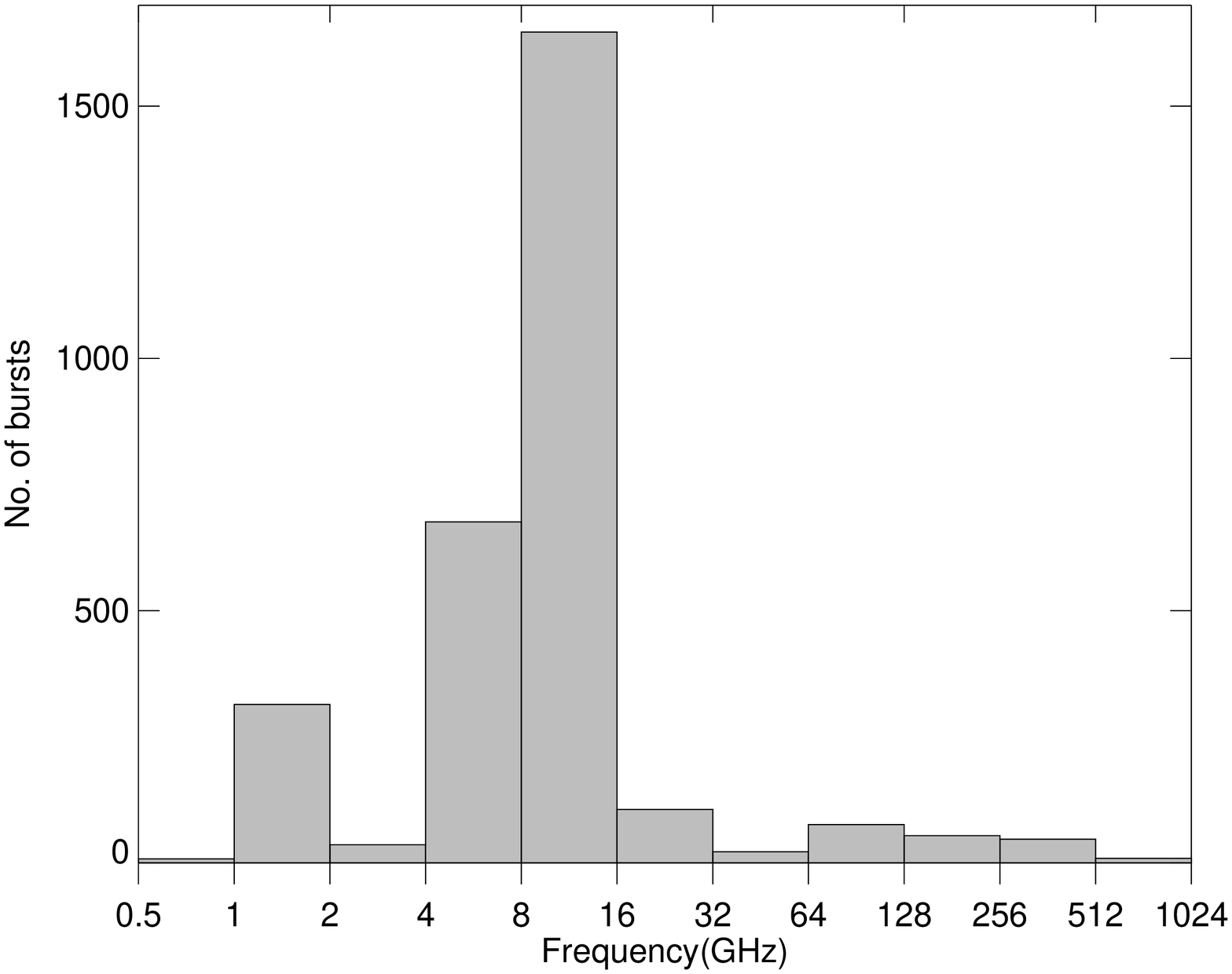}
\includegraphics[width=0.48\textwidth]{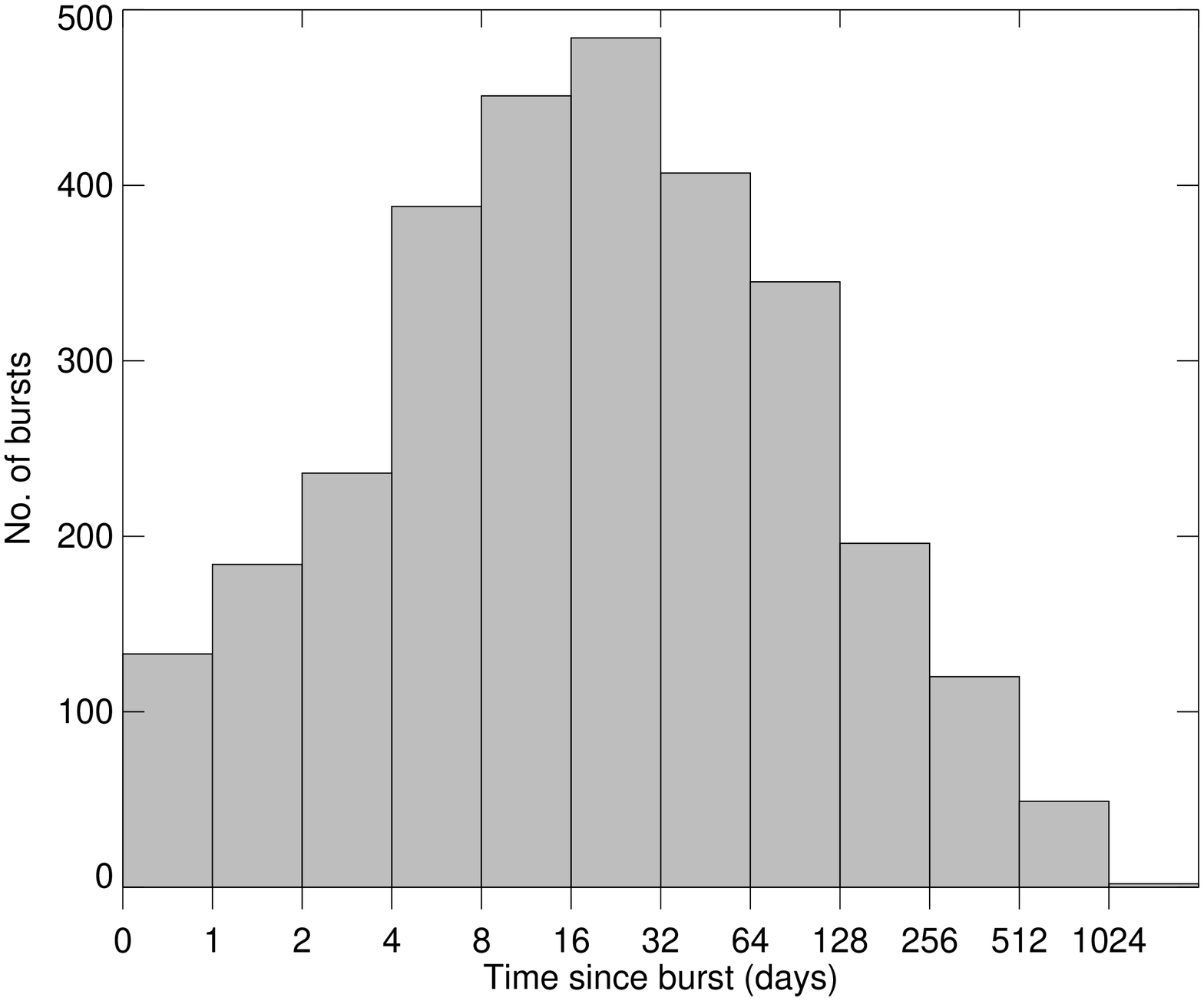}
\caption{Histograms summarizing the distribution of data of the radio selected 
sample  catalog as a function of 
observing frequency (upper panel) and of the time
elapsed since the burst
(lower panel). }
\label{sample-statistics}
\end{figure}

Our compiled sample consists of 304 GRBs observed with the radio
telescopes between January 1997 and January 2011, along with the 2011
April 28 Fermi burst, GRB\,110428A.  The sample consists of a total of
2995 flux density measurements taken between the frequencies 0.6 GHz
to 660 GHz bands spanning a time range from 0.026 to 1339 days.  In
Figure~\ref{sample-statistics}, we plot the histogram summarizing the
distribution of the data in our catalog, as a function of frequency
and time elapsed since the burst.  A total of 1539 measurements were
taken in 8.5 GHz frequency band, while 657 measurements were taken in
5 GHz frequency band. 

The sample includes 33 Short-hard bursts (SHBs), 19 X-ray flashes
(XRFs) and 26 GRBs with possible supernova associations (SN/GRBs). The
SN/GRB category includes all the GRBs with confirmed associations as
well as likely supernova (SN) associations \citep[see][]{hb11}. The
SN/GRBs with known supernovae (SNe) associated with them in our sample
are: GRB 980425 (SN 1998bw), GRB 011121 (SN 2001ke), GRB 021211 (SN
2002lt), GRB 030329 (SN 2003dh), GRB 031203 (SN 2003lw), GRB 050525A
(SN 2005nc), GRB 060218 (SN 2006aj), GRB 091127 (SN 2009nz) and GRB
101219B (SN 2010ma). The spectroscopically confirmed SN associations
among these are GRBs 980425, 030329, 031203, 060218 and 091127. Of the 26
SN/GRBs associations, 17 and 9 were discovered in the pre-\emph{Swift}
and post-\emph{Swift} epoch, respectively.  Among the 33 SHBs, 28 were
detected by the \emph{Swift} alone.  We also have one candidate
galactic transient in our sample \citep[GRB 070610,][]{kck+08,castro08}.

Most of the afterglows (270 in total) in our sample were observed as
part of the VLA radio afterglow programs, whereas, 15 bursts were observed
by the Expanded VLA (EVLA), and 19 southern bursts with the Australia
Telescope Compact Array (ATCA). Among the total of 285 VLA/EVLA
bursts, 8 bursts were followed by both the VLA/EVLA and the ATCA; and
38 bursts were observed by the Westerbork Synthesis Radio Telescope
(WSRT) in addition to the VLA/EVLA.  A total of 19 VLA/EVLA bursts
were followed by the Ryle telescope, and 11 bursts in our sample were
observed by the Giant Metrewave Radio Telescope (GMRT). The Very Long
Baseline Array (VLBA) with its sub-milliarcsecond angular resolution
was used to observe five bursts (see 
Table~\ref{MasterTable}).

For the VLA observations, most of the afterglows were first observed in the
8.5 GHz frequency band, since it is the most sensitive VLA band. 
Once a GRB was detected in the 8.5 GHz band, we
followed it at other VLA frequencies to measure the continuum
spectrum. However, in cases of bright, nearby or high-redshift 
(high-$z$) GRBs or GRBs with good amount
of multiwaveband data, we
carried out follow-up observations for few epochs even for the initially
non-detected afterglows, depending upon its interest in other wavebands, VLA
time availability factors etc.
The data were taken in standard interferometric mode for an
integration time of typically 30 minutes (including the time on the calibrators
and the GRB integration time). The total bandwidth used was 100 MHz.
The data was analysed using standard {\em AIPS} routines.

In Table~\ref{MasterTable}, we list the comprehensive properties of
our sample. In column 1, the GRB names are given. Column 2 lists the
missions under which these GRBs were discovered (please refer to the
table notes for the abbreviations of the mission names).  We tabulate
the J2000 positions of the GRBs in columns 3 and 4.  Columns 5, 6 and
7 tabulate the detection statistics in X-ray, optical and radio bands,
respectively. Symbol `Y' indicates a detection, while `N' indicates a
non-detection. The symbol `X' means the burst was not observed in that
particular band, whereas `Y?' indicates that the observation was made
but a detection could not be confirmed. Column 8 tabulates all the
radio telescopes used to observe the corresponding burst. We report
the $T_{90}$ durations (the time interval over which 90\% of the total
background subtracted counts are observed, with the interval starting
at 5\% of the total counts have been observed \citep{kpk+95}) in the
observer's frame and the redshifts $z$ in columns 9 and 10,
respectively. Here  $T_{90}$ durations are mentioned in the
energy band of the specific detector which detected a GRB. For example,
for {\em Swift} bursts, the $T_{90}$ durations are in 15--350~keV 
energy range. Column 11 shows the 15--150 keV fluences ($S_{15-150}$)
of our sample and column 12 indicates the $k$-corrected isotropic
bolometric $\gamma$-ray energy ($E_{iso}^{bol}$, energy range
1--10,000 keV in the rest frame). We report the X-ray fluxes at 11 hr
($F_{X}^{11h}$) in 0.3--10 keV range in column 13. Here the fluxes for
the \emph{BeppoSAX} bursts are quoted in 1.6--10 keV range. We also
report the R-band (0.7 $\mu$m) optical flux densities at 11 hr
($F_{R}^{11h}$, in $\mu$Jy) in column 14. Column 15 tabulates the jet
break times $t_j$ in cases where a clear jet break was seen and column
16 indicates the number density ($n$) of the circumburst medium.  For
the GRBs in which the number density of the medium was not known, we
have assumed a density of 1 cm$^{-3}$ and have indicated them with
``[1]''. We estimate the collimation angle $\theta_j$ (degrees) and
the beaming-corrected bolometric energy ($E_{true}^{bol}$) of GRBs
with observed $t_j$ using \citet{fks+01,bfk03} and tabulate them in
columns 17 and 18, respectively.  Column 19 lists the references.
Each cell has 8 references in the following order: $T_{90}$, $z$,
$S_{15-150}$, $E_{iso}^{bol}$, $F_{X}^{11h}$, $F_{R}^{11h}$, $t_j$,
and $n$.  If there is no data in a particular cell, the corresponding
reference is indicated with [].  However, a reference indicated by []
for a non-empty corresponding cell implies that that particular value
has been calculated by us and this paper is the reference.

\subsection{Multiwaveband detection statistics of our sample}\label{venn}

In Table~\ref{tab:multi}, we summarize the detection statistics in 
radio, X-ray and optical bands for all the bursts in our sample.  We
subdivide the sample into pre-\emph{Swift} and post-\emph{Swift} bursts.
This split was made to investigate possible selection biases.
The pre-\emph{Swift} GRB event rate was low enough that radio
follow-up was undertaken for nearly all events \citep[see][]{fkbw03},
while only a subset could be observed among the \emph{Swift} bursts.

\begin{deluxetable*}{l|rrrrrr|rrrrrr|rrrrrr}
\tablecaption{Multiwaveband statistics of the GRB afterglow sample
\label{tab:multi}}
\tablewidth{0pt}
\tablehead{
\colhead{} \vline & \multicolumn{6}{c}{Complete sample} \vline & 
\multicolumn{6}{c}{Pre-\emph{Swift} sample} \vline & 
\multicolumn{6}{c}{Post-\emph{Swift} sample}\\
\colhead{} \vline & \multicolumn{2}{c}{Radio} &  \multicolumn{2}{c}{Optical}  &
\multicolumn{2}{c}{X-ray} \vline 
& \multicolumn{2}{c}{Radio} & \multicolumn{2}{c}{Optical} & 
 \multicolumn{2}{c}{X-ray} \vline  &
\multicolumn{2}{c}{Radio}  & \multicolumn{2}{c}{Optical} & 
\multicolumn{2}{c}{X-ray} \\
\colhead{} \vline 
& \colhead{No.} & \colhead{frac.} & \colhead{No.} & \colhead{frac.} & \colhead{No.} & \colhead{frac.} \vline & \colhead{No.} &
\colhead{frac.} & \colhead{No.} & \colhead{frac.} & \colhead{No.} & \colhead{frac.} \vline & \colhead{No.} & \colhead{frac.} &
\colhead{No.} & \colhead{frac.} & \colhead{No.} & \colhead{frac.}
}
\startdata
Detection  &    95&     0.31&   196&    0.65&   221&    0.73 &   42& 0.34&       60& 0.49&       52&0.42 &        53& 0.29 &
136&    0.75 & 169 & 0.93\\             
Non-detection & 206&    0.68&   94&     0.32&   13&     0.05 &   81& 0.66&       55&  0.45&      6& 0.05 &        125&    0.69 & 39&
0.22& 7 & 0.04\\
Not-observed  & 0&      0&      10&     0.03 &  68&     0.23 &   0& 0&   8& 0.06&        63& 0.51 &       0& 0&   2&
0.01& 5 & 0.03\\
Unconfirmed   & 3&      0.01&   4&      0.02&   2&      0.01 &   0& 0&   0&0&    2& 0.02 & 3& 0.02&       4& 0.02&        0 & 0\\
\tableline
 & & & & & & & & & & & & & & & & & & \\
Total   &       304&1   &       304&1   &       304&1 &  123& 1& 123&1 & 123&1 &  181&    1& 181&1        & 181 &1
\enddata
\end{deluxetable*}

\begin{figure}
\centering
\includegraphics[width=0.23\textwidth]{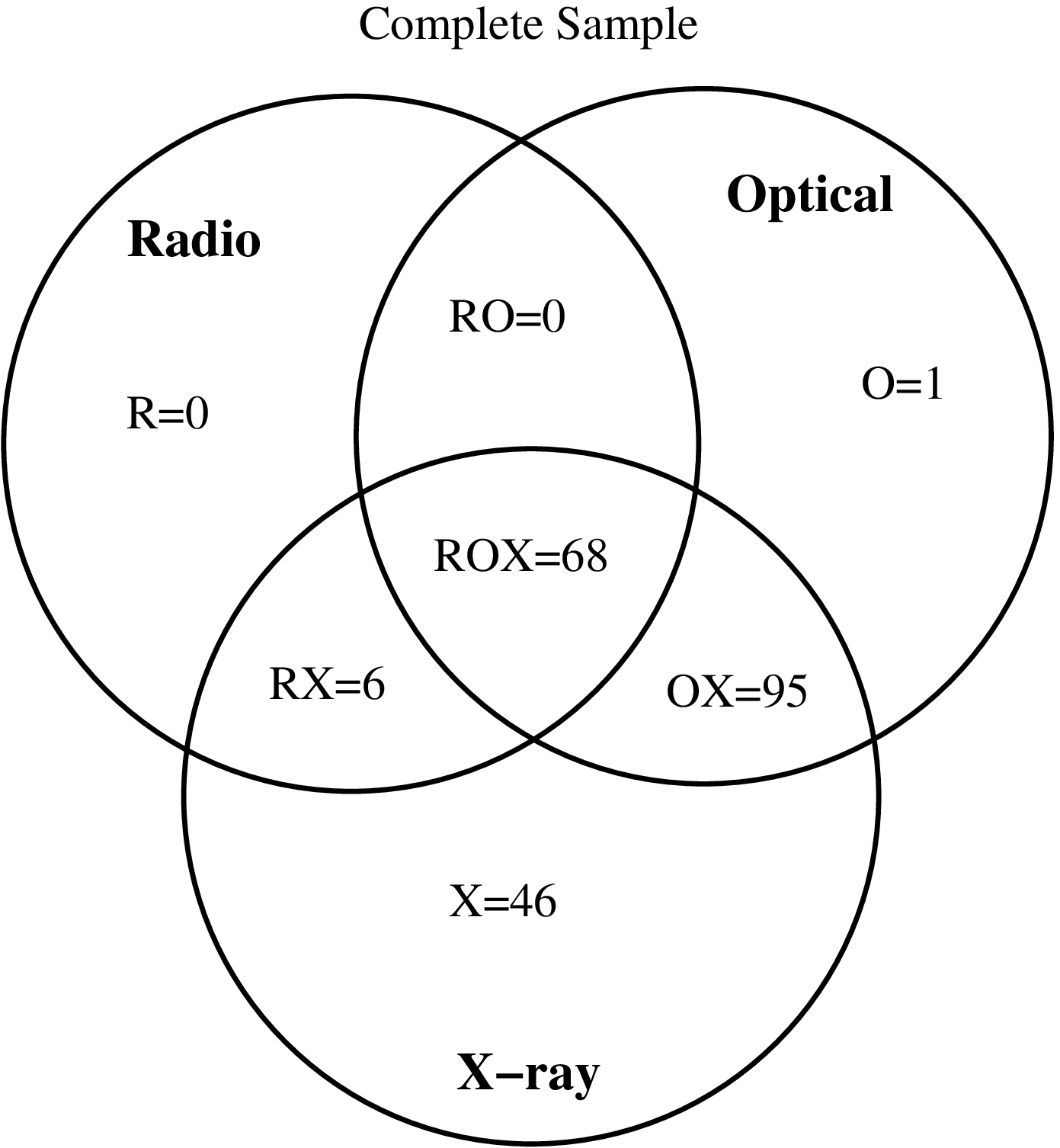}
\includegraphics[width=0.23\textwidth]{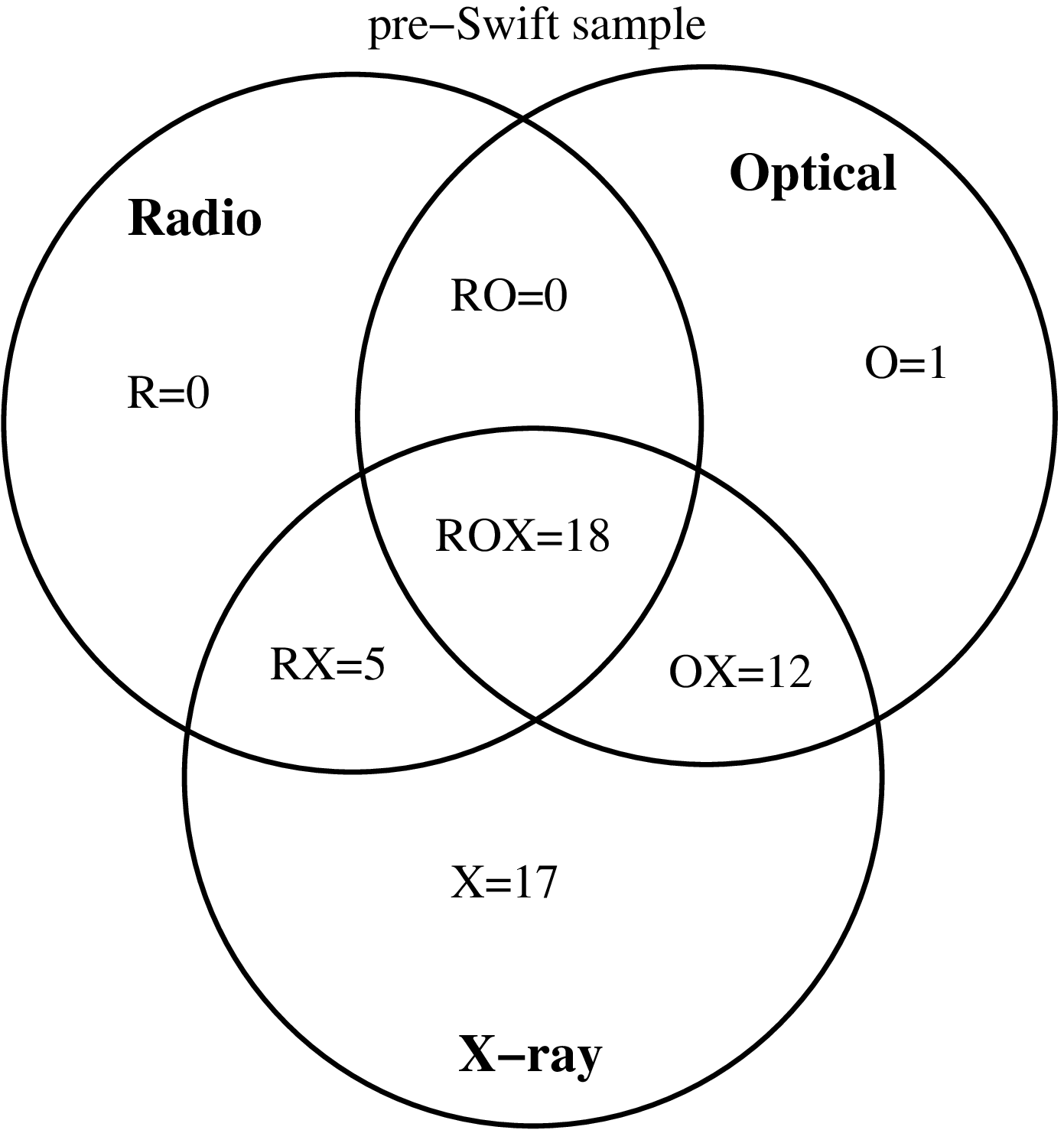}
\includegraphics[width=0.23\textwidth]{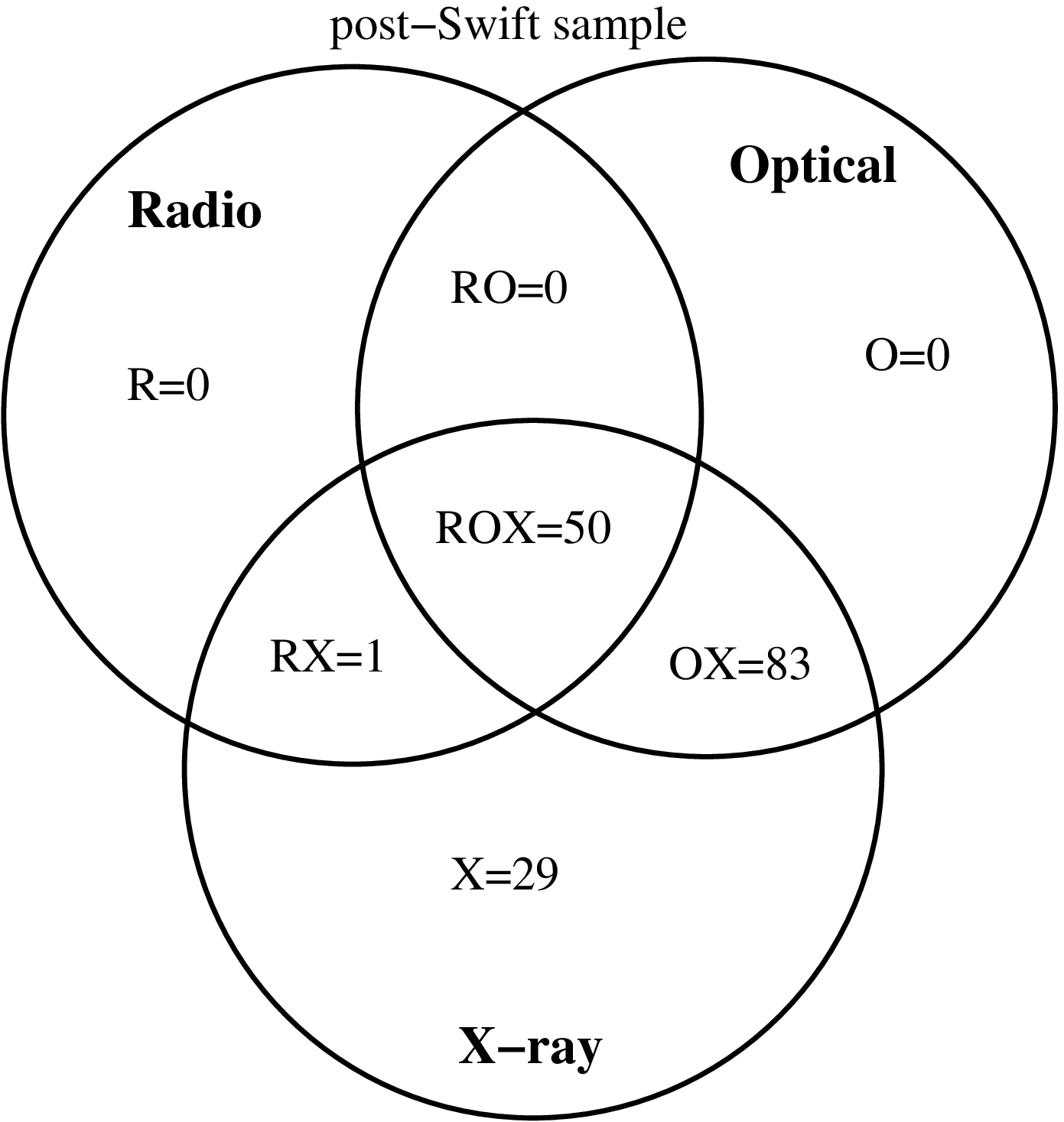}
\caption{Venn diagrams illustrating the different relationships 
  between the radio, optical, and X-ray afterglows for the entire
  sample, the pre-{\em Swift} and the post-{\em Swift} sample. The
  symbols X, O and R indicates detections in X-ray, optical and radio
  bands, respectively. The combinations of these letters indicate
  detections in those respective bands.  For example, RX indicates
  detection in radio and X-ray but not in optical, while ROX indicate
  afterglows seen in all three bands.  }
\label{fig:venn}
\end{figure}

Out of the 304 bursts, 123 bursts were observed in the pre-\emph{Swift}
epoch from 1997 until 2004. The remaining 181 bursts were observed
between 2005 and April 2011.  A total of 95 bursts resulted in radio
detections, while 206 were non-detections. For three bursts an initial
detection could  not be confirmed. Thus the total detection rate for our
sample in the radio band is 31\%.  The radio detection statistics in
pre-\emph{Swift} and post-\emph{Swift} samples are 42/123 (34\%) and
53/181 (29\%), respectively.  It is striking how similar the
detection ratios are in both samples.
In contrast, the X-ray detection rates increase from 42\% to 93\%
after the launch of \emph{Swift}.  This is because of the dedicated
X-ray telescope (XRT) on-board \emph{Swift}, which autonomously slewed
to Burst Alert Transient (BAT; onboard {\em Swift})
detected bursts.  Note also that 51\% of the bursts were not
observed in the X-ray bands in the pre-\emph{Swift} sample. The optical
detection rates also increased from 48\% to 75\% between the
pre-\emph{Swift} and the post-\emph{Swift} epoch. This too was due to the
on-board UV/optical telescope (UVOT) which was autonomously slewed
along with the XRT, and also due to the availability of rapid,
well-localized positions for ground-based follow-up. We refer the
reader to Table~\ref{tab:multi} for the detailed statistics of the
multiwaveband observations of our sample.

In Figure~\ref{fig:venn}, we plot a Venn diagram for our entire sample
illustrating the different relationships between the radio, optical
and X-ray observations of the afterglows. We also indicate this
relationship for the pre-\emph{Swift} and the post-\emph{Swift} sample
separately. For the Venn diagrams, if a GRB did not have an
observation in any band (i.e. 'X' in Table~\ref{tab:multi}), we
excluded them.  We also excluded those events which had a unconfirmed
detection in any bands (i.e. 'Y?'). Thus, we included a total of 226
bursts, out of which 10 bursts had no detection in any band.  Among
those 10 bursts with no afterglow detection, 6 are SHBs (GRBs 021201,
020531, 050911, 051105A, 071112B and 090417A)
 and the remaining are normal long GRBs (GRBs 970111,
 990217, 000615A and 071018).  Hence, the final Venn diagram of
the complete sample includes 216 bursts. The X-ray detections are
marked with X, optical detections are marked with O and radio
detections are marked with R. ROX indicates the bursts which were
detected in all three bands, whereas a combination of two letters
indicates a joint detection in those respective wavelengths. For
example, RX indicates bursts detected in radio and X-ray bands but not
in optical.  A single letter, X for example, indicates an afterglow
detected in the X-ray band only.  Our sample consists of 68 bursts
which were detected in all three bands.  There were 95 bursts which
had detections in X-ray and optical but not in radio.  There was no
burst which was detected in only the radio band (i.e. R), nor were
there any bursts detected in only the radio and the optical bands
without being detected in the X-ray band (i.e.  RO). However, 6 bursts
were detected in the radio and the X-ray bands but not in the optical.
These form a subset of the dark bursts.  Only one optical burst could
not be detected in any other band. The number of only X-ray detected
afterglows is 46. In the post-\emph{Swift} and the pre-\emph{Swift}
samples the X-ray only afterglows are 29 and 17, respectively.  In the
pre-\emph{Swift} sample the GRBs detected in all three bands was only
18, rising to 50 in the post-\emph{Swift} bursts.

\section{Radio afterglow sample analysis}\label{sec:det}

\subsection{Radio flux density distributions}\label{sec:radioafterglows}

Since our data consists of detections as well as upper limits, we
incorporated the Kaplan--Meier Product Limit method \citep[K--M;][]{fn85} in
deriving the flux density distributions and 
mean flux density estimates of our sample. The K--M estimator
was applied to a sample of bursts for which flux density measurements
had been made at one frequency over a fixed time interval. For
detections, we averaged multiple observations of the same GRB to
obtain the mean flux density. This is perfectly acceptable
 since over the short periods the flux density
variations at radio frequencies are dominated by interstellar
scintillation. Table~\ref{KM} summarizes our results for 5 and 8.5 GHz
bands. Here, the mean estimates are given for only detections 
as well as by including the upper limits. The KM$_{25}$, KM$_{50}$ and 
KM$_{75}$ are the K--M method estimates of the
minimum flux density above which 25\%, 50\%, and 75 \% of the total
radio afterglows will lie. 

\begin{deluxetable*}{lccc|cccc|cccc}
\tabletypesize{\scriptsize}
\tablecaption{K--M estimates of the flux density distribution
\label{KM}}
\tablewidth{0pt}
\tablehead{
\colhead{} &  \colhead{} &\colhead{} & \colhead{} 
\vline & \multicolumn{4}{c}{Detections Only} \vline &  \multicolumn{4}{c}{Detections+Upper
limits}\\
\colhead{Freq.} & \colhead{Time range} & \colhead{Total} & 
\colhead{Total} 
\vline & \colhead{KM$_{25}$ $^\mathrm{a}$} &  \colhead{KM$_{50}$ $^\mathrm{b}$} &  
\colhead{KM$_{75}$ $^\mathrm{c}$} & \colhead{Mean flux} \vline
& \colhead{KM$_{25}$ $^\mathrm{a}$} &  \colhead{KM$_{50}$ $^\mathrm{b}$} &  
\colhead{KM$_{75}$ $^\mathrm{c}$} & \colhead{Mean flux}\\
\colhead{GHz} & \colhead{Days} & \colhead{Detections} & \colhead{Data points} \vline & 
\colhead{$\mu$Jy} & \colhead{$\mu$Jy} & \colhead{$\mu$Jy} & \colhead{$\mu$Jy} \vline
& \colhead{$\mu$Jy} & \colhead{$\mu$Jy} & \colhead{$\mu$Jy} & \colhead{$\mu$Jy}
}
\startdata
4.86 & $2-8$    & 27 & 63 & 362 & 251 & 199 & $290\pm25$ & 243 & 155 & 64 & $197\pm17$\\
4.86 & $5-10$  & 17 & 40 & 377 & 248 & 198 & $290\pm35$ & 238 & 88 & 44 & $193\pm23$\\
4.86 & $7-14$  & 20 & 39 & 419 & 228 & 152 & $285\pm42$ & 237 & 132 & 82 & $206\pm27$\\
\tableline
  & & & & & & & & & & & \\
8.46 & $2-8$    & 65 & 157 & 389 & 219 & 156 & $287\pm26$ & 199 & 107 & 31 & $162\pm15$\\
8.46 & $5-10$  & 51 & 107 & 343 & 197 & 126 & $275\pm34$ & 196 & 92 & 55 & $170\pm20$\\
8.46 & $7-14$  & 43 & 99 & 418 & 203 & 123 & $302\pm42$ & 185 & 92 & 26 & $173\pm22$ 
\enddata
\tablecomments{
\tablenotetext{a}{K-M estimates that 25\% of all bursts will
have radio afterglows above this value.} 
\tablenotetext{b}{K-M estimates that 50\% of all bursts will
have radio afterglows above this value.} 
\tablenotetext{c}{K-M estimates that 75\% of all bursts will
have radio afterglows above this value.}
}
\end{deluxetable*}

\begin{figure}
\centering
\includegraphics[width=0.48\textwidth]{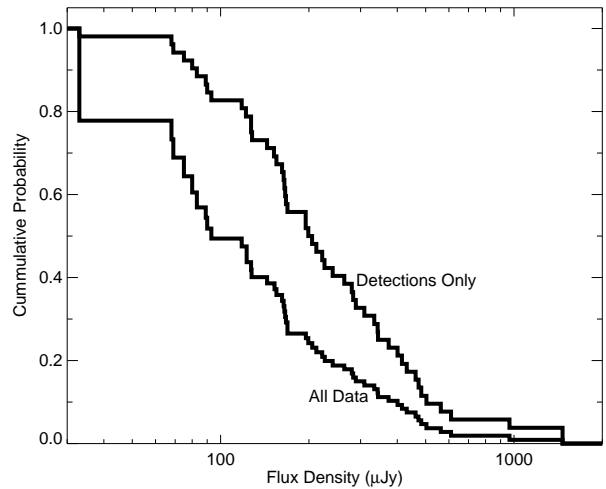}
\caption{The Kaplan-Meier probability distribution of the radio
  flux densities at 8.5 GHz for a complete sample of 107 GRBs toward
  which measurements or upper limits were made between 5 and 10 days. 
The distribution of the 51 detections is shown for
  comparison. }
\label{fig:kmplot}
\end{figure}

In Figure~\ref{fig:kmplot}, we plot the 
Kaplan--Meier probability distribution of the radio
  flux densities at 8.5 GHz for a complete sample of 107 GRBs toward
  which measurements or upper limits were made between 5 and 10 days. 
The distribution of the 51 detections is shown for
  comparison.
The overall detection rate in this time range is 48\% ({\em i.e.,} 
  51/107).
  The mean of the 51 detections is 275$\pm$34 $\mu$Jy while the mean
  of the entire sample (upper limits included) is 170$\pm$20 $\mu$Jy.
  Half of all GRBs produce radio afterglows with flux densities at 8.5
  GHz in excess of 92 $\mu$Jy.

\begin{figure}
\centering
\includegraphics[width=0.48\textwidth]{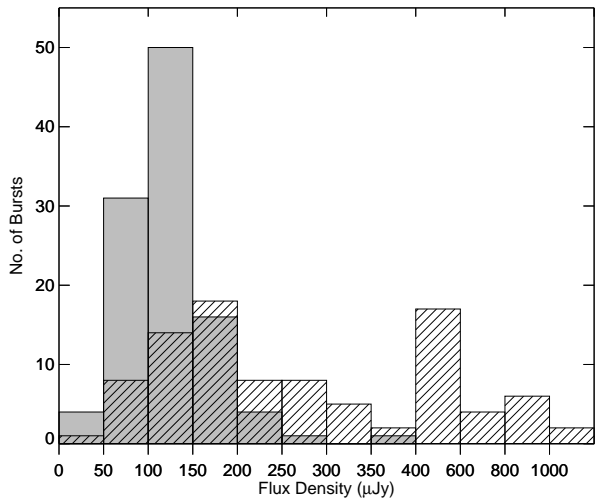}
\caption{The filled histogram represents the distribution for 
  the 8.5 GHz 3-$\sigma$ upper limits for GRBs in our sample between days 5--10.
  The hatched histogram shows the distribution for the detections
  between days 5--10. The small range of peak flux densities between
  the radio detections and non-detections suggests that the detection
  fraction of radio afterglows is largely determined by the
  instrumental sensitivities.}
\label{fig:grb-histro}
\end{figure}

Before making any detailed comparisons, it is important to understand
any possible biases that might exist in these data. Specifically, we
ask whether a radio non-detection is the result of an inadequate
measurement (i.e., not observing at a correct time or with enough
sensitivity), or whether it is the result of an underlying physical
cause.  
In Figure~\ref{fig:grb-histro}, we plot the distributions for the radio
3-sigma upper limits as well as for the detections between 5--10 days.
The upper limits peak in 100--150 $\mu$Jy
range, whereas the detections peak in 150--200 $\mu$Jy range with a
tail extending to 1 mJy. As first noted by \citet{frail05}, the
difference in the upper limits and the detections is not highly
significant (see also Table~\ref{tab:radio-peak}). There is only about
a factor of 50 difference between the radio flux density of the
brightest and the faintest detected cosmological bursts (i.e.
$z\ge0.4$). This low dynamic range stands in sharp contrast to the
orders of magnitude difference between the bright and faint events
seen in either X-ray or afterglow samples \citep{kkz06,ros+11}.  This
narrow flux density range suggests that the detection
fraction of our radio afterglow sample is determined by the instrumental
sensitivities.

\begin{figure}
\centering
\includegraphics[width=0.48\textwidth]{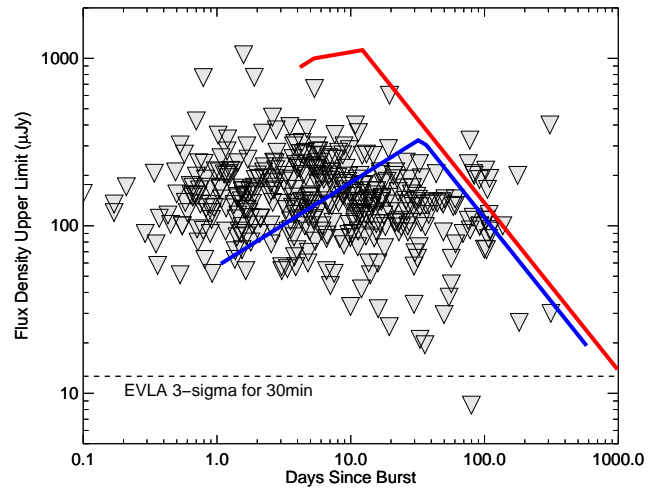}
\caption{Upper limits at 8.5 GHz frequency band for all GRBs for which 
  no afterglow was detected.  The red line represents the light curve
  of a rare, bright event GRB\,980703 and the blue line represents the light
  curve of a more typical event GRB\,980329. The detection fraction of
  radio afterglows in the first 10 days
certainly appears to be mainly limited by the sensitivity. Black dashed 
line indicates 3-sigma flux density of the EVLA in its full capacity for a
30 min integration time.}
\label{fig:flux-UL}
\end{figure}

In Figure~\ref{fig:flux-UL}, we plot all those GRBs which were never
detected in the radio band. We also overlay one rare, bright GRB light
curve GRB\,980703 (red curve) and a more typical GRB\,980329 (blue
curve). The plot shows that our upper limits do not cut off at a fixed
flux density but are tightly clustered within one order of magnitude
(between 30 $\mu$Jy to around 400 $\mu$Jy). As illustrated in
Figure~\ref{fig:flux-UL}, most of the observations would have been
capable of detecting a bright GRB\,980703-like afterglow, while up to
the first 10 days (when most of the afterglow observations take
place), only around 50\% observations would have been capable of
detecting an average event like GRB\,980329.  This result reinforces
our suggestion that radio afterglow searches are strongly sensitivity
limited. In this figure, we also plot the 3-sigma flux density of EVLA
for a 30min integration in its full capacity. 

\subsection{$k$-corrected radio spectral luminosity}\label{sec:kcorrected}

\begin{figure*}
  \centering \includegraphics[width=0.98\textwidth]{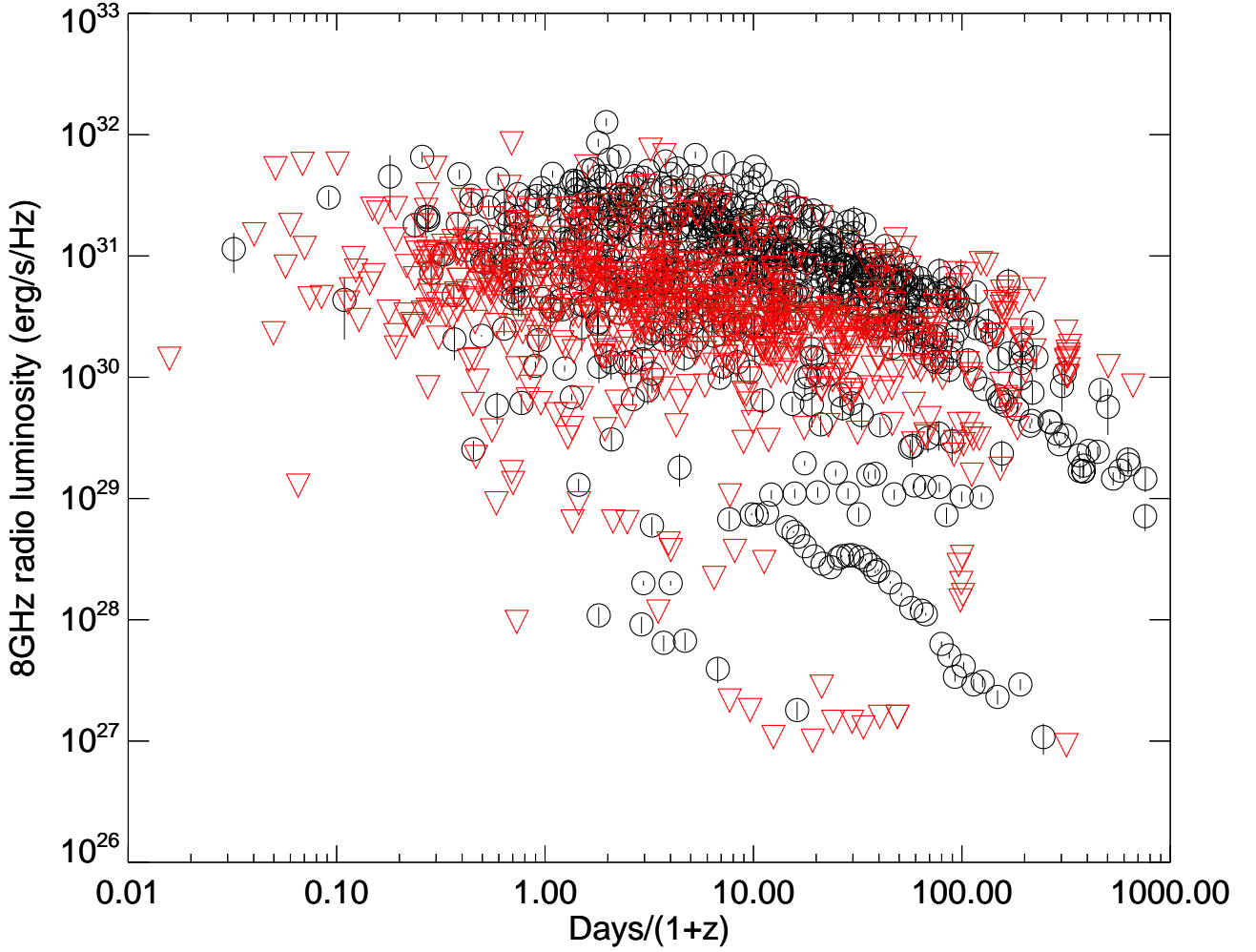}
\caption{The $k$-corrected 
  radio spectral luminosities at 8.5 GHz for radio-detected afterglows
  (black circles) versus non-detected 3-sigma luminosity upper limits
  (red triangles) with respect to the rest frame time. The
  luminosity curve for the average cosmological burst varies over a
  small range, but there are also a number of low-luminosity events.}
\label{fig:det-nondet}
\end{figure*}

In Figure~\ref{fig:det-nondet}, we plot the $k$-corrected radio spectral
luminosities  with the 3-sigma upper
limits. To convert the flux density into
spectral luminosity $L$, we use $L=4 \pi F d_L^2/(1+z)$, where $F$ is the
radio flux density and $d_L$ is the luminosity distance corresponding to the
redshift $z$. To incorporate $k$-correction, we multiply
this luminosity with a $k$-correction factor of $(1+z)^{\alpha-\beta}$
(where $\alpha$ and $\beta$ are the time and frequency indices in 
$F \propto t^\alpha \nu^\beta$). Thus the $k$-corrected radio
spectral luminosity is $L=4 \pi F d_L^2 (1+z)^{\alpha-\beta-1}$.  
Here we choose $\alpha=0$ and $\beta=1/3$
corresponding to an optically thin, flat, post-jet-break light curve
\citep{fck+06}. For the
GRBs with unknown redshift, we assume the average redshifts of
pre-\emph{Swift} and \emph{Swift} bursts, i.e. $\langle z \rangle_{pre}=1.3$ 
for the pre-\emph{Swift}
bursts and $\langle z \rangle_{post}=2.0$ for the post-\emph{Swift} bursts
(see \S\ref{sec:red}).  
In our sample we have 147 GRBs with known
redshifts, while 157 GRBs don't have a measured redshift.
This figure, similar to Figures~\ref{fig:grb-histro} and \ref{fig:flux-UL},
 also shows that for most of the sample, the
difference in the spectral luminosities between the detections and the
upper limits is not very significant.  The average luminosity for
detections is $1.1 \times 10^{31}$ erg s$^{-1}$ Hz$^{-1}$, whereas, the average
luminosity 3-$\sigma$ upper limit for non-detections is $6.4 \times 10^{30}$
erg s$^{-1}$ Hz$^{-1}$, i.e., the difference between them is not even a factor of
two. Figure~\ref{fig:det-nondet} does show, however, the existence of the
low-luminosity events which are several orders of magnitude fainter
than the normal cosmological events. We will discuss these populations
in more detail in \S\ref{sec:different}.

\subsection{Canonical light curves of radio afterglows}\label{sec:cannon}

\begin{figure}
\centering
\includegraphics[width=0.48\textwidth]{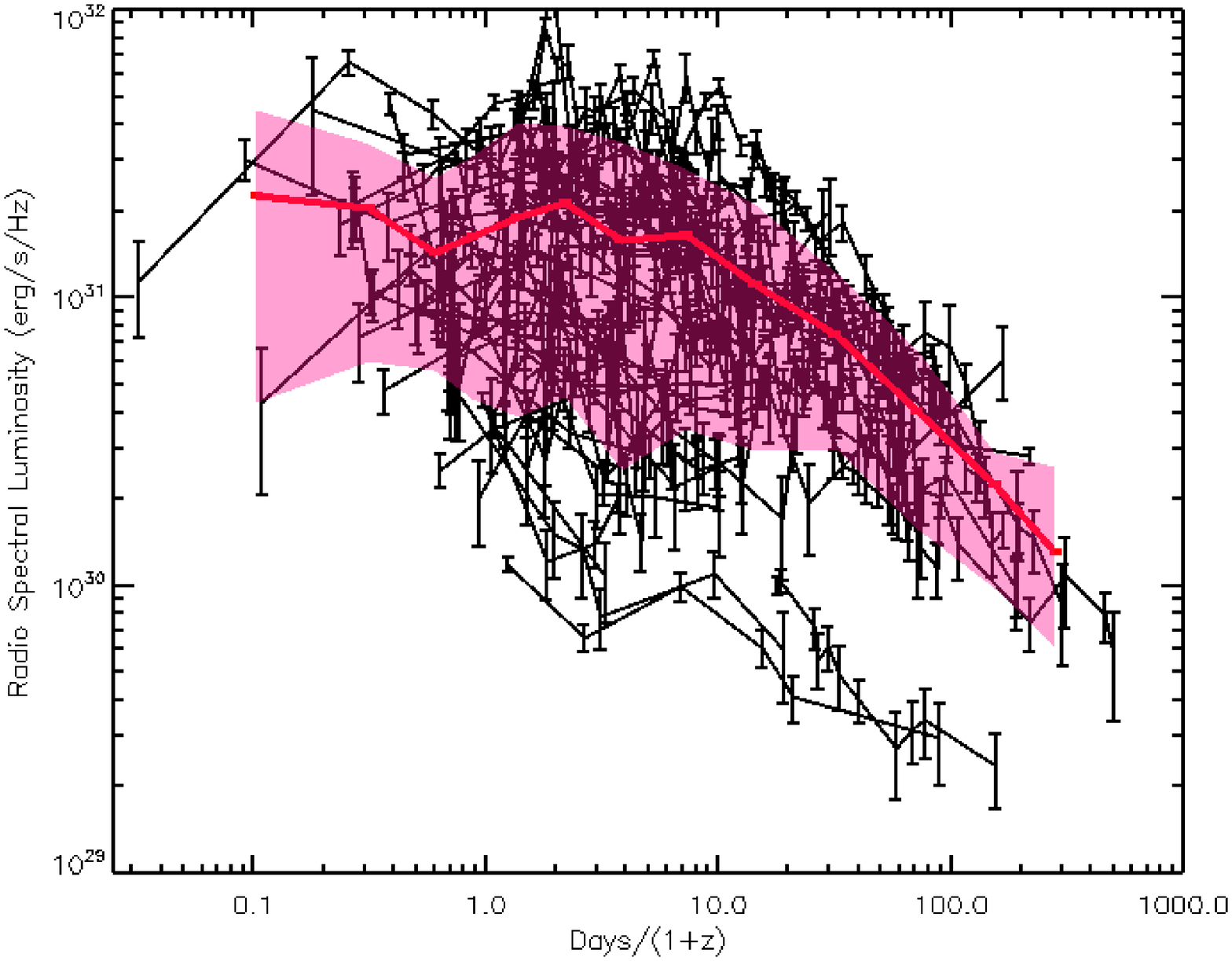}
\includegraphics[width=0.48\textwidth]{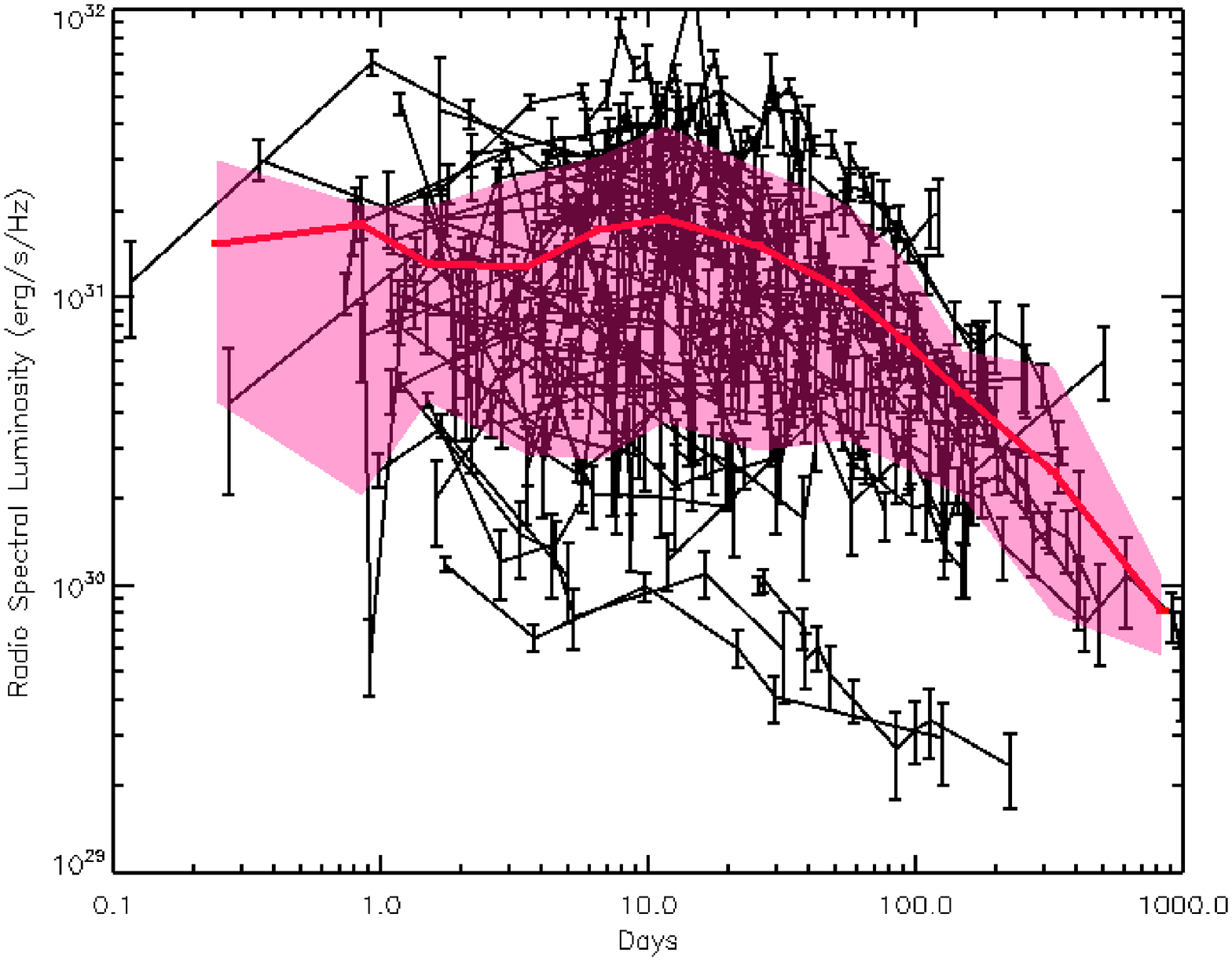}
\caption{{\it Upper panel:}
  The radio light curves at 8.5 GHz for the long-duration,
  cosmological GRBs in the rest frame time.  The red solid line
  represents the mean light curve. The pink shaded area represents the
  75\% confidence band. Lower panel shows the same plot for time in
  the observer's time frame.}
\label{fig:canonical}
\end{figure}

In Figure~\ref{fig:canonical}, we plot light curves of all the
detected radio afterglows at 8.5 GHz in their rest frames as well as
the observer frame, excluding the GRB sub-classes (SHB, XRF and
SNe/GRB) from which the low luminosity events in
Figure~\ref{fig:det-nondet} originate. We average these data to
produce a mean light curve for the long-duration, cosmological GRBs
(LGRB) along with the 75\% confidence bands. A canonical LGRB if
observed from beginning to end will have a mean luminosity of
$\sim2\times 10^{31}$ erg s$^{-1}$ Hz$^{-1}$ until about 3--6 d
(10--20 d) in the rest-frame (observer-frame). After this time there
is a gradual power-law decay with an index of approximately $-1$.  The
mean light curve shows two peaks (rest frame), one between day
0.1--0.2 and one around day 2, and a dip around day 1. The first peak
is not significant since there are too few data points. The peak after
day 2 is significant but it is likely the same peak as identified in
Figure~\ref{fig:Peaktime-GRB}. The dip near day 1 appears real and may
signify a transition between two different emission components
(reverse shock and forward shock).

\subsection{Bursts of different classes}\label{sec:different}

\begin{figure}
\centering
\includegraphics[width=0.48\textwidth]{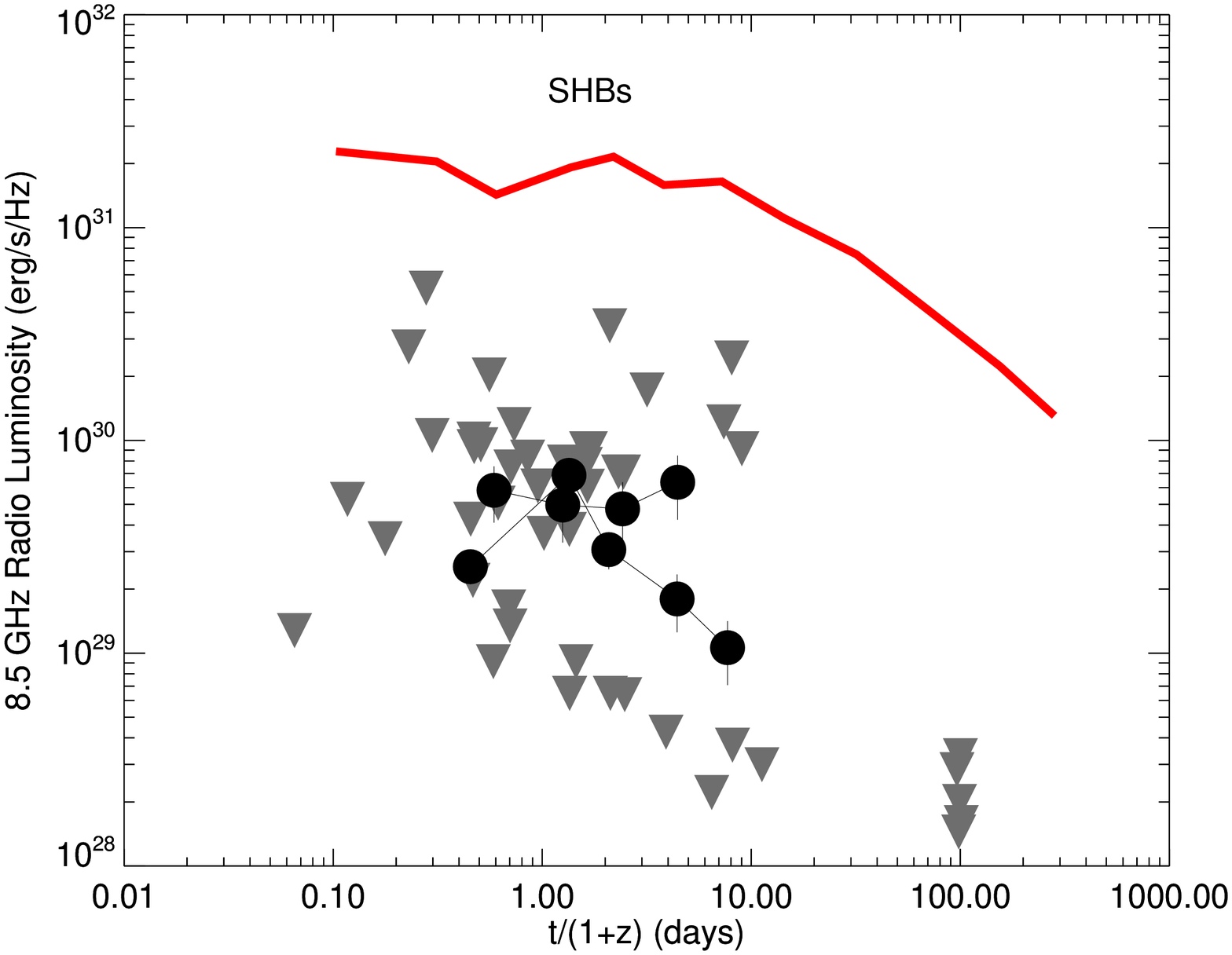}
\includegraphics[width=0.48\textwidth]{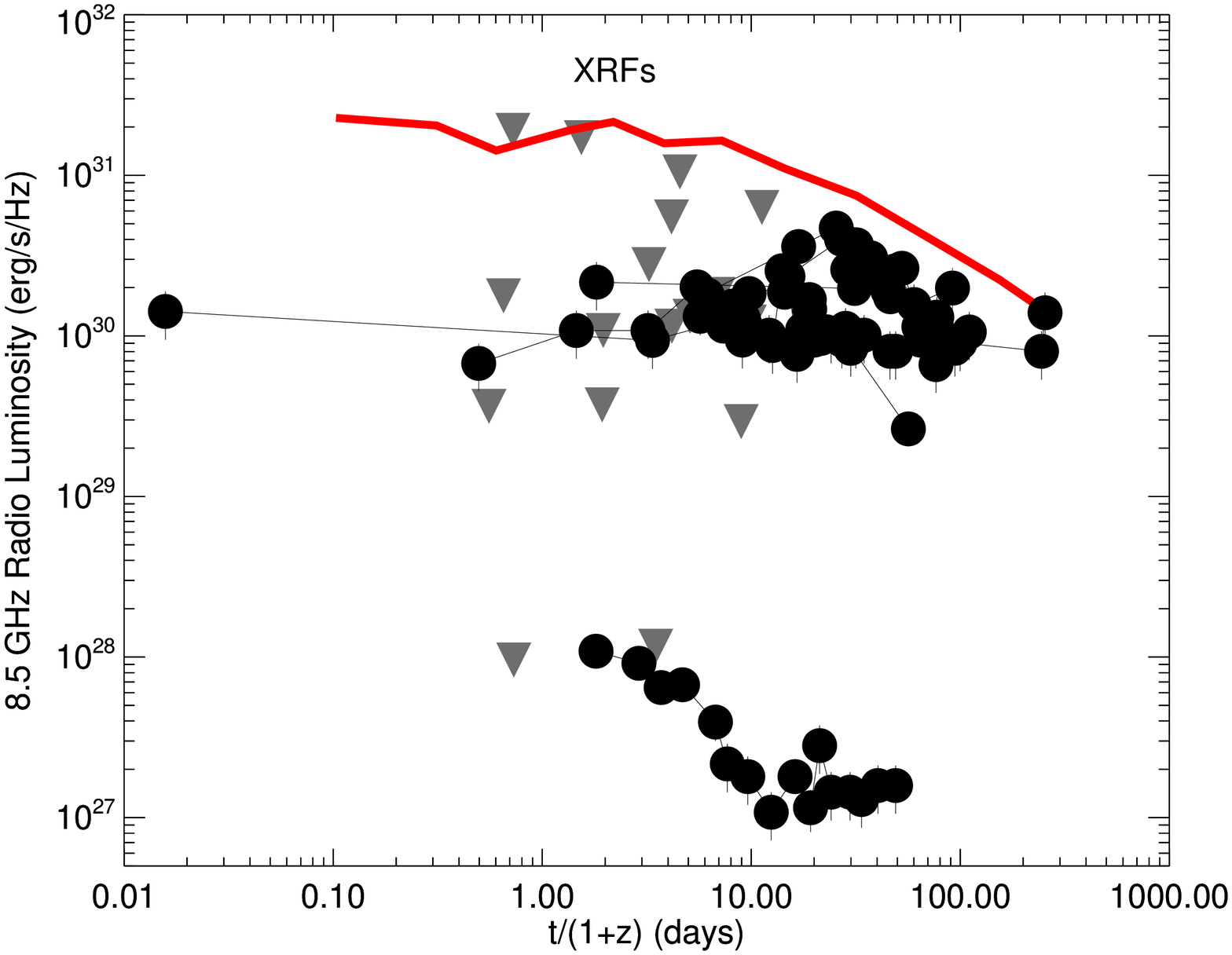}
\includegraphics[width=0.48\textwidth]{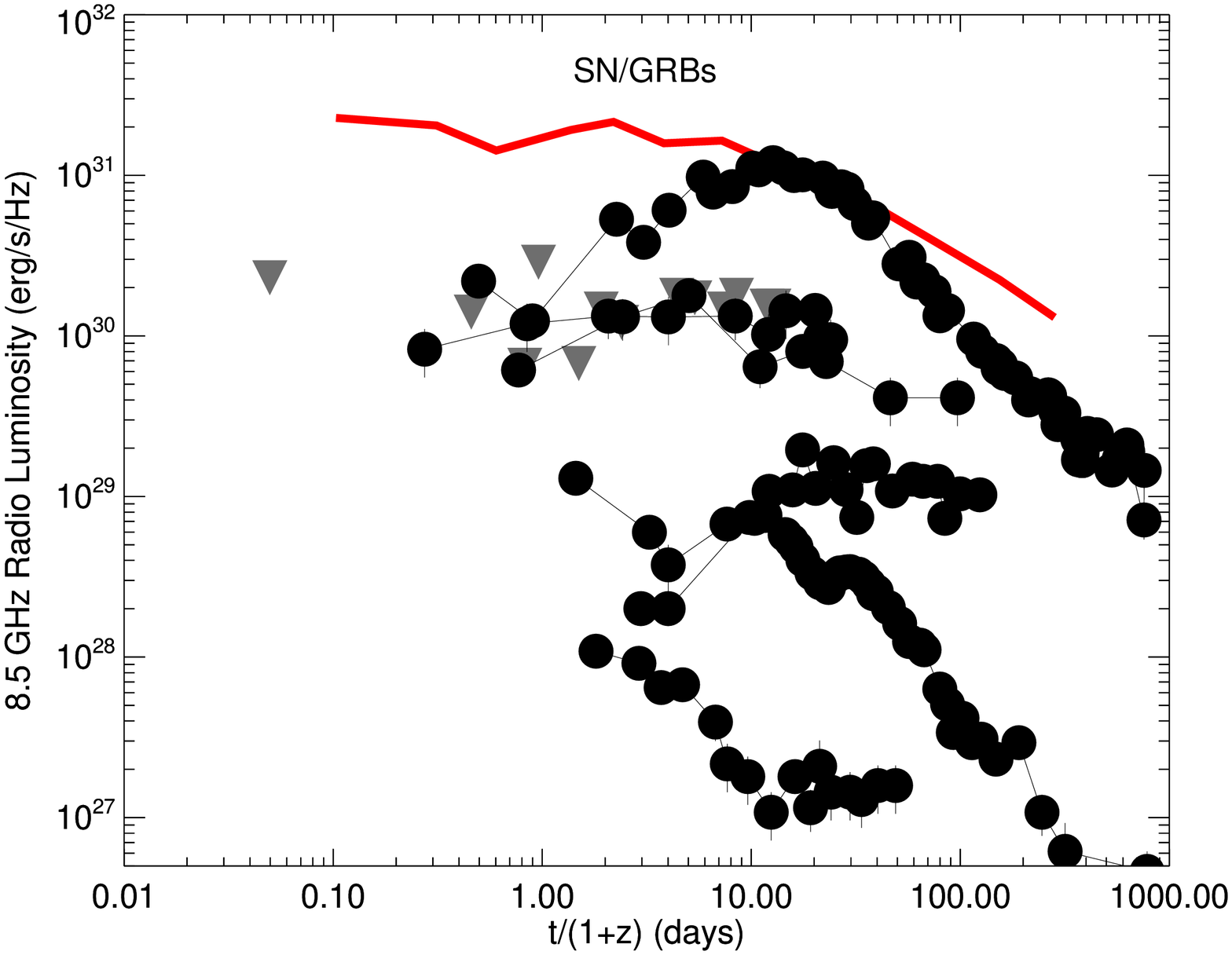}
\caption{A compilation of radio luminosity curves in the rest frame 
  for three different populations of GRBs: short, hard bursts (top),
  X-ray flashes (middle), and supernova-GRB associations (bottom).
Here black filled circles represent the detections and
the grey filled triangles represent the upper limits.
  The red solid line represents the mean light curve for the
  long-duration, cosmological sample from Figure~\ref{fig:canonical}.}
\label{fig:diff}
\end{figure}

We explore the differences in the luminosity light curves of SHBs,
XRFs and SN-GRBs.  In Figure~\ref{fig:diff}, we plot the radio
luminosity light curves for the different GRB classes in the 8.5 GHz
band. The top plot is the light curve for SHBs.
Our sample included 35 SHBs but there are only two detections
GRB\,050724 and GRB\,051221A.  For 13 SHBs the redshift was not known.
We averaged the redshifts of known-$z$ SHBs and assigned this value
($\langle z\rangle$=0.55) to the unknown-$z$ SHBs. Here we also
overlay the mean light curve of LGRBs obtained from the
Figure~\ref{fig:canonical}.  It is clear that SHBs are intrinsically
radio dim objects. They are more than an order of magnitude fainter
than a LGRB.  In the middle frame of Figure~\ref{fig:diff} we plot the
light curves of the 19 XRFs in our sample which includes seven
detections. Most XRFS are usually only about a factor of ten fainter
than LGRB (red curve). The extreme outlier is GRB\,060218, which is
4-5 orders of magnitude dimmer than a typical event. In the bottom
frame of Figure~\ref{fig:diff} we selected only those 9 GRBs for which
the known SN associations is exist.  The SNe/GRBs nearly fully
populate this radio luminosity plot. Some SNe/GRBs such as GRB\,980425
and GRB\,060218 are dim, while others like GRB\,030329 are as bright
as a typical LGRB.

%

\subsection{Peak of radio afterglow detections}\label{sec:radioafterglows}

Out of the 95 radio-detected afterglows (see \S\ref{venn}), 63 had
radio light curves (i.e. 3 or more detections in a single radio band),
whereas 32 bursts had less than 3 detections.  For the GRBs for which
the light curves were available, we determined the peak flux density
and the time of the peak at the VLA frequency bands 
(i.e. 1.4 GHz, 4.9 GHz, 8.5 GHz, 15 GHz and 22.5 GHz bands)
by fitting the forward shock formula of the
form \citep{frail05}.

\begin{equation}
 f(t) =
\begin{cases}
 F_m t_m^{-1/2} t^{1/2}, 
        &  t < t_m \\
F_m t_m t^{-1}, 
        &  t > t_m 
\end{cases}\label{eqn:simple}
\end{equation}

This formula may not accurately represent the full complexity of the
radio light curve evolution. However, it is good enough to determine
the approximate values for the peak flux density $F_m$ and the time of the peak $t_m$.
These fitted values of peak flux density and peak times are provided in 
Table~\ref{tab:radio-peak}.  All of the values quoted are in the observer's
frame.  In our sample there are 24 radio afterglows for which we
detect a small excess of radio emission at early times ($t<3$ days)
compared to the fit in Equation~\ref{eqn:simple}.  We suspect that this
excess may be the result of a separate reverse shock component. We will
investigate this hypothesis in a separate paper. To avoid a possible
reverse shock peak, we took the following precautions: If the light curve indicated
two peaks, one before 3 days and one after that, we fit the later
peak; If the best fit peak flux density was within first 3 days, we exclude it
as this could be the possible reverse shock emission.  The
early-peaked light curve of GRB\,060218 is included as the lone
exception since it has a very low redshift, and hence the peak before day 3 
could still come from the forward shock. 
For the afterglows which had
less than 3 detections, we chose the value of highest flux density.  In the
case of single detection point at a particular frequency, we simply chose that value. 
There were 3 afterglows in 22.5 GHz band, 4 afterglows in 15 GHz band, 
7 afterglows each in 8.5 and 4.9 GHz bands and 1 afterglow in 1.4 GHz band with
single detection points. The
second set of entries in Table~\ref{tab:radio-peak} (i.e. after the
horizontal line) represent the flux density values taken directly from the
data, and hence do not not have the best-fit errors.  Rest
frame peak times were obtained for the bursts with known redshift by
using $t_m/(1+z)$.  We do not estimate the rest frame peak time for
bursts with unknown redshifts.

\tabletypesize{\scriptsize}
\begin{longtable}{lrrrr}
\tablecaption{The radio peak flux densities and the time of peak  for radio afterglows
\label{tab:radio-peak}}
\tablewidth{0pt}
\tablehead{
\colhead{GRB} & \colhead{Freq.} & \colhead{Peak Flux ($F_{m}$)}  & \colhead{Peak time ($t_{m}$)}
& \colhead{$t_{m}/(1+z)$} \\
\colhead{} & \colhead{GHz} & \colhead{$\mu$Jy} & \colhead{days} & \colhead{days}
}
970508  &       1.43    &$      381   \pm     19    $&$     179.1   \pm     7.6     $&$
97.6    \pm     4.1     $\\
970508  &       4.86    &$      780   \pm     13    $&$     57.6    \pm     0.9     $&$
31.4    \pm     0.5     $\\
970508  &       8.46    &$      958   \pm     11      $&$     37.2    \pm     0.4     $&$
20.3    \pm     0.2     $\\
970828  &       8.46    &$      144   \pm     31    $&$     7.8     \pm     1.8     $&$
4.0     \pm     0.9     $\\
980329  &       4.86    &$      171   \pm     14      $&$     91.5    \pm     11.3    $&$
22.9    \pm     2.8     $\\
980329  &       8.46    &$      332   \pm     11    $&$     33.5    \pm     1.4     $&$
8.4     \pm     0.4     $\\
980425  &       1.38    &$      22260 \pm     195   $&$     47.1    \pm     0.5     $&$
46.7    \pm     0.5     $\\
980425  &       2.5     &$      31621 \pm     265   $&$     32.7    \pm     0.3     $&$
32.4    \pm     0.3     $\\
980425  &       4.8     &$      38362 \pm     337   $&$     18.3    \pm     0.1     $&$
18.1    \pm     0.1     $\\
980425  &       8.64    &$      39360 \pm     557   $&$     12.7    \pm     0.1     $&$
12.6    \pm     0.1     $\\
980519  &       4.86    &$      330   \pm     47    $&$     17.9    \pm     2.4     $&$
\ldots          $\\
980519  &       8.46    &$      205   \pm     23    $&$     12.6    \pm     1.3     $&$
\ldots          $\\
980703  &       1.43    &$      263   \pm     81    $&$     25.4    \pm     5.5     $&$
12.9    \pm     2.8     $\\
980703  &       4.86    &$      1055    \pm   30    $&$     9.1     \pm     0.2     $&$
4.6     \pm     0.1     $\\
980703  &       8.46    &$      1370  \pm     30    $&$     10      \pm     0.2     $&$
5.1     \pm     0.1     $\\
981226  &       8.46    &$      137   \pm     34    $&$     8.2     \pm     1.6     $&$
3.9     \pm     0.8  $\\
990510  &       8.46    &$      255   \pm     34    $&$     4.2     \pm     0.5     $&$
1.6     \pm     0.2     $\\
991208  &       1.43    &$      263   \pm     49    $&$     8.9     \pm     1.8     $&$
5.2     \pm     1.1     $\\
991208  &       4.86    &$      845   \pm     48      $&$     12.8    \pm     0.6     $&$
7.5     \pm     0.4     $\\
991208  &       8.46    &$      1804  \pm     24    $&$     7.8     \pm     0.1     $&$
4.6     \pm     0.1     $\\
000301C &       8.46    &$      520   \pm     24      $&$     14.1    \pm     0.5     $&$
4.6     \pm     0.2     $\\
000418  &       4.86    &$      897   \pm     39    $&$     27      \pm     1       $&$
12.7    \pm     0.5     $\\
000418  &       8.46    &$      1085    \pm     22    $&$     18.1    \pm     0.3     $&$
8.5     \pm     0.1     $\\
000911  &       8.46    &$      263   \pm     33    $&$     3.1     \pm     0.3     $&$
1.5     \pm     0.1     $\\
000926  &       4.86    &$      460   \pm     31    $&$     16.9    \pm     1.5     $&$
5.6     \pm     0.5     $\\
000926  &       8.46    &$      629   \pm     24    $&$     12.1    \pm     0.5     $&$
4.0     \pm     0.2     $\\
001018  &       8.46    &$      590   \pm     68    $&$     4.7     \pm     0.6     $&$
        \ldots  $\\
010222  &       8.46    &$      93    \pm     25    $&$     16.8    \pm     3.5     $&$
6.8     \pm     1.4     $\\
010921  &       4.86    &$      161   \pm     20    $&$     31.5    \pm     3.3     $&$
21.7    \pm     2.3     $\\
011030  &       4.86    &$      121   \pm     19    $&$     19.4    \pm     2.5     $&$
        \ldots  $\\
011030  &       8.46    &$      139   \pm     9     $&$     23.7    \pm     1.4     $&$
        \ldots  $\\
011121  &       8.7     &$      655   \pm     40      $&$     8.1     \pm     0.7     $&$
5.9     \pm     0.5     $\\
011211  &       8.46    &$      162   \pm     13    $&$     13.2    \pm     1.6     $&$
4.2     \pm     0.5     $\\
020405  &       8.46    &$      113   \pm     17    $&$     18.2    \pm     3.7     $&$
10.8    \pm     2.2     $\\
020819B &       8.46    &$      291   \pm     21    $&$     12.2    \pm     1.1     $&$
8.7     \pm     0.8     $\\
020903  &       4.86    &$      782   \pm     28    $&$     36.7    \pm     1.3     $&$
29.4    \pm     1.0     $\\
021004  &       4.86    &$      470   \pm     26    $&$     32.2    \pm     1.4     $&$
9.7     \pm     0.4     $\\
021004  &       8.46    &$      780     \pm   23    $&$     18.7    \pm     0.5     $&$
5.6     \pm     0.2     $\\
021004  &       15      &$      1308    \pm     260     $&$     4.1     \pm     0.8     $&$
1.2     \pm     0.2     $\\
021004  &       22.5    &$      1614  \pm     52    $&$     8.7     \pm     0.3     $&$
2.6     \pm     0.1     $\\
021206  &       4.86    &$      480   \pm     69    $&$     7       \pm     0.8     $&$
\ldots  $\\
030226  &       8.46    &$      171   \pm     23      $&$     6.7     \pm     1       $&$
2.2     \pm     0.3     $\\
030329  &       1.43    &$      2232  \pm     30    $&$     78.6    \pm     2.1     $&$
67.3    \pm     1.8     $\\
030329  &       4.86    &$      10337 \pm     33    $&$     32.9    \pm     0.1     $&$
28.2    \pm     0.1     $\\
030329  &       8.46    &$      19567   \pm   28    $&$     17.3    \pm     0.1     $&$
14.8    \pm     0.1     $\\
030329  &       15      &$      34042 \pm     84    $&$     10.9    \pm     0.1     $&$
9.3     \pm     0.1     $\\
030329  &       22.5    &$      49596 \pm     66    $&$     8.4     \pm     0.1     $&$
7.2     \pm     0.1     $\\
030329  &       43      &$      59318 \pm     177     $&$     5.8     \pm     0.1     $&$
5.0     \pm     0.1     $\\
030723  &       8.46    &$      204   \pm     40    $&$     75      \pm     10.6    $&$
\ldots  $\\
031203  &       1.43    &$      929   \pm     60    $&$     65.5    \pm     5.3     $&$
59.3    \pm     4.8     $\\
031203  &       4.86    &$      828   \pm     28    $&$     58.4    \pm     2.1     $&$
52.9    \pm     1.9     $\\
031203  &       8.46    &$      724   \pm     19      $&$     48      \pm     1.3     $&$
43.4    \pm     1.2     $\\
041219A  &       4.86    &$      473   \pm     28    $&$     6.3     \pm     0.9     $&$
\ldots  $\\
050416A &       4.86    &$      485   \pm     36    $&$     48.5    \pm     3.5     $&$
29.4    \pm     2.1     $\\
050416A &       8.46    &$      373   \pm     36    $&$     49      \pm     3.6     $&$
29.7    \pm     2.2     $\\
050509C &       8.46    &$      344   \pm     45    $&$     44.1    \pm     5.7     $&$
\ldots  $\\
050603  &       8.46    &$      377   \pm     53      $&$     14.1    \pm     1.8     $&$
3.7     \pm     0.5     $\\
050713B &       4.86    &$      424   \pm     34    $&$     21      \pm     2.2     $&$
\ldots  $\\
050713B &       8.46    &$      343     \pm     24    $&$     14.1    \pm     0.9     $&$
\ldots  $\\
050820A &       8.46    &$      150   \pm     31    $&$     9.4     \pm     1.9     $&$
2.6     \pm     0.5     $\\
050904  &       8.46    &$      76      \pm     14      $&$     35.3    \pm     1.5     $&$
4.8     \pm     0.2     $\\
051022  &       8.46    &$      268   \pm     32    $&$     5.2     \pm     0.7     $&$
2.9     \pm     0.4     $\\
060218  &       4.86    &$      245   \pm     50    $&$     3.8     \pm     0.7     $&$
3.7     \pm     0.7     $\\
060218  &       8.46    &$      471   \pm     83    $&$     2       \pm     0.3     $&$
1.9     \pm     0.3     $\\
070125  &       8.46    &$      1028  \pm     16    $&$     84.1    \pm     2       $&$
33.0    \pm     0.8     $\\
070125  &       15      &$      1250  \pm     81    $&$     18.6    \pm     1.4     $&$
7.3     \pm     0.5     $\\
070125  &       22.5    &$      1778  \pm     75    $&$     13.6    \pm     0.5     $&$
5.3     \pm     0.2     $\\
070612A &       4.86    &$      580   \pm     20    $&$     140.3   \pm     5.8     $&$
86.8    \pm     3.6     $\\
070612A &       8.46    &$      1028  \pm     16    $&$     84.1    \pm     2       $&$
52.0    \pm     1.2     $\\
071003  &       8.46    &$      616   \pm     57    $&$     6.5     \pm     0.5     $&$
2.5     \pm     0.2     $\\
071010B &       4.86    &$      227   \pm     114   $&$     12.5    \pm     4.4     $&$
6.4     \pm     2.3     $\\
071010B &       8.46    &$      341   \pm     41    $&$     4.2     \pm     0.5     $&$
2.2     \pm     0.3     $\\
080603A &       8.46    &$      207   \pm     26      $&$     5.2     \pm     1.8     $&$
1.9     \pm     0.7     $\\
081221  &       8.46    &$      174   \pm     27    $&$     8.9     \pm     1.7     $&$
\ldots  $\\
090313  &       8.46    &$      435   \pm     22      $&$     9.4     \pm     0.5     $&$
2.1     \pm     0.1     $\\
090323  &       8.46    &$      243   \pm     13      $&$     15.6    \pm     1       $&$
3.4     \pm     0.2     $\\
090328  &       8.46    &$      686   \pm     26    $&$     16.1    \pm     0.7     $&$
9.3     \pm     0.4     $\\
090423  &       8.46    &$      50    \pm     11      $&$     33.1    \pm     6.3     $&$
3.6     \pm     0.7     $\\
090424  &       8.46    &$      236   \pm     37   $&$     5.2     \pm     0.7     $&$
3.4     \pm     0.5     $\\
090715B &       8.46    &$      191   \pm     36   $&$     9.2     \pm     2.3     $&$
2.3     \pm     0.6     $\\
090902B &       8.46    &$      84    \pm     16   $&$     14.1    \pm     2.6     $&$
4.9     \pm     0.9     $\\
091020  &       8.46    &$      399   \pm     21   $&$     10.9    \pm     0.6     $&$
4.0     \pm     0.2     $\\
100414A &       8.46    &$      524   \pm     19   $&$     8       \pm     0.3     $&$
3.4     \pm     0.1     $\\
100418A &       4.86    &$      522   \pm     83   $&$     70.3    \pm     8.1     $&$
43.4    \pm     5.0     $\\
100418A &       8.46    &$      1218  \pm     12   $&$     47.6    \pm     0.6     $&$
29.4    \pm     0.4     $\\
100814A &       4.5     &$      496   \pm     24   $&$     13      \pm     1       $&$
5.3     \pm     0.4  $\\
100814A &       7.9     &$      613   \pm     23   $&$     10.4    \pm     0.4     $&$
4.3     \pm     0.2  $\\
\tableline
 & & & & \\
990510  &       4.86    &       177     &       9.2     &       3.5     \\
991208  &       15      &       3100    &       5.4     &       3.2     \\
991216  &       4.86    &       126     &       17.4    &       8.6     \\
991216  &       15      &       1100    &       1.3     &       0.6     \\
000131  &       8.7     &       207     &       10.8    &       2.0     \\
000210  &       8.46    &       93      &       8.6     &       4.6     \\
000301C &       4.86    &       240     &       4.3     &       1.4     \\
000301C &       15      &       660     &       3.6     &       1.2     \\
000301C &       22.5    &       884     &       4.3     &       1.4     \\
000418  &       15      &       1350    &       12.3    &       5.8     \\
000418  &       22.5    &       1100    &       14.6    &       6.9     \\
000911  &       4.86    &       71      &       11.1    &       5.4     \\
000926  &       15      &       820     &       4.7     &       1.5     \\
000926  &       22.5    &       1415    &       7.2     &       2.4     \\
001007  &       8.46    &       222     &       7.1     &       \ldots  \\
010222  &       4.86    &       144     &       2.3     &       0.9     \\
010921  &       22.5    &       330     &       25.9    &       17.9    \\
010921  &       8.46    &       229     &       27      &       18.6    \\
011121  &       4.8     &       510     &       6.9     &       5.1     \\
020305  &       8.46    &       76      &       22.9    &       \ldots  \\
020903  &       1.43    &       294     &       91      &       72.8    \\
020903  &       8.46    &       1058    &       23.8    &       19.0    \\
021206  &       8.46    &       137     &       68.6    &       \ldots  \\
030115A &       4.9     &       89      &       6       &       1.7     \\
030115A &       8.46    &       83      &       7.3     &       2.1     \\
030226  &       22.5    &       328     &       74      &       24.8    \\
030227  &       8.46    &       64      &       6.6     &       \ldots  \\
030323  &       22.5    &       530     &       3.5     &       0.8     \\
030418  &       8.46    &       69      &       27.7    &       \ldots  \\
030723  &       4.86    &       217     &       77.9    &       \ldots  \\
031203  &       22.5    &       483     &       13.5    &       12.2    \\
050401  &       8.46    &       122     &       5.7     &       1.5     \\
050525A &       8.46    &       164     &       13.5    &       8.4     \\
050603  &       4.86    &       196     &       6.5     &       1.7     \\
050730  &       8.46    &       212     &       9.1     &       1.8     \\
050824  &       8.46    &       152     &       7.4     &       4.0     \\
051022  &       4.86    &       209     &       57      &       31.5    \\
060116  &       8.46    &       363     &       28      &       \ldots  \\
060218  &       1.43    &       198     &       4.9     &       4.7     \\
060218  &       22.5    &       250     &       3       &       2.9     \\
060418  &       8.46    &       216     &       4.3     &       1.7     \\
070125  &       4.86    &       308     &       27.9    &       11.0    \\
071003  &       4.86    &       224     &       3.8     &       1.5     \\
071020  &       8.46    &       141     &       3.2     &       1.0     \\
081203B &       8.46    &       162     &       7       &       \ldots  \\
081221  &       4.86    &       142     &       14.4    &       \ldots  \\
090313  &       4.86    &       361     &       7.9     &       1.8     \\
100413A &       8.46    &       80      &       8.8     &       \ldots  \\
100414A &       4.86    &       420     &       38      &       16.0    \\
100901A &       4.5     &       331     &       4.9     &       2.0  \\
100901A &       7.9     &       440     &       4.9     &       2.0  \\
100901A &       33.6    &       378     &       11.9    &       4.4  \\
100906A &       8.46    &       215     &       5       &       1.8  \\
\end{longtable}

\begin{figure}
  \centering
  \includegraphics[width=0.48\textwidth]{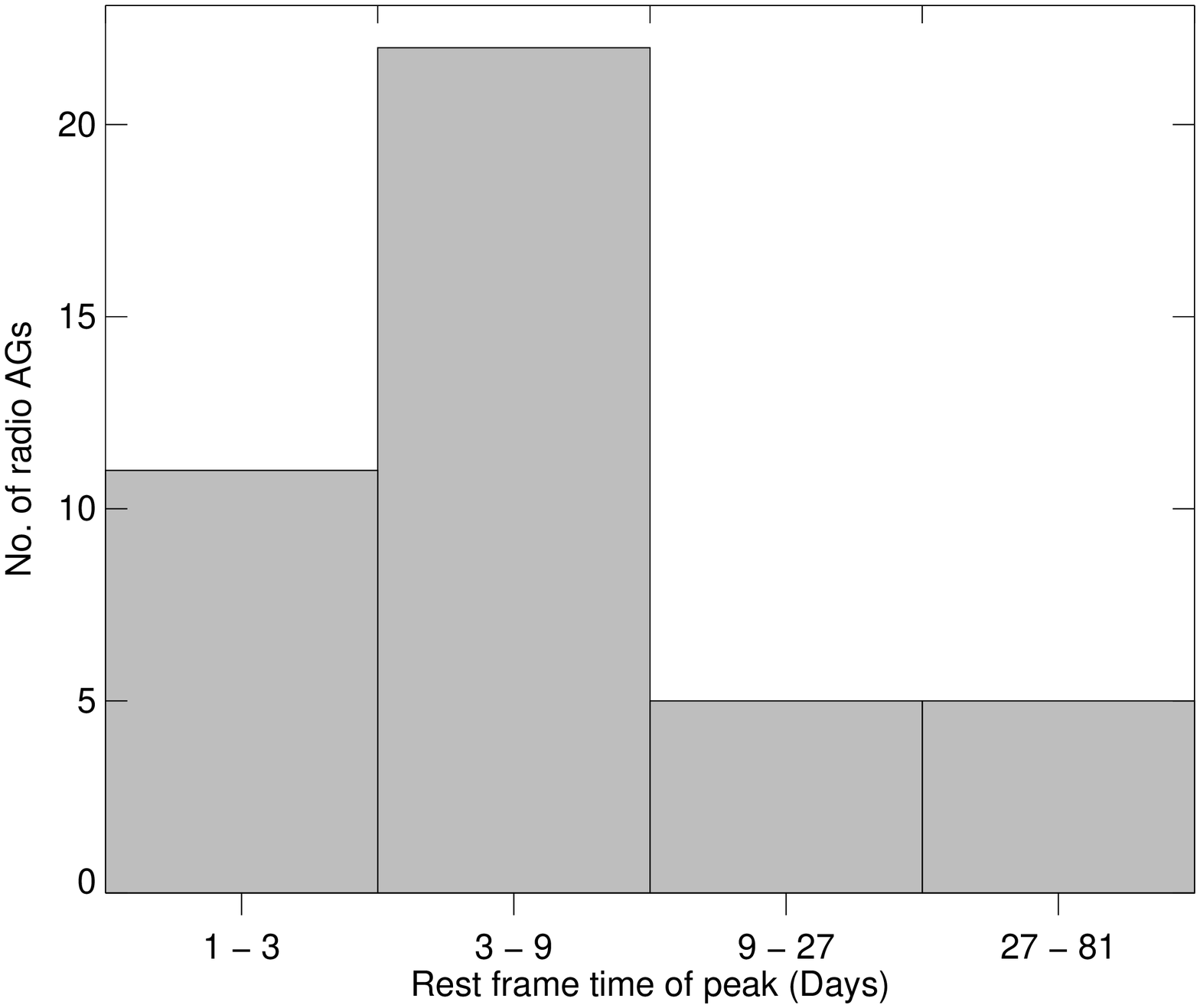}
  \includegraphics[width=0.48\textwidth]{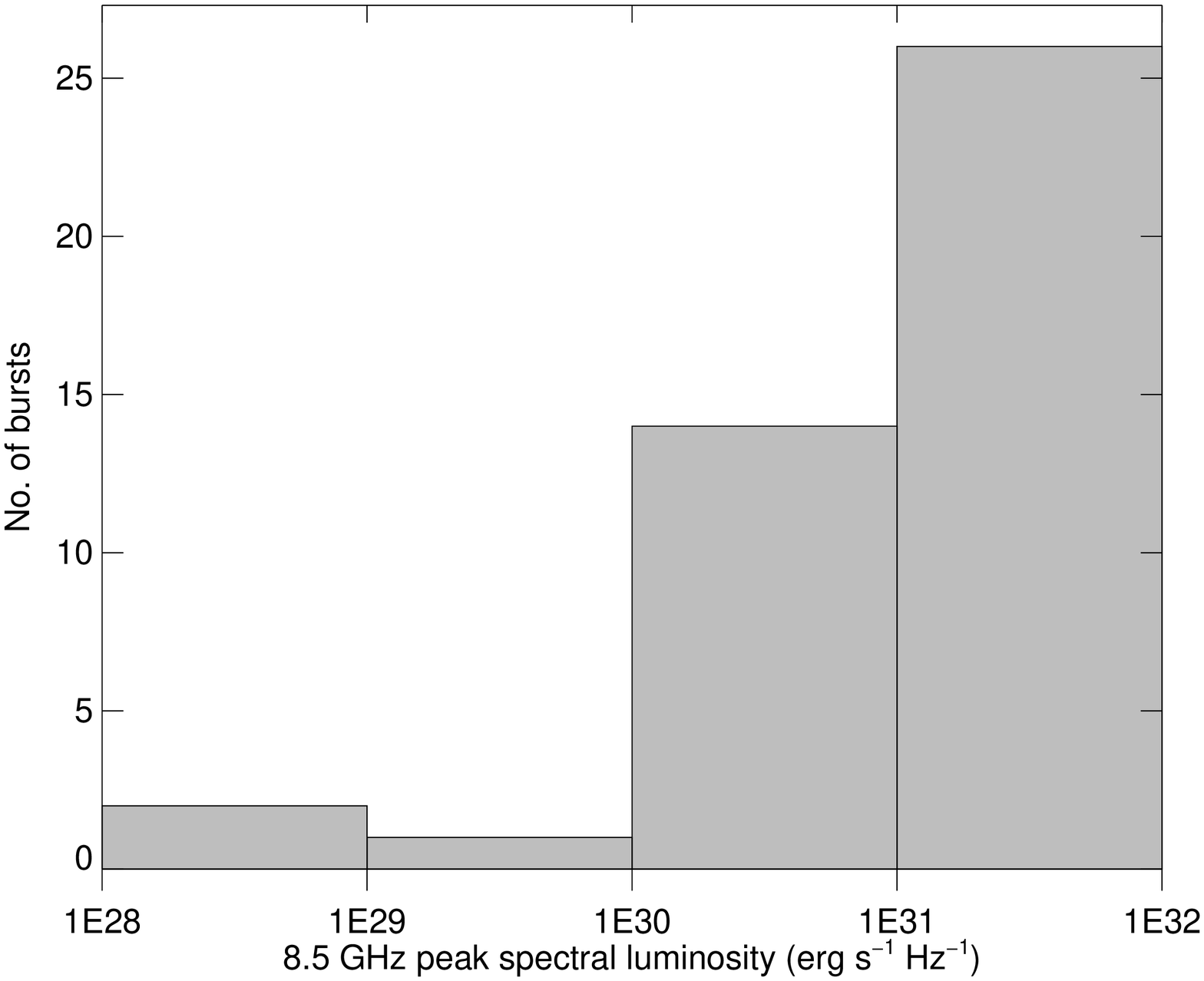}
\caption{{\bf Upper panel:} 
  Histogram for the time-to-peak distribution in the rest frame for
  radio afterglows at 8.5 GHz. Most of the afterglows peak between
  $3-6$ days.  {\bf Lower panel:} Distribution of peak radio
  luminosity at 8.5 GHz.  Most of the bursts peak in the range
  $10^{31}-10^{32}$ erg s$^{-1}$ Hz$^{-1}$.}
\label{fig:Peaktime-GRB}
\end{figure}

In Figure~\ref{fig:Peaktime-GRB}, we plot the histogram for rest frame
peak times at 8.5 GHz band.  It is clear that most of the radio afterglows peak
between $3-9$ days in the rest frame. This could be due to the 
$\nu_m$ passing the radio frequencies around this time.
Another possibility is that jet breaks in GRBs occur mostly around this time
(see table \ref{MasterTable} and references therein) and radio stops rising after the occurrence of
jet break.
In the Figure~\ref{fig:Peaktime-GRB}, we also plot the peak luminosity distribution of all
the detected afterglows at all epochs in 8.5 GHz band. The luminosity
distribution peaks in the range $10^{31}-10^{32}$ erg s$^{-1}$ Hz$^{-1}$. 

\subsection{1.4 GHz analysis}
\label{sec:lband}

While there is much less data at 1.4 GHz than at 8.5 GHz, obtaining
the afterglow properties at this frequency is still useful. There are
a number of wide-field imaging instruments coming on line later 
this decade which are optimized to work at this frequency. This
includes telescopes such as Australian Square Kilometre Array
Pathfinder (ASKAP), the phased array feed Apertif on the Westerbork
Synthesis Radio Telescope (WSRT), and the South African MeerKAT array.

A total of 310 measurements were made at this frequency for 55 GRBs.
Out of 310 measurements, 125 resulted in detections corresponding to
12 GRBs. Thus in terms of number of GRBs detected, the detection rate is 
22\%, while the detection rate in terms of measurements made
is 40\%. However, this detection rate is
likely an overestimate since our 1.4 GHz sample is strongly biased.
Observations at this frequency were not taken unless a bright radio
afterglow had been detected at 8.5 GHz.

\begin{figure}
\centering
\includegraphics[width=0.48\textwidth]{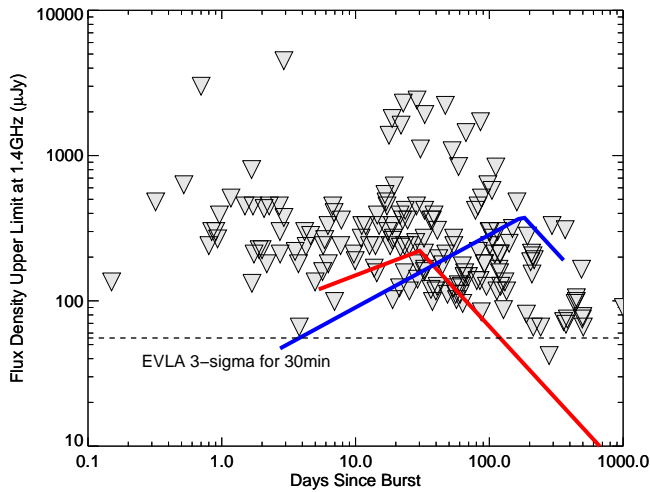}
\caption{3-$\sigma$ upper limits at 1.4 GHz frequency 
  band for all GRBs for which no afterglow was detected.  The red line
  represent the light curve of event GRB 980703 and the blue line
  represents the light curve of event GRB 970508. We also plot the 1.4 GHz EVLA
3-$\sigma$ upper limit for a 30 minute integration time.}
\label{fig:Lflux-UL}
\end{figure}

In Figure~\ref{fig:Lflux-UL}, we plot our non-detections as 3-$\sigma$
upper limits with respect to the epoch since explosion.  We also
overlay 1.4 GHz light curves of GRB 980703 (red curve) and GRB 970508
(blue curve). The plot shows that almost all the GRB upper limits are
tightly clustered within one to two orders of magnitude (between 60 $\mu$Jy
to around 3000 $\mu$Jy).  As illustrated in Figure~\ref{fig:Lflux-UL},
most of the observations would not have been capable of detecting
afterglows from these representative GRBs. This figure suggests that
radio searches are severely sensitivity limited at 1.4 GHz.  With
better sensitivity or more telescope time (see 
Figure~\ref{fig:Lflux-UL}), some improvement is expected detectable but as we
show in \S\ref{sec:synthetic}, the optimal search strategy is to use
higher observing frequencies.

\begin{figure}
  \centering \includegraphics[width=0.48\textwidth]{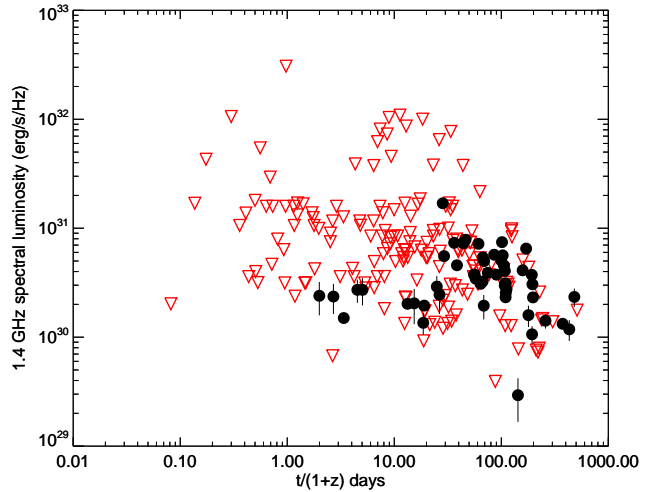}
\caption{The detected $k$-corrected 
  radio spectral luminosities at 1.4 GHz band (black circles) versus
  non-detected 3-sigma luminosity upper limits (red triangles) with
  respect to the rest frame time since burst. The luminosity curve for
  the average cosmological burst is tightly clustered between $10^{30}-10^{31}$
erg s$^{-1}$ Hz$^{-1}$.}
\label{fig:Ldet-nondet}
\end{figure}

In Figure~\ref{fig:Ldet-nondet}, we plot the $k$-corrected 1.4 GHz radio
spectral luminosities for the LGRBs and compare them
with the 3-sigma upper limits for the non-detections.  The figure
again clearly shows that no difference in the spectral luminosities
between the detections and the upper limits, reinforcing the severe
sensitivity limitation in resulting in various non-detections. The
mean 1.4 GHz luminosity for the detected bursts is $2\times 10^{30}$
erg s$^{-1}$ Hz$^{-1}$, and the mean 3-$\sigma$ upper limit for the non-detections
is $10^{31}$ erg s$^{-1}$ Hz$^{-1}$. Due to our strong selection bias these numbers
likely overestimate the luminosity of a typical GRB afterglow at this
frequency.

\section{Parameter Distributions}\label{sec:multi}


\subsection{Redshift distribution}\label{sec:red}

\begin{figure}
  \centering \includegraphics[width=0.48\textwidth]{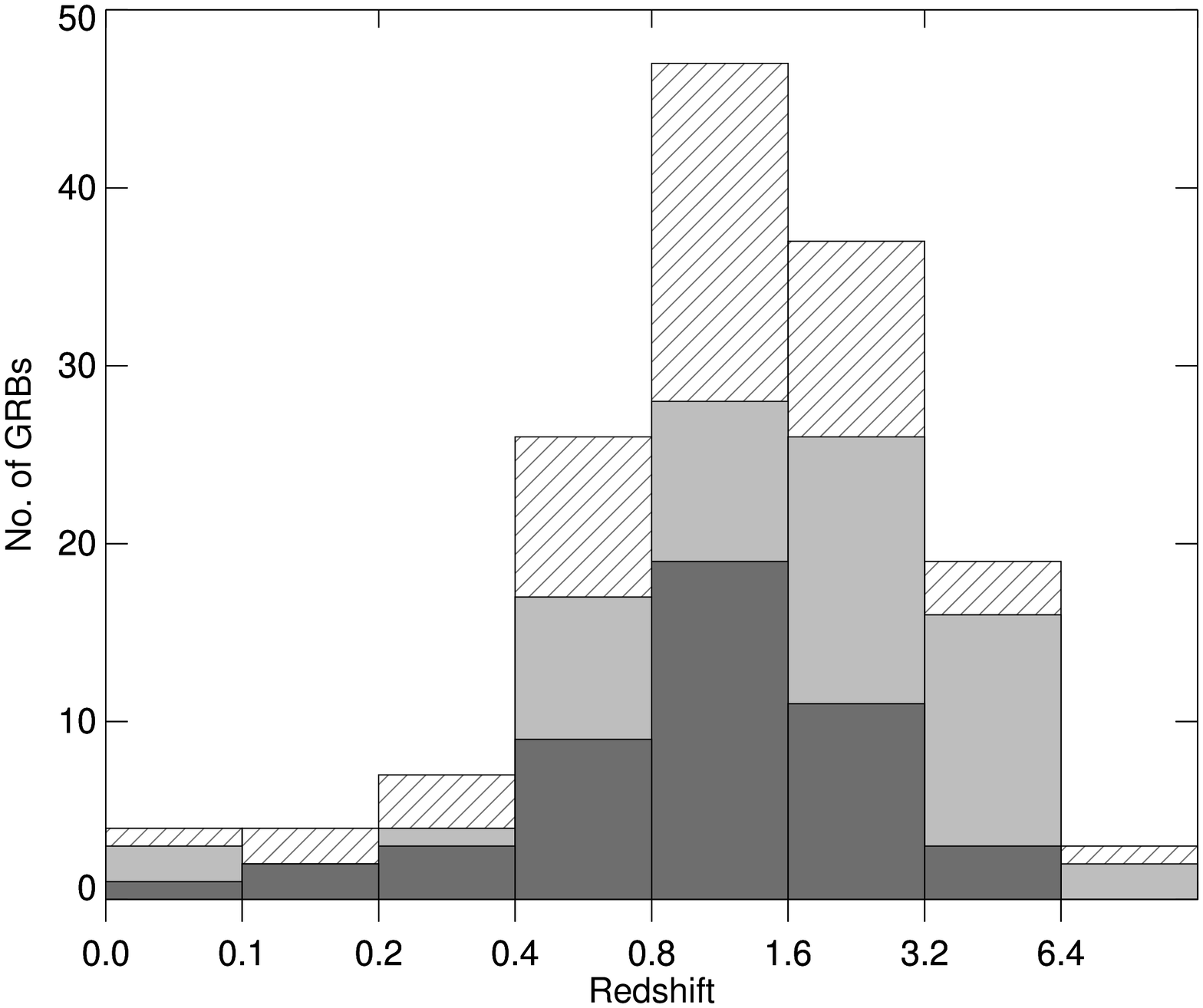}
\includegraphics[width=0.48\textwidth]{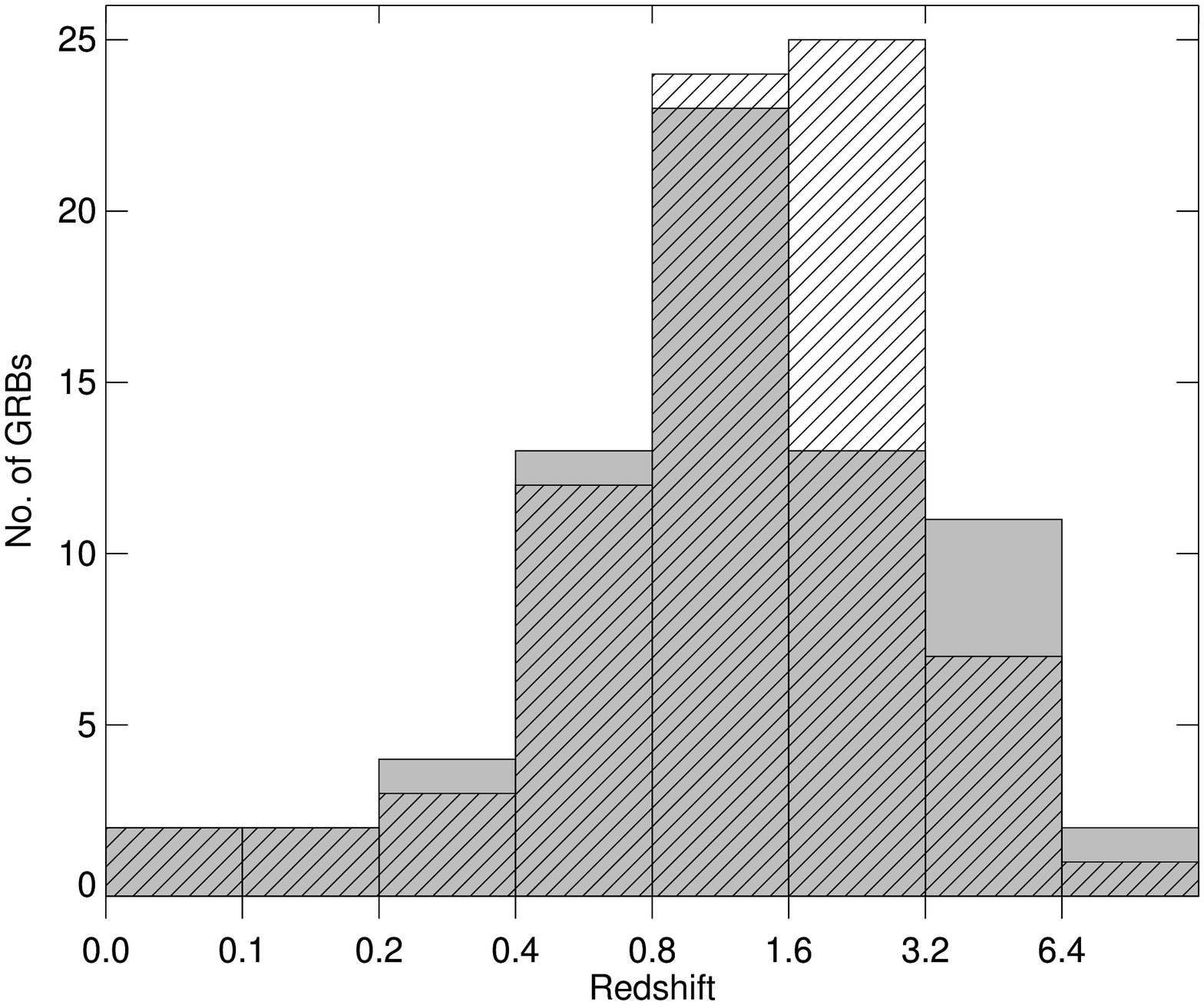}
\caption{{\bf Upper panel:} Redshift distributions for the post-\emph{Swift} (light 
  gray histogram) and the pre-\emph{Swift} bursts (dark gray histogram). 
  The average redshift of our sample is $z_{av}=1.8$. The average
  redshifts in the pre- and post-\emph{Swift} bursts are
  $\langle z \rangle_{pre}=1.3$ and $\langle z \rangle_{post}=2.0$, respectively.
The redshift distribution of the complete sample
is shown with the hatched histogram. 
{\bf Lower panel:} The redshift distribution of the
detected (hatched histogram) versus non-detected (filled histogram)
radio afterglows for the complete sample.
}
\label{fig:z}
\end{figure}

In our radio-selected sample of 304 GRBs we have 147 GRBs with known
redshifts.  This includes two bursts for which only a redshift range
is known (GRB\,980326 with $z=0.9-1.1$ and GRB\,980329 with $z=2-3.9$.
We have 99 GRBs with redshifts available in the post-\emph{Swift}
epoch as compared to just 48 bursts with known redshifts in
pre-\emph{Swift} epoch. The remaining 157 bursts had no redshift
estimates, although 9 of them do have upper limits. In our sample, 59
bursts had $z\le1$ and 88 bursts had $z>1$.  GRB\,090423 is the object
with highest confirmed spectroscopic redshift of $z=8.26$
\citep{tfl+09,sal09}, whereas the highest photometric redshift in our sample
($z=$9.4) is for GRB\,090429B \citep{clf+11}. The highest redshift in
pre-\emph{Swift} sample is 4.5 for GRB\,000131 \citep{anderson00}.  GRB\,980425 is the
object with the lowest redshift with $z=0.0085$ \citep{tinney98}.  In
Figure~\ref{fig:z}, we plot the redshift distribution of our sample.
We plot the pre-\emph{Swift} and the post-\emph{Swift} bursts for
comparison. The figures indicate that the post-\emph{Swift} bursts are
mostly found in the redshift range of $0.8-6.4$ (71 bursts out of 99), whereas the maximum
number of pre-\emph{Swift} bursts lies in the redshift range of
$0.4-3.2$ (39 bursts out of 44).  Here we also plot the redshift distribution of
radio-detected and non-detected sample. Most of the non-detected radio
afterglows lie in range $z=$0.8--1.6 and the detected radio afterglows lie
between redshift of 0.8--3.2. There is obviously no correlation between
the redshift of the burst and its detectability in the radio band.
Further support to the fact that the redshifts of the sample of radio
detected GRBs and non-detected GRBs are not from different populations
come from the Kolmogorov--Smirnov (K--S) test \citep{press}. 
Comparing the redshifts of the two sample gives $P=0.61$, strengthening the
fact that the two samples are not from the different populations.

\begin{figure}
  \centering
  \includegraphics[width=0.48\textwidth]{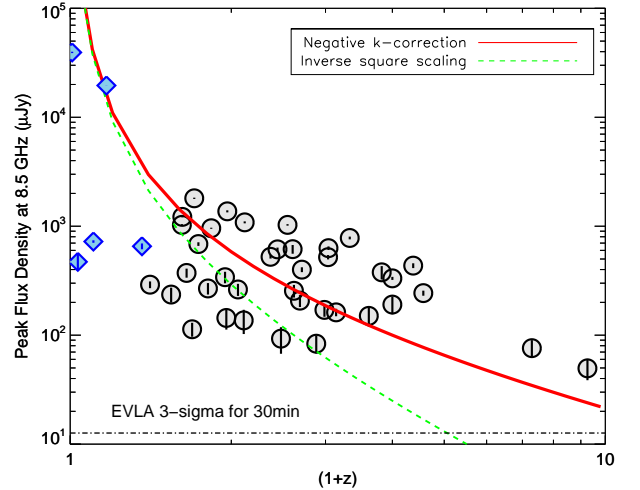}
\caption{Radio peak flux density versus ($1+z$) plot for radio afterglows 
  with known redshifts.  The radio measurements are in the 8.5 GHz
  band (in observer's frame).  Blue diamonds denote SNe/GRBs, while all
  the gray circles denote long-duration, cosmological GRBs.  Red and
  green lines are the symbolic lines (with random normalization)
  indicating the two different scenarios of the flux density scaling.
  The green dashed line indicates if the flux density scales as simply
  the inverse square of the luminosity distance. The red thick line is
  the flux density scaling in the canonical afterglow model which
  includes a negative-$k$ correction effect, offsetting the diminution
  in distance. 3-$\sigma$ EVLA sensitivity is also plotted to show the
  EVLA capability to go deeper in $z$.}
\label{fig:flux-redshift}
\end{figure}

The average redshift of our sample is $\langle z \rangle_{total}=1.8$.  The average
redshifts in pre- and post-\emph{Swift} bursts are $\langle z \rangle_{pre}=1.3$
and $\langle z \rangle_{post}=2.0$, respectively. The mean redshift for the
pre-\emph{Swift} bursts is similar to the value $\langle z\rangle$=1.4
derived by \citet{jlf+06}, however, they measure $\langle
z\rangle$=2.8 for a post-\emph{Swift} optically-selected sample. 
Instrumental sensitivity may explain the origin of this bias in our
radio-selected sample to slightly lower redshifts for
post-\emph{Swift}.  However, \citet{fjp+09} has measured the
average redshift of {\em Swift} sample to be $\langle z\rangle$=2.2,
based on a sample of 146 {\em Swift} bursts, which is 
closer to mean redshift of our radio selected sample.
 As first noted by \cite{cl00} and shown by
\citet{fck+06}, the detection rate of radio afterglows is largely
insensitive to redshift. This is shown in Figure~\ref{fig:flux-redshift},
where we plot peak radio flux densities at 8.5 GHz band as a function
of ($1+z)$.  The red line shows the effect of the negative
$k$-correction factor or radio afterglows, caused by the effects of
spectral and temporal redshift, offsetting the diminution in distance.
However, despite this effect we note that the mean peak flux density
at $z\sim$2.8 is close to the sensitivity limit of existing
instruments.
Figure~\ref{fig:flux-redshift}, however, shows that more
sensitive telescopes, such as EVLA, will be able to detect higher-$z$
bursts and enable us to test this hypothesis. In
the mean time when discussing any distance-dependent effects for our
sample we will adopt our derived redshifts rather than those in the
literature.

\subsection{Fluence distribution}
\label{sec:distribution}

\begin{figure}
\centering
\includegraphics[width=0.48\textwidth]{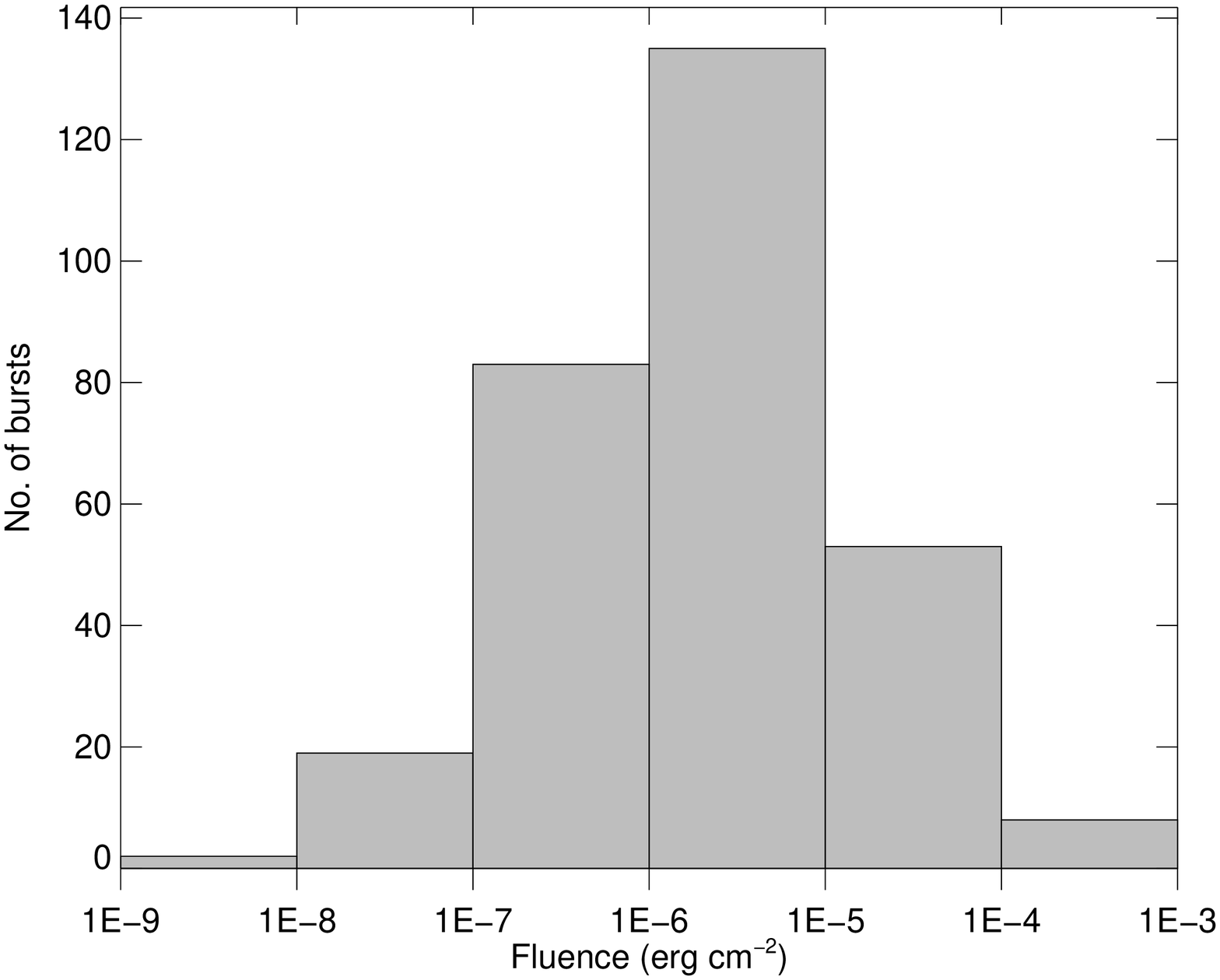}
\includegraphics[width=0.48\textwidth]{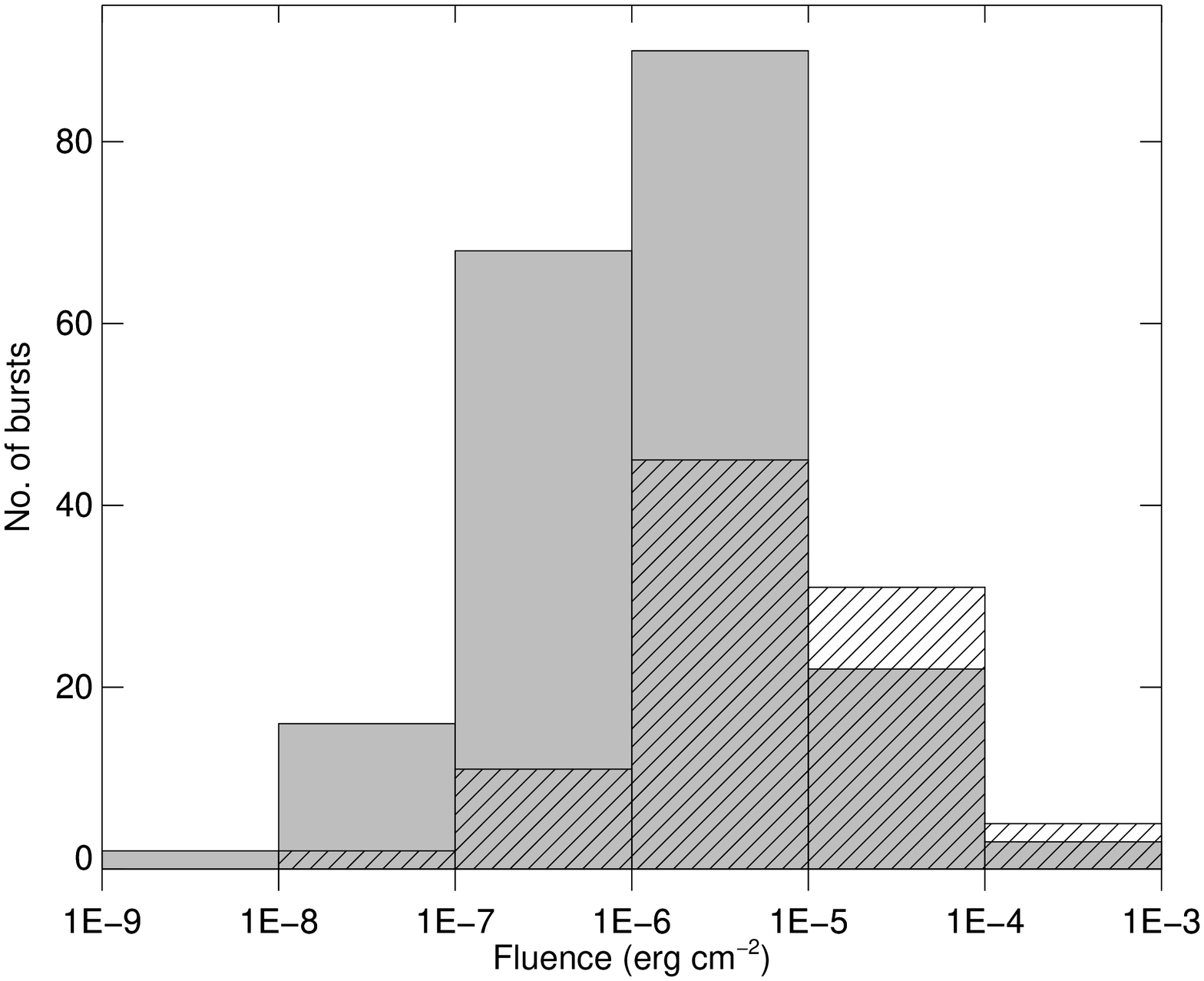}
\caption{{\it Upper Panel:} 
  Histogram of the fluence distribution for our entire sample.
{\it Lower panel:} Fluence  distribution of the radio-detected sample (hatched histogram) 
versus the non-detected
sample (filled histogram).
}
\label{fig:fluence1}
\end{figure}

Since the fluence of \emph{Swift} bursts is available in the
15--150~keV energy range, all the fluences quoted in
Table~\ref{MasterTable} are given in this range. We have converted
fluences from the non-\emph{Swift} telescopes to this energy range
(15--150~keV) adopting the method explained in \citet{nf09}. The
fluences quoted in Table~\ref{MasterTable} are the fluences in the
observer's frame. In Figure~\ref{fig:fluence1}, we plot the fluence
distribution of the entire sample as well as for the radio-detected
and non-detected sample.  Even though both the samples peak in the range of
$10^{-6}-10^{-5}$ erg cm$^{-2}$, 176 out of 206 ($\sim$85\%)  
non-detected radio afterglows have 
the fluence values $\le 10^{-6}$ erg cm$^{-2}$, while  86\% of the 
radio detected, i.e. 82 out of 95 bursts, have fluence
values $\ge 10^{-6}$ erg cm$^{-2}$. Thus there is a clear overall trend of 
non-detected radio afterglows favoring  the lower values of
fluence as compared to the higher fluence values favored by the detected.
The K--S test also gives the $P$ value of $P=2.61\times10^{-7}$
supporting the fact that the
radio-detected and
non-radio-detected GRB fluences are being derived from the different populations.

\subsection{Energy distribution of the prompt GRB emission}

\begin{figure}
\centering
\includegraphics[width=0.48\textwidth]{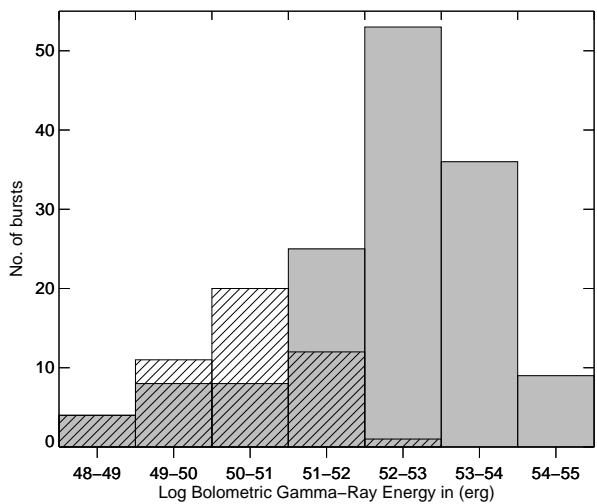}
\caption{The isotropic energy distribution of our complete sample. 
  Overlapped hatched histogram shows the beaming-corrected
  energy distribution.  }
\label{fig:EnergyDistribution}
\end{figure}

We calculated the isotropic-equivalent $\gamma$-ray energies
$E_{iso}^{bol}$ for all bursts with known redshifts.  The isotropic
energies given in Table~\ref{MasterTable} are $k-$corrected bolometric
energies. We adopt the rest-frame bandpass in the energy range
1~keV--10~MeV.  We quote all the original references from which the
energy values are taken.  However, for those bursts in which this
energy was not known, we obtained the information about the best-fit
model of the prompt GRB spectrum from the GCN circulars
archive\footnote{http://gcn.gsfc.nasa.gov/gcn3\_archive.html}. The
prompt emission is best fit by either band function \citep{bmf+93} or
a cut-off power law function (CPL) or a simple power law (see
\citet{rlb+09} for details).  We used these best fit parameters and
adopted the method of \citet{bfs01} to estimate the $k$-corrected
energy in 1~keV--10~MeV range.  Since GRBs are collimated events, the
beaming-corrected energy is smaller than the isotropic-equivalent
energy. To obtain this, we identified all the GRBs with known jet
breaks ($t_j$) from the literature. We also obtained all available
circumburst densities $n$ from the literature.  If a GRB had a known
jet break but no available density measurement, we assumed density $n=
1$ cm$^{-3}$ (we denote it as [1] in Table~\ref{MasterTable}).  Once
having values of $t_j$ and $n$, we estimated the collimation angle
$\theta_j$ using \citep{fks+01}.  The beaming fraction $f_b=(1-\rm
cos\theta_j)\approx\theta_j^2/2$ was estimated to calculate the
beaming-corrected bolometric energies $E_{true}^{bol}$, where
$E_{true}^{bol}=f_b E_{iso}^{bol}$.

In Figure~\ref{fig:EnergyDistribution}, we plot the isotropic energy
$E_{iso}^{bol}$ distribution available for 144 GRBs of our sample (46 GRBs in
pre-\emph{Swift} epoch and 98 GRBs in post-\emph{Swift} epoch). While
maximum GRBs lie in the energy range of $10^{52}-10^{53}$ erg, the
energy range spread of the sample is between
$1.6\times10^{48}-4.1\times10^{54}$ erg,
 i.e. 7 orders of magnitude (this includes all the GRBs including
SHBs, XRFs and SN-GRBs). 
 This energy distribution and
wide spread to lower energies is characteristic of other large samples
\citep{kkz+10}.
The figure also shows the beaming corrected energy distribution (hatched histogram).
There are
only 48 GRBs in our sample (excluding upper limits) which had
beaming-corrected energies estimated.
Even though most of the GRBs lie between energy range
$10^{50}-10^{51}$ erg, the total spread in energy range is from
$2.4\times10^{48}-1.4\times10^{52}$ erg.  This distribution is
similar to the previous studies \citep[e.g.][]{rlb+09}.

\begin{figure}
\centering
\includegraphics[width=0.48\textwidth]{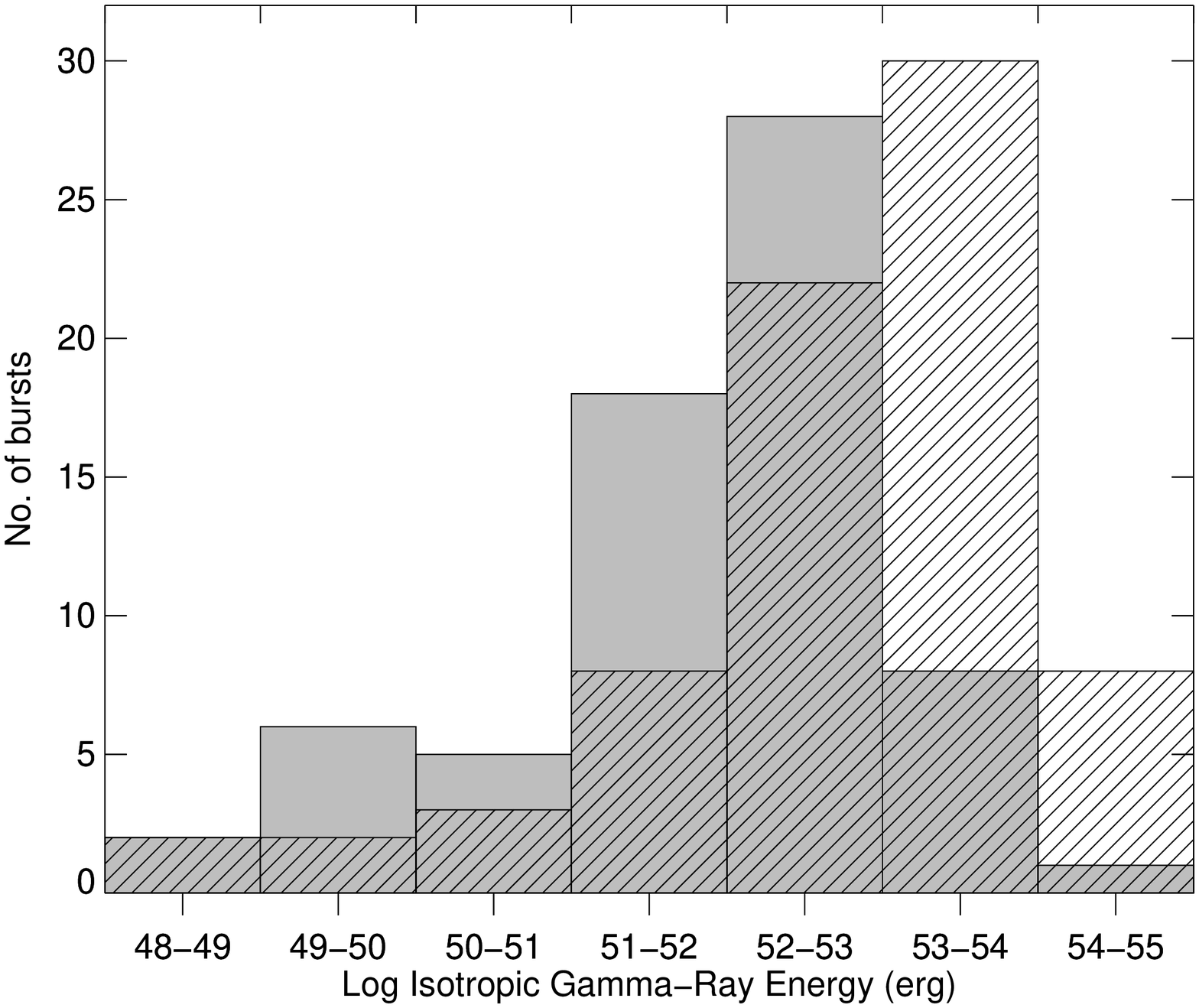}
\includegraphics[width=0.48\textwidth]{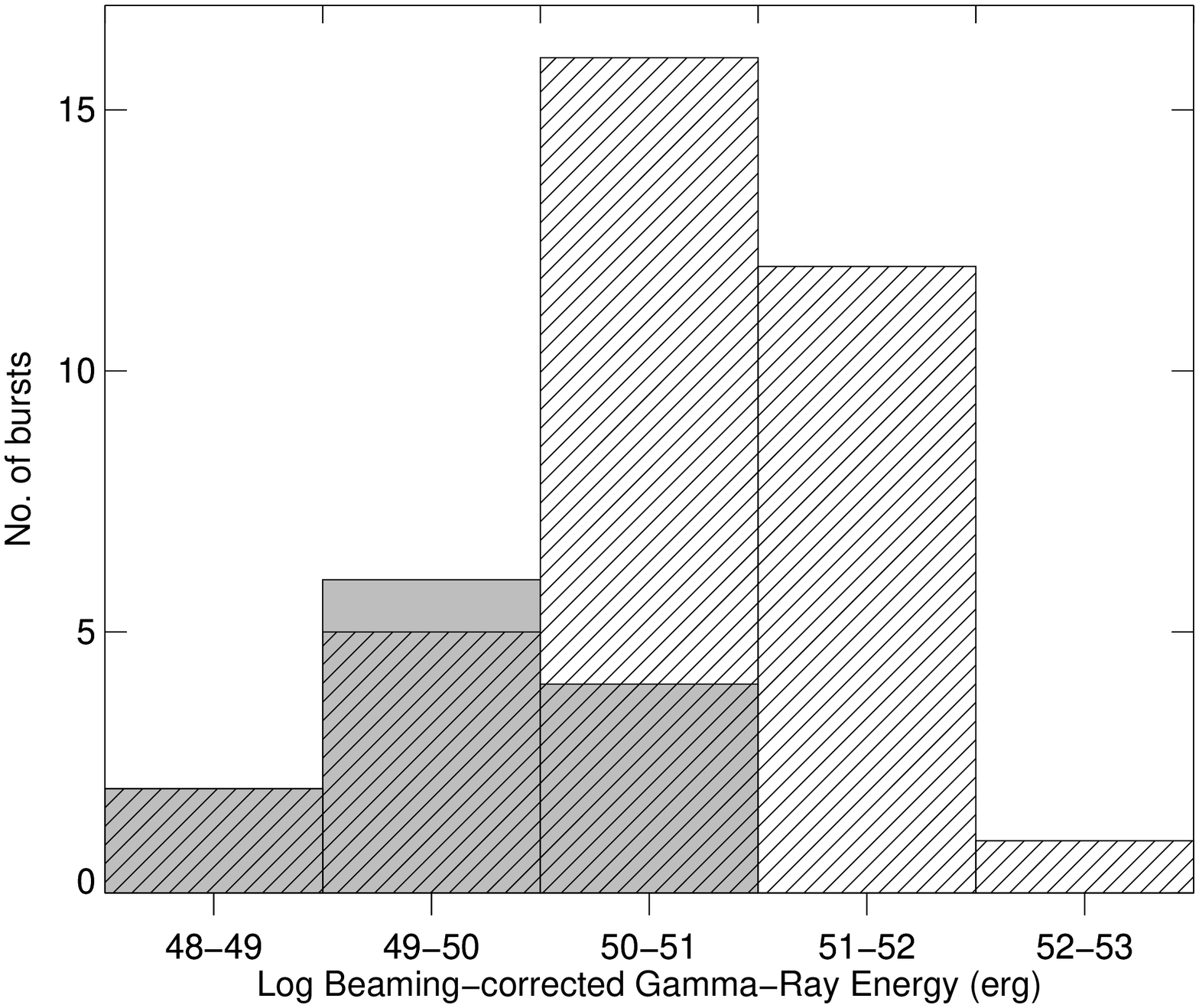}
\caption{Upper panel shows the isotropic-equivalent $\gamma$-ray energy distribution 
of the radio-detected sample (hatched histogram) versus the non-detected
sample (gray filled histogram). 
Lower panel shows the same distribution for the beaming corrected energies.
}
\label{E-DND}
\end{figure}

In Figure~\ref{E-DND}, we plot the isotropic-equivalent and beaming
corrected $\gamma$-ray energies for the radio-detected and
non-detected sample. In case of isotropic energies, the non-detected
sample peaks at $10^{52}-10^{53}$ erg, whereas, the detected sample
peaks at $10^{53}-10^{54}$ erg range.  While 60 out of 95 radio
detected bursts have isotropic-equivalent $\ge10^{53}$ erg, only
9 out of 206 radio non-detected bursts have energies above 
$10^{53}$ erg.  For beaming corrected energies,
the non-detected sample peaks at $10^{49}-10^{50}$ erg, whereas, the
detected sample peaks at $10^{50}-10^{51}$ erg range.  This indicates
that bolometric energies can indeed be a good indicator of the radio
detectability of a burst. The K--S test gives the $P$ values for
the isotropic and beaming corrected energies of the two samples
to be $P=9.97\times10^{-7}$ and $P=3.47\times10^{-3}$,
respectively (see Table 4). This implies that
the two samples are inconsistent with being drawn from the same distribution
and $\gamma-$ray energetics is indeed a good indicator of the 
radio detectability of 
a burst, with isotropic-equivalent energy being the stronger indicator.

\subsection{Distribution of X-ray and optical fluxes}

The X-ray fluxes ($F_X^{11h}$) quoted in Table~\ref{MasterTable} are
in the energy range 0.3--10~keV. The X-ray fluxes of \emph{BeppoSAX}
bursts are given in the energy range of 1.6--10~keV range. All X-ray
fluxes are corrected for Galactic extinction and we quote the X-ray
fluxes at a time of 11 hr after the burst. This is partly because most
GRBs have measurements at this time but also because it is expected
that the cooling frequency at this time will usually lie between the
optical and X-ray frequencies \citep{nf09}. The afterglow flux will
then be independent of the circumburst density for observing
frequencies above the cooling frequency.  We obtained most of our
X-ray fluxes at 11 hr from \citet{gbb+08,dpg+06,sbb+08,sbb+11}.
However, for the bursts in which the flux at 11 hr was not available,
we obtained the X-ray light curve from the \emph{Swift} XRT Lightcurve
Repository \citep{ebp+07} and derived the flux as close to 11 hr as
possible (i.e. within $\pm0.5$ hr). 
 If the flux at $\sim11$ hr was not available, we fit the
X-ray light curve with a simple powerlaw and extrapolated the flux to
11 hr. This was done for 33 afterglows. Only in 5 cases, when the X-ray light curve was not
available, did we extrapolated the available X-ray flux to 11 hr using
temporal index ($\alpha_X$, in $F_X\propto t^{\alpha_X}$)
$\alpha_X=-1.17$ for long bursts and $\alpha_X=-1.22$ for SHBs
\citep{nf09}.

\begin{figure}
\centering
\includegraphics[width=0.48\textwidth]{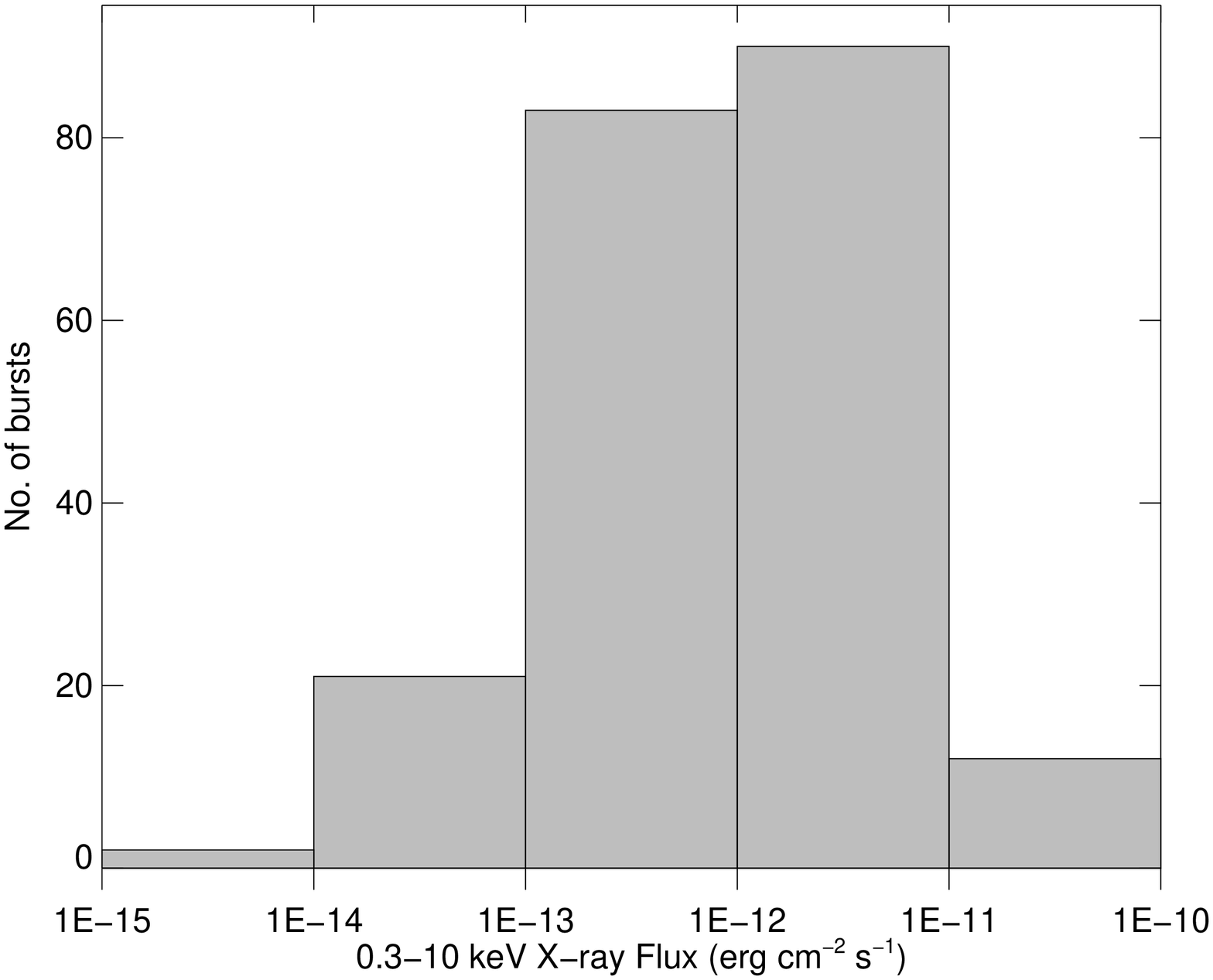}
\includegraphics[width=0.48\textwidth]{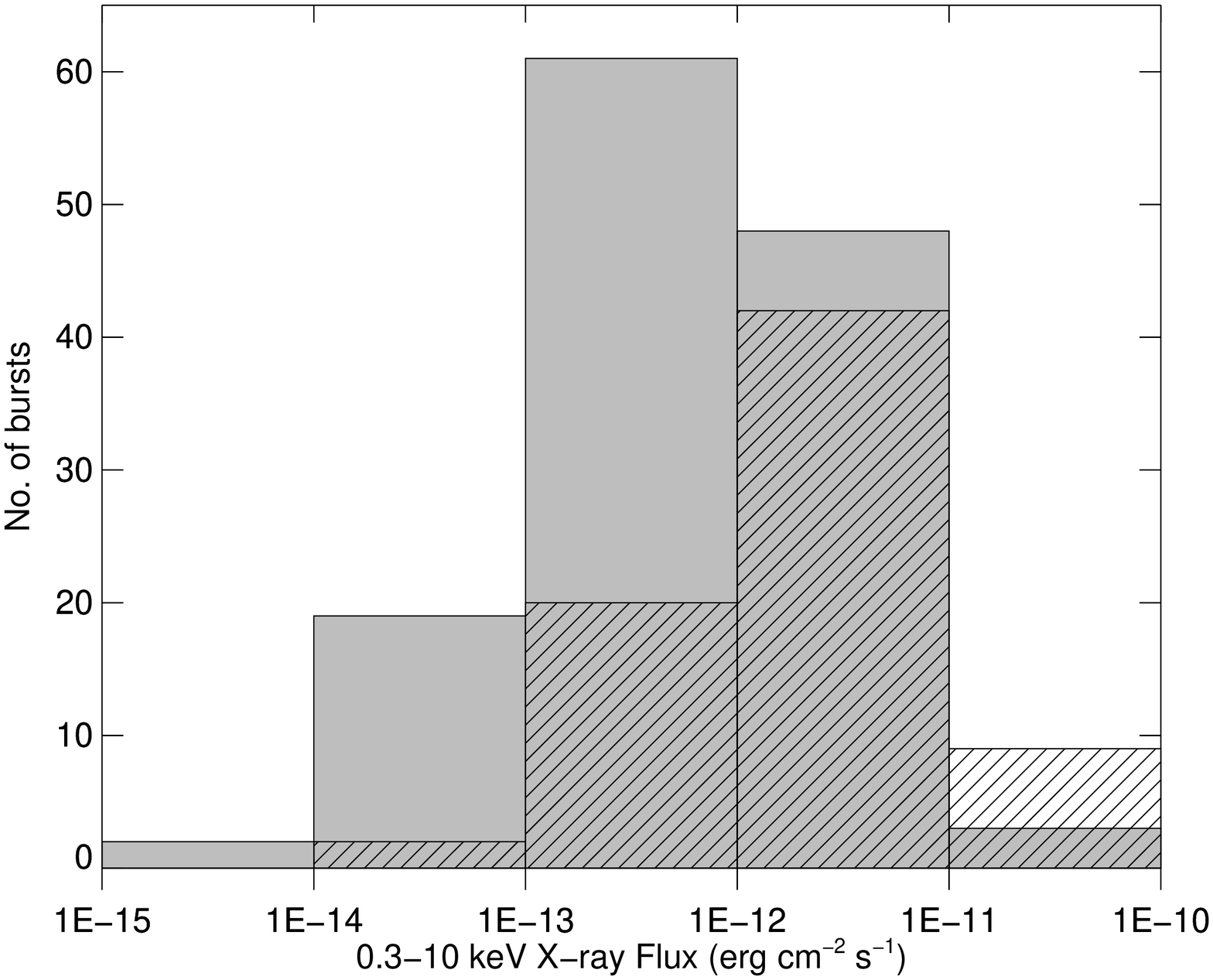}
\caption{{\it Upper panel:} The distribution of the X-ray 
  fluxes for our sample. {\it Lower panel:} 
This  shows X-ray flux at 11 hr distribution 
of the radio-detected sample (hatched histogram) versus the non-detected
sample (gray filled histogram). 
}
\label{fig:fluxhisto}
\end{figure}

Optical R band flux densities ($F_R^{11h}$) are also obtained at 11 hr
in units of $\mu$Jy. For those bursts in which we could not find the
desired values from the published references, we used the GCN
circulars and then extrapolated to the reported values at 11 hr using
the best fit time decay index available. If a decay index was not
available, we used a temporal slope of $\alpha_R=-0.85$ (in $F_R
\propto t^{\alpha_R}$) for long bursts and $\alpha_R=-0.68$ for SHBs.
The bursts for which an R-band magnitude was not available, we used
spectral index of $\beta_R=-1$ (in $f_R \propto \nu^{\beta_R}$) to
convert the magnitudes into R band. We also corrected the flux
densities for Galactic extinction using \citet{sfd98}. All the
magnitudes were converted into $\mu$Jy using online NICMOS unit
conversion
tool\footnote{http://www.stsci.edu/hst/nicmos/tools/conversion\_form.html}.

\begin{figure}
\centering
\includegraphics[width=0.48\textwidth]{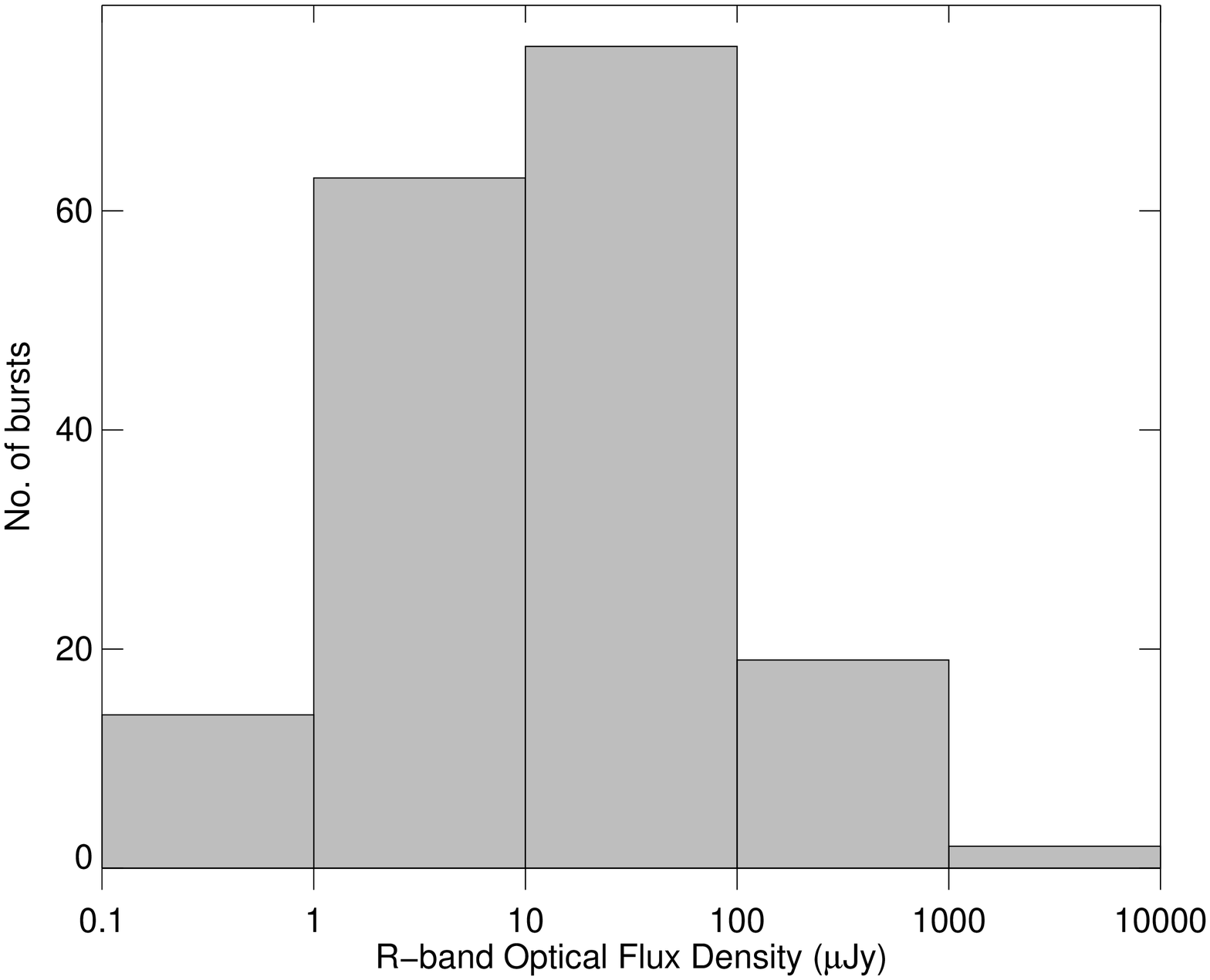}
\includegraphics[width=0.48\textwidth]{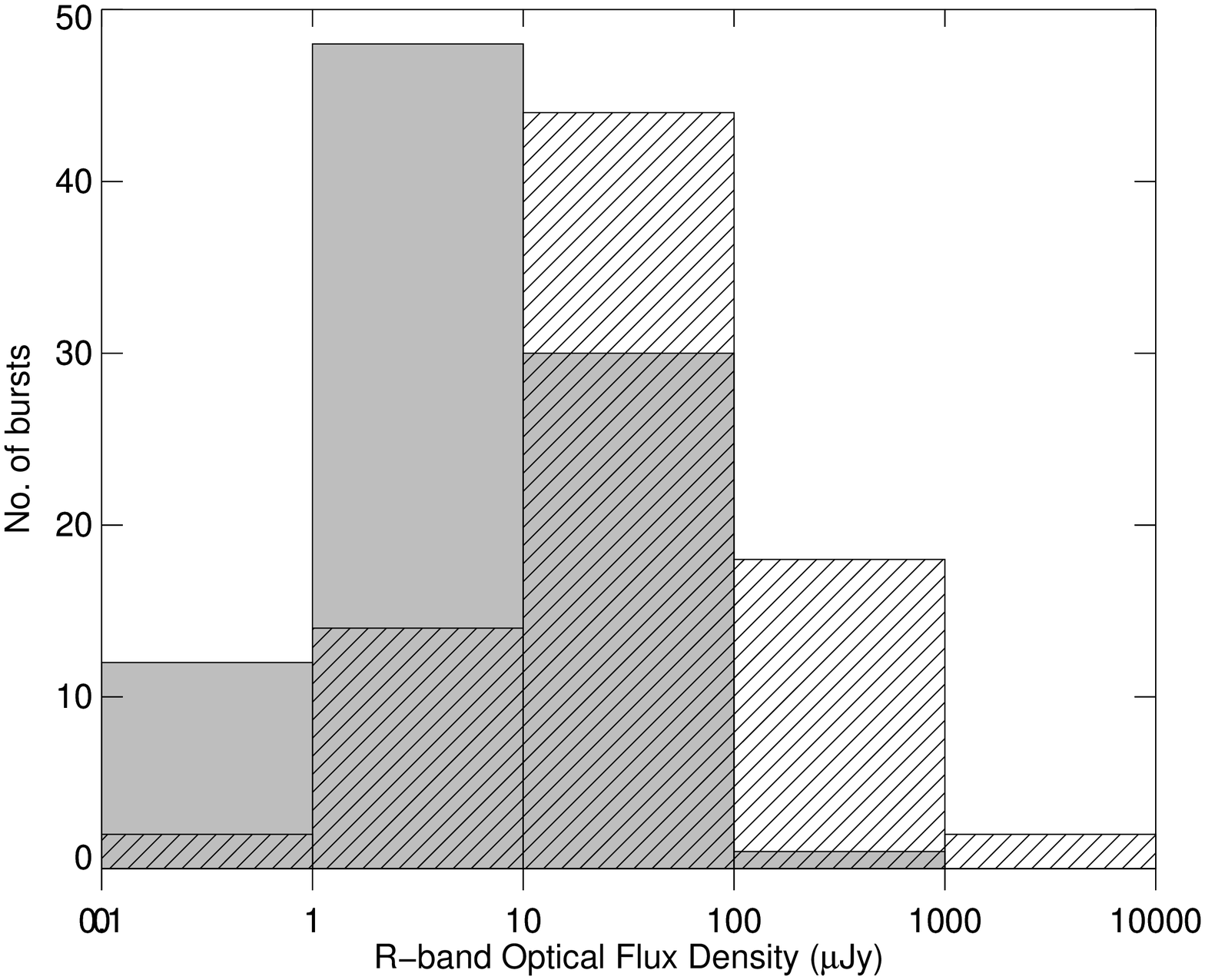}
\caption{
Upper panel shows the distribution of
the  optical flux densities for our sample.
 Lower panel
shows the optical flux density distribution 
of the radio-detected sample (hatched histogram) versus the non-detected
sample (gray filled histogram). 
}
\label{fig:fluxhisto2}
\end{figure}

\begin{deluxetable}{lc}
\tablecaption{$P$ values for Kolmogorov-Smirnov test for the radio detected versus non-detected GRB sample 
\label{tab:correl1}}
\tablewidth{0pt}
\tablehead{
\colhead{Parameter} &  \colhead{$P$ value}
}
\startdata
Redshift & 0.61\\
Isotropic equivalent $\gamma$-ray energy & $9.97\times10^{-7}$\\
Beamed $\gamma$-ray energy & $3.47\times10^{-3}$\\
Fluence & $2.61\times10^{-7}$\\
X-ray flux (0.3--10 keV) & $3.61\times10^{-6}$\\
Optical $R$-band flux density & $1.30\times10^{-9}$
\enddata
\end{deluxetable}

In Figure~\ref{fig:fluxhisto}, we plot the X-ray flux distribution of
our sample which is spread between $10^{-15}-10^{-10}$ erg cm$^{-2}$
s$^{-1}$.  The majority of bursts have X-ray fluxes in the range
$10^{-12}-10^{-11}$ erg cm$^{-2}$ s$^{-1}$.  Lower panel shows the
X-ray flux at 11 hr of the radio-detected and non-detected samples.
While the radio-detected sample peaks between $10^{-12}-10^{-11}$ erg
cm$^{-2}$ s$^{-1}$ range, the non-detected samples peaks at an order
of magnitude less flux between the range $10^{-13}-10^{-12}$ erg
cm$^{-2}$ s$^{-1}$.  In Figure~\ref{fig:fluxhisto2}, we plot the
optical flux density distribution of our sample which ranges from 0.1
$\mu$Jy to $10^4$ $\mu$Jy. Most of the optical bursts lie in 10--100
$\mu$Jy range.  In this figure, we also plot the optical R-band flux
density at 11 hr for the radio-detected and non-detected sample.  The
non-detected sample peaks at 1--10 $\mu$Jy, whereas, the detected
sample peaks at 10-100 $\mu$Jy.

On average, the X-ray fluxes of bursts are an indicator of radio
detectability.  Likewise, the dust-extinction corrected optical
magnitude of the burst is also a potential indicator of the radio
detectability of the burst.
The K--S test also supports this fact with $P$ values of
$P=3.61\times10^{-6}$ and 
$P=1.30\times10^{-9}$ for the X-ray and the optical fluxes, respectively (Table~\ref{tab:correl1}.

\section{Correlative properties}
\label{corr}

Here we proceed to look for correlations between the peak radio flux
density (or luminosity) at 8.5 GHz band with various GRB properties
(such as fluence, isotropic and beaming-corrected energy) and the
X-ray and optical afterglow properties (redshift, flux density). We
estimate the correlation in terms of the Pearson's correlation
coefficient, or R-index.  The R-index varies between -1 to +1. A
negative correlation coefficient indicates negative correlation
between the quantities.  An R-index close to 0 means no correlation or
a weak correlation. R-index values of $\ge0.5$ indicates a significant
correlation between the parameters.  We use only LGRBs to obtain the
correlation coefficient to maintain homogeneity for this correlation
analysis.

\begin{figure}
\centering
\includegraphics[width=0.48\textwidth]{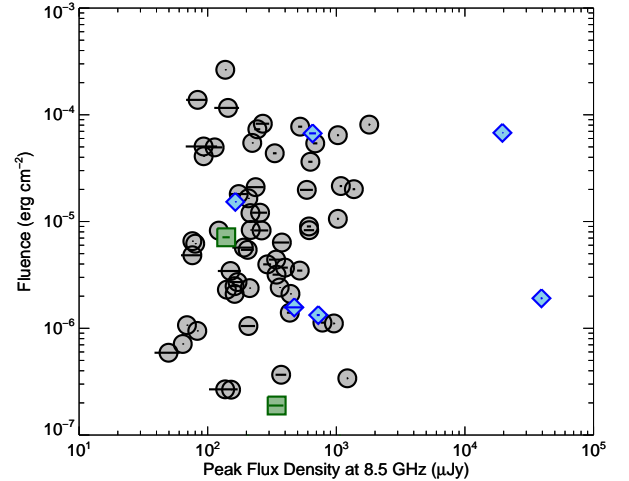}
\caption{Peak radio flux densities at 8.5 GHz band taken from 
  Table~\ref{tab:radio-peak} on the X-axis and their corresponding
  fluences in 15--150 keV band are plotted on the Y-axis. Blue diamonds
  denote SN/GRBs and green squares denote XRFs, while all the gray
  circles denote long cosmological GRBs.}
\label{fig:fluence}
\end{figure}

In Figure~\ref{fig:fluence}, we plot the 15--150 keV fluences of radio
detected GRBs against their corresponding peak radio flux densities at
8 GHz band from Table~\ref{tab:radio-peak}. There is no correlation
between the two.  The Pearson's correlation coefficient, R-index,
between the two is 0.02. The figure also shows that while the spread
in fluence is four orders of magnitude, radio peak flux densities are
tightly clustered. 

\begin{figure}
  \centering \includegraphics[width=0.48\textwidth]{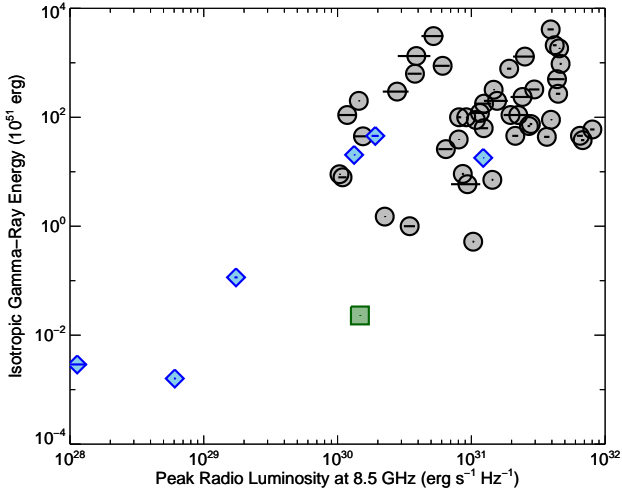}
\caption{The peak radio luminosity at 8.5 GHz versus isotropic-equivalent 
  $\gamma$-ray energies for various bursts.  Here blue diamonds
  correspond to the SN/GRB events, green squares XRF events and gray
  circles are long cosmological GRBs.}
\label{fig:isor}
\end{figure}

We also plot the isotropic $\gamma$-ray energies $E_{iso}^{bol}$ against their respective peak
radio luminosities In Figure~\ref{fig:isor}. 
There in no correlation between the radio spectral luminosity and the
isotropic energy of GRBs with R-index of 0.12.  
 
\begin{figure}
\centering
\includegraphics[width=0.48\textwidth]{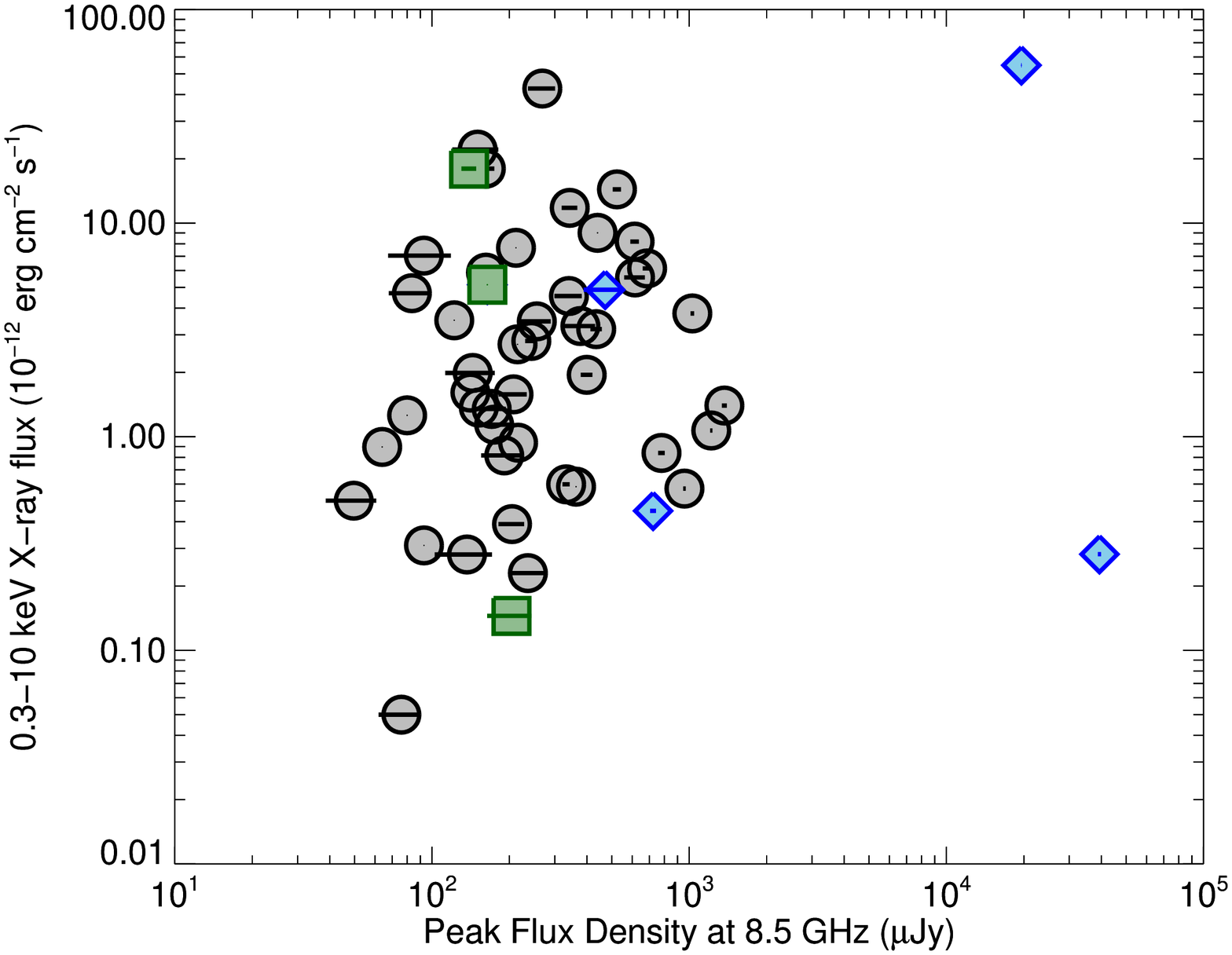}
\includegraphics[width=0.48\textwidth]{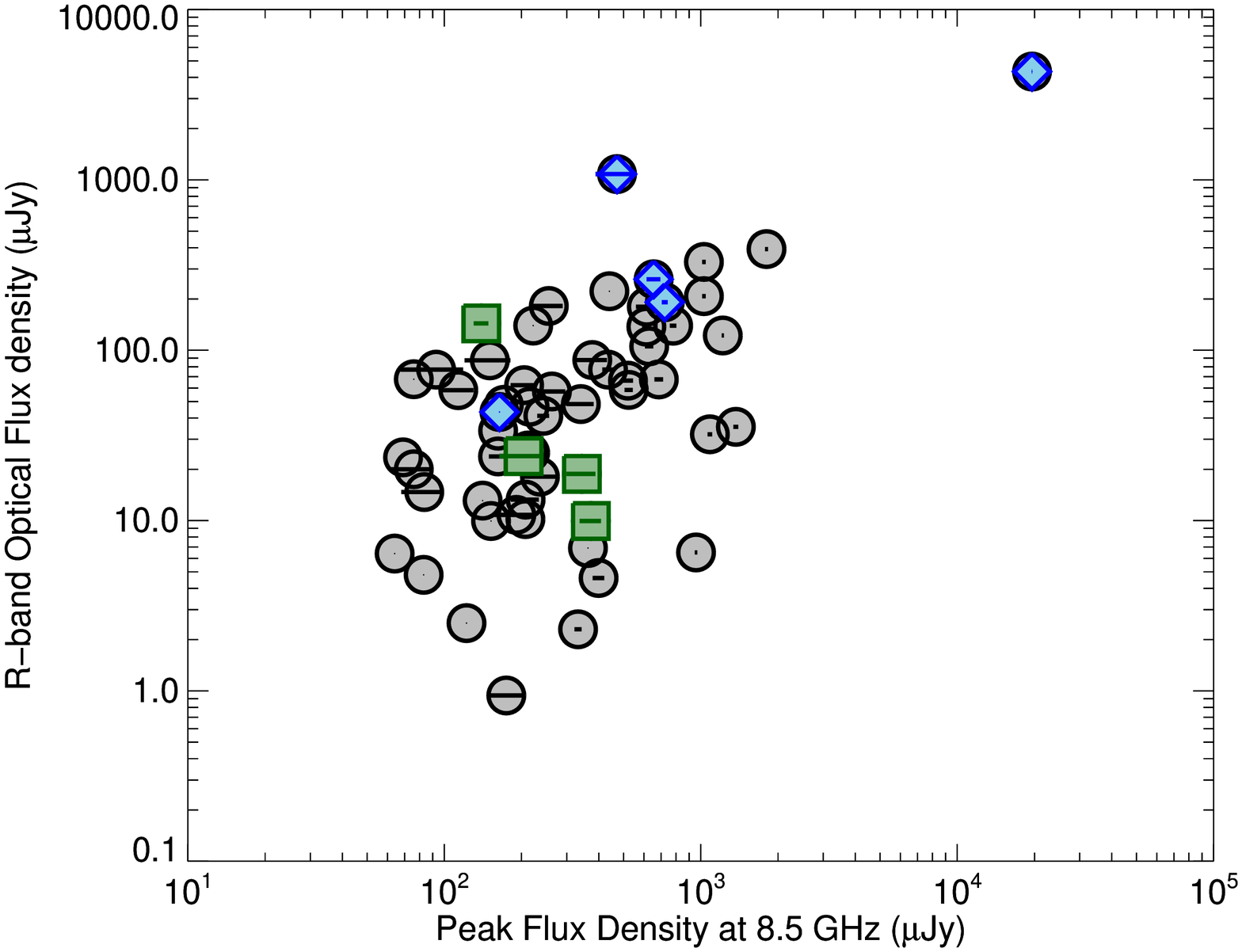}
\caption{{\it Upper panel:} 
  Peak radio flux densities at 8.5 GHz taken from
  Table~\ref{tab:radio-peak} are on the X-axis and their corresponding
  X-ray fluxes at 11 hr in 0.3--10 keV band are plotted on the Y-axis.
  Blue diamonds denote SN/GRBs and green squares denote XRFs, while all
  the gray circles denote long cosmological GRBs. {\it Lower panel:}
  Peak radio flux density versus the R-band optical flux at 11 hrs.}
  \label{fig:fluxcor}
\end{figure}

In Figure~\ref{fig:fluxcor}, we plot the X-ray fluxes at 11 hr for
radio-detected GRBs against their corresponding peak radio flux
densities at 8.5 GHz band from Table~\ref{tab:radio-peak}. There is no
correlation between the two with a R-index value of -0.05.
We also plot the
R-band optical flux densities at 11 hr for the radio-detected GRBs
against their corresponding peak radio flux densities at 8.5 GHz band
from Table~\ref{tab:radio-peak}.
The two parameters seem to be correlated. With a correlation index of
0.62, $F_R^{11h}$ is the only parameter correlated with the peak
strength of a radio afterglow.  Despite this correlation,
Figure~\ref{fig:fluxcor} still illustrates the clustering of the radio
points over a narrow flux density range.

In Table~\ref{tab:correl}, we tabulate the R-index correlation between
various GRB parameters.  The correlation coefficients vary between -1
to +1.
The correlation coefficient values clearly indicate that the strength
of the peak radio flux density does not depend upon the fluence,
isotropic $\gamma$-ray energy or the X-ray flux. Only the optical flux
density at 11 hr has a positive correlation with the peak strength of
the radio afterglows.

\begin{deluxetable}{llr}
\tablecaption{R-index correlation coefficient between various parameters of GRB afterglows
\label{tab:correl}}
\tablewidth{0pt}
\tablehead{
\colhead{Parameter 1} & \colhead{Parameter 2} & \colhead{R-index}
}
\startdata
Peak radio flux density & Fluence & 0.02\\
Peak radio luminosity & Isotropic bol. energy & 0.12\\
Peak radio flux density & X-ray flux & $-0.05$\\
Peak radio flux density & Optical flux & 0.62
\enddata
\end{deluxetable}

\section{Synthetic radio light curves}
\label{sec:synthetic}

\begin{figure}
\centering
\includegraphics[width=0.48\textwidth]{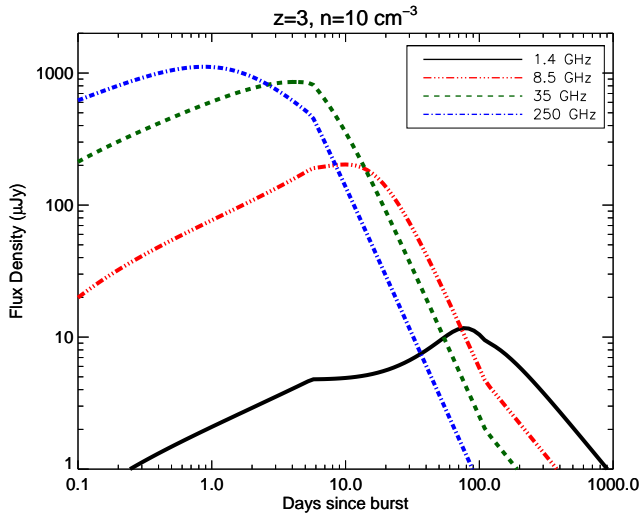}
\caption{Plot of radio flux density light curves for a ``standard'' GRB
  afterglow at various frequencies for a redshift $z=$3 and density of
  10 cm$^{-3}$. See text for more details.}
\label{fig:density2}
\end{figure}

In order to compare the properties of our radio-sample with
predictions from the basic afterglow theory, we plot synthetic light
curves of GRB radio afterglows and determine their dependence on
various afterglow parameters such as density, kinetic energy, redshift
and microscopic shock parameters.  In Figure~\ref{fig:density2}, we
plot the radio afterglow flux density at the median GRB redshift $z=3$
\citep{jlf+06} and density $n=10$ cm$^{-3}$ for different frequencies
i.e. 1.4 GHz, 8.5 GHz, 35 GHz (EVLA bands) and 250 GHz (ALMA band).
Here we fix other parameters such as the isotropic kinetic energy
$E_{KE,iso}=10^{53}$ erg, the beaming angle $\theta_j=0.2$ rad, the
electron energy density $\epsilon_e=0.1$, the magnetic energy density
$\epsilon_B=1$\%, and the electron spectral index $p=2.2$.  The GRB
parameters that we use here are good averages from the afterglow
broadband modeling \citep{pk01,yhsf03}.

While not unique, the 8.5 GHz light curve in Figure~\ref{fig:density2}
reproduces the average properties of our radio-selected sample derived
in \S\ref{sec:det}. The mean flux density (Table~\ref{KM}), the
time-to-peak (\S\ref{sec:radioafterglows}) and the overall timescale
of the decay is as seen in the data. The initial rise of the synthetic
light curve at 8.5 GHz is steeper than the canonical long-duration
event (Figure~\ref{fig:canonical}), but as we have already noted (e.g.
\S\ref{sec:cannon}), this may be the result of an additional early
emission component, possibly a reverse shock. The light curves in this
figure are generated with only a forward shock component.

With this close agreement between the model and the data, we can now
investigate the effects of changing the observing frequency. As
expected, there is a clear trend of increasing peak flux density and
decreasing time-to-peak with increasing frequency. For illustration we
compute the EVLA 3-$\sigma$ sensitivities at 1.4 GHz, 8.5 GHz and 35
GHz bands and ALMA band-6 for a 1-hr integration. These are 40 $\mu$Jy, 9 $\mu$Jy and
17$\mu$Jy, and 42$\mu$Jy, respectively.
The challenges of detecting typical
long-duration event events below 1.4 GHz are immediately apparent. For
the above given set of parameters, which are used to generate these
light curves, EVLA, ASKAP, and WSRT/Apertif will not be able to detect
radio afterglows at 1.4 GHz unless they are observed for long
integration times. Higher frequencies are clearly favored for
detecting radio afterglows.  We expect a large increase in the
fraction of detectable radio afterglows with ALMA and the EVLA, should
higher observing frequencies be used than in the past.

\begin{figure*}
\centering
\includegraphics[width=0.48\textwidth]{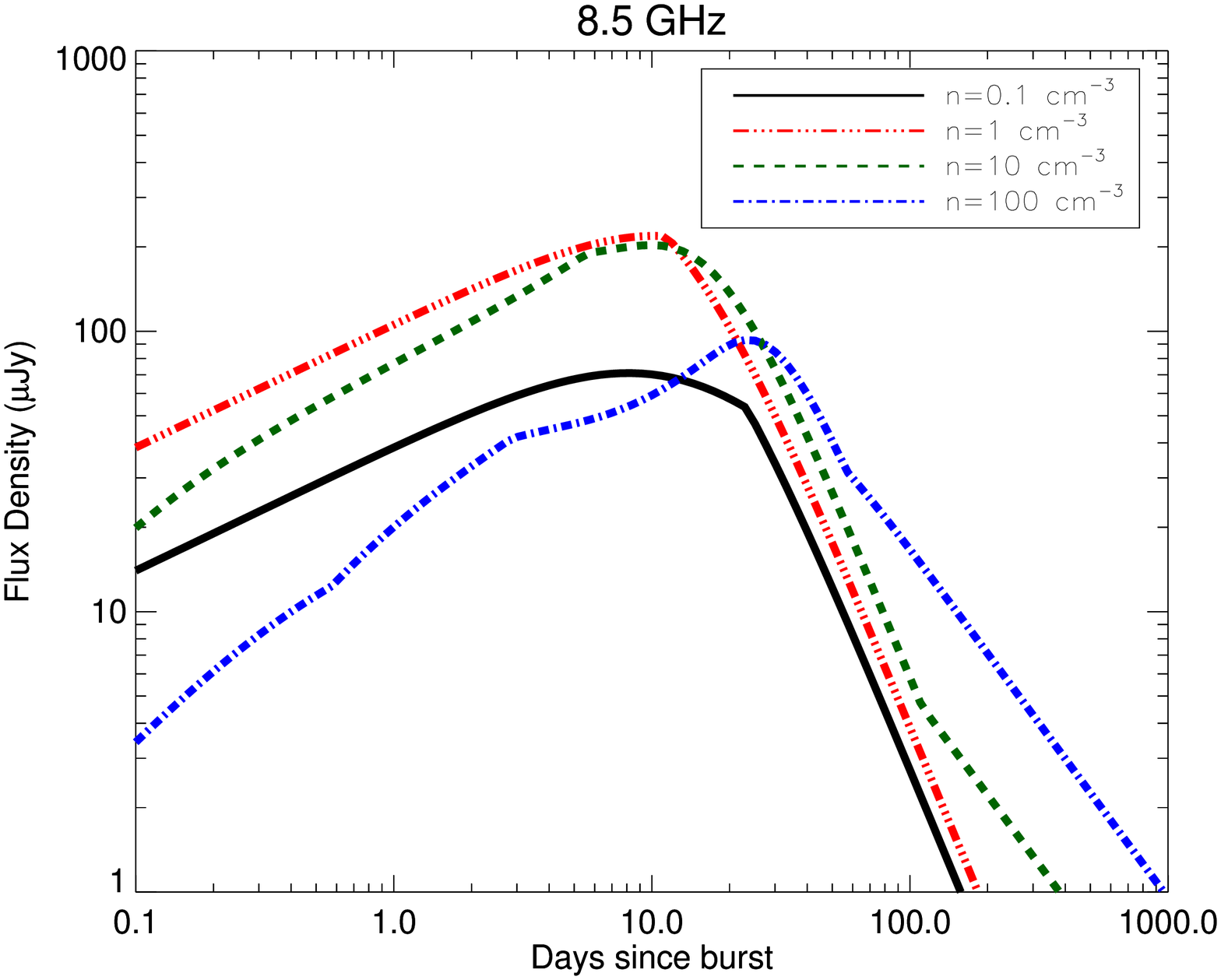}
\includegraphics[width=0.48\textwidth]{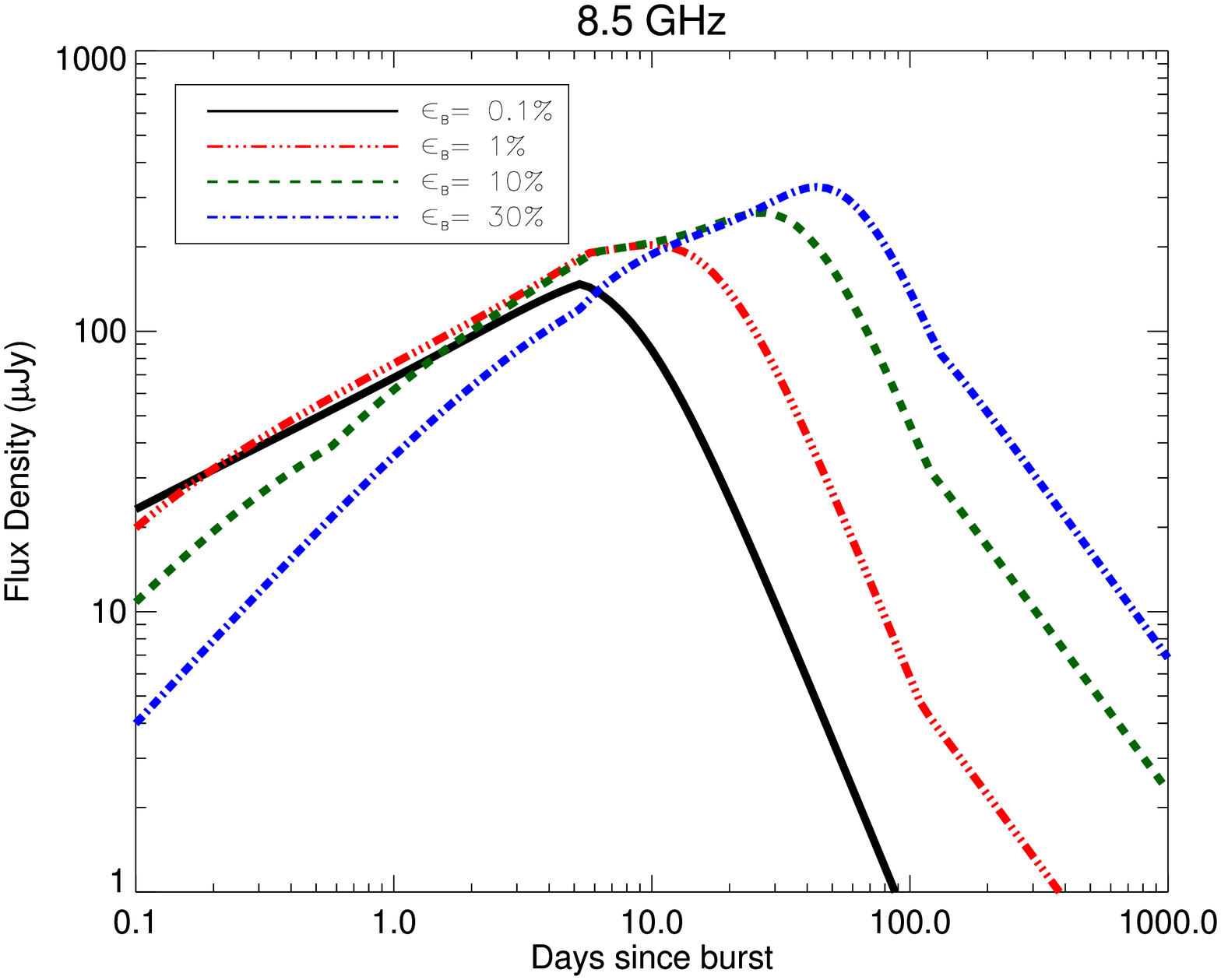}
\includegraphics[width=0.48\textwidth]{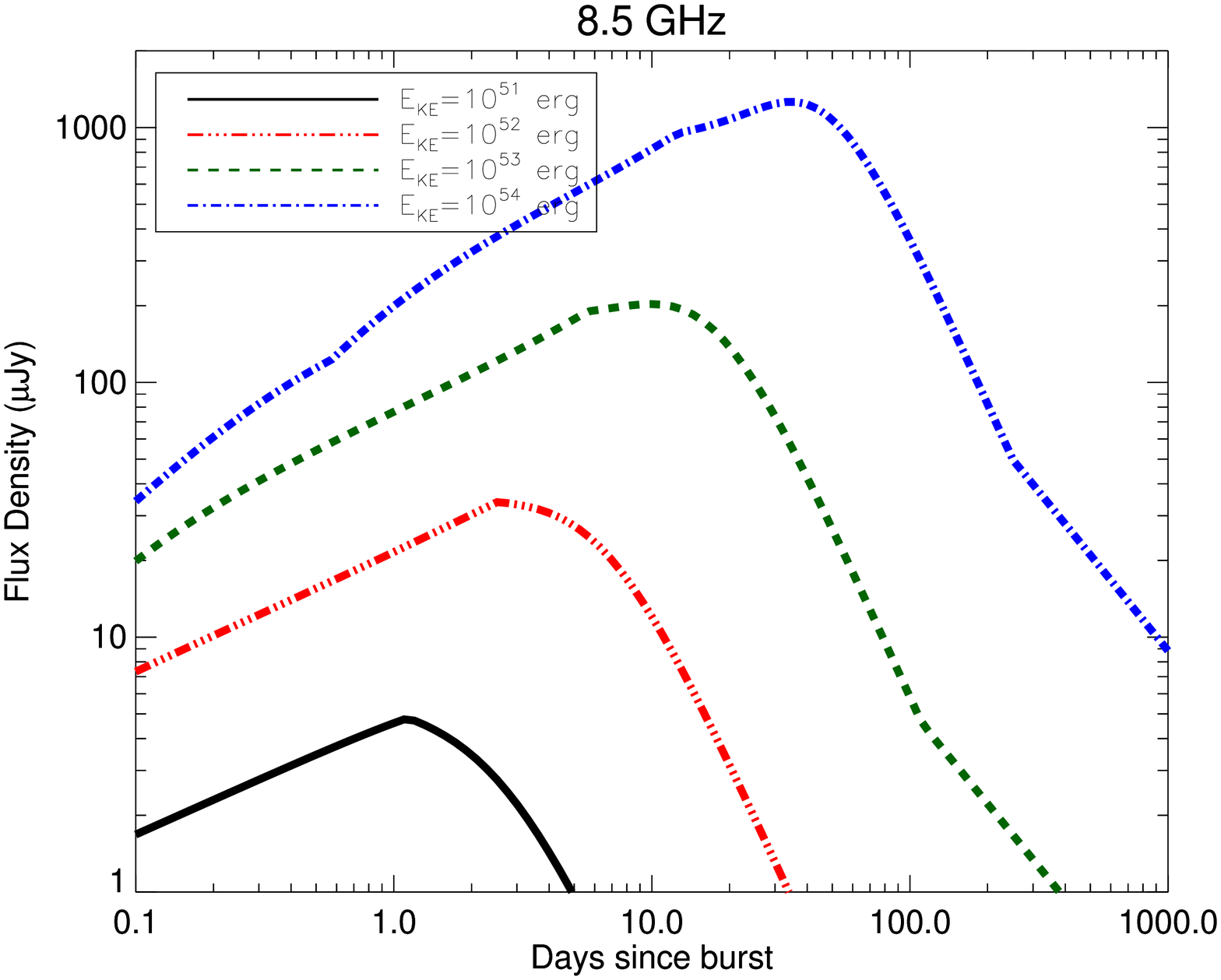}
\includegraphics[width=0.48\textwidth]{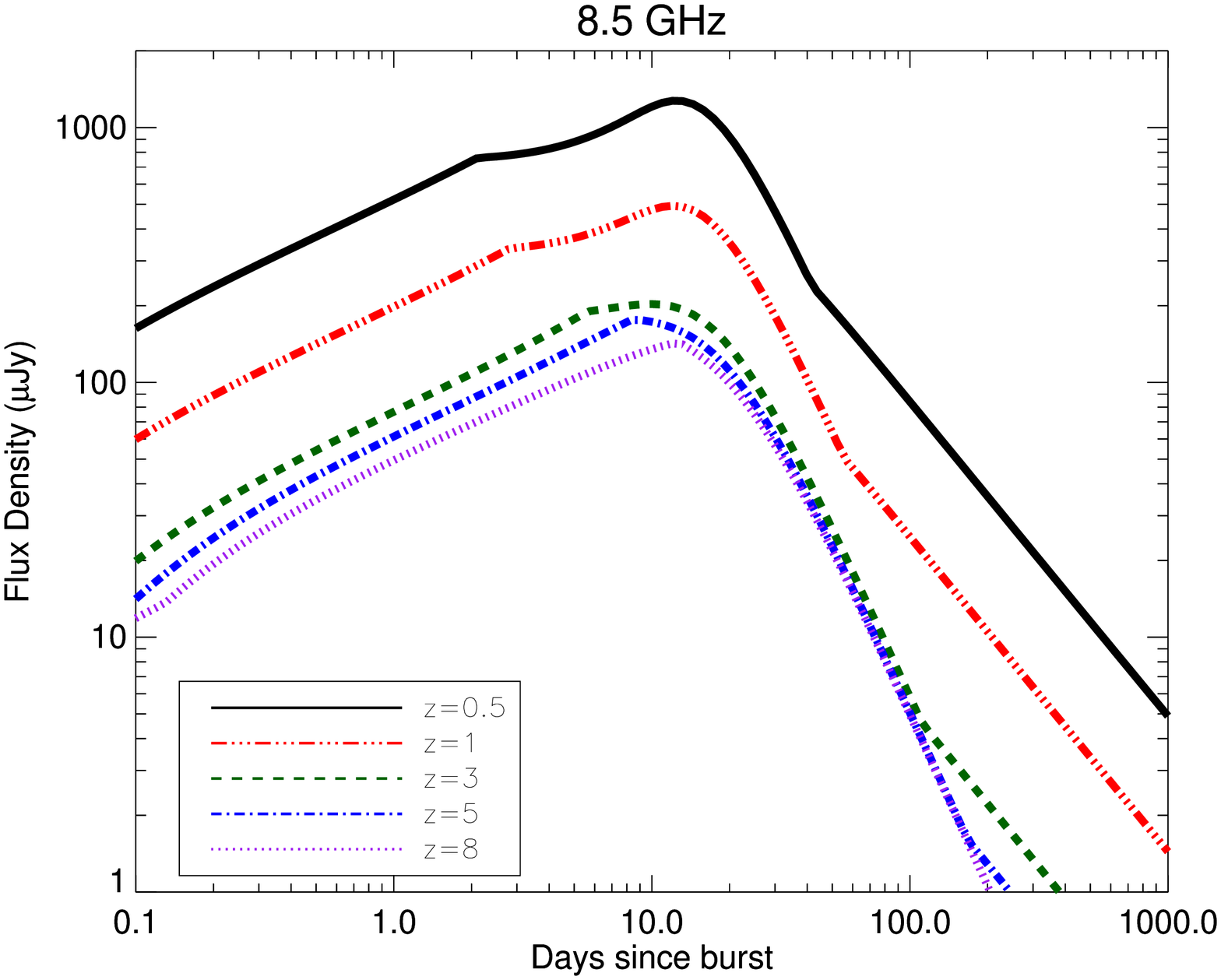}
\caption{8.5 GHz synthetic radio light curves for various values of density (top left),
magnetic energy density (top right), isotropic kinetic energy (lower left)
and redshift (lower right). EVLA sensitivity in this band for 1 hr integration
time is 9 $\mu$Jy.}
\label{fig:synall}
\end{figure*}

In Figure~\ref{fig:synall}, we derive the dependence of the radio
afterglow light curve on various parameters, such as density,
microscopic parameters and kinetic energy. Since there is only a weak
dependence on the electron energy density ($\epsilon_e$) and electron
spectral index ($p$), we do not show these plots here.

We adopt the constant density model for the circumburst medium to
generate these synthetic light curves. The plots are made for a
frequency of 8.5 GHz to enable comparison with our existing sample. As
noted above, future EVLA observations would better to be conducted at
higher frequencies. Increasing the centimeter radio frequency modifies
the flux density and timescales somewhat but it does not affect the
dependencies that we see below.

In the top left panel of Figure~\ref{fig:synall} we plot the synthetic
light curves of radio afterglows for various density values $n$ at a
redshift $z=3$.  We fix other parameters as $E_{KE,iso}=10^{53}$ erg,
$\theta_j=0.2$ rad, $\epsilon_e=0.1$, $\epsilon_B=1$\%, and $p=2.2$.
The radio afterglow brightness is a strong function of density.  The
flux density depends upon two competing effects.  There is an initial
increase in radio emission due to the enhanced synchrotron emission as
the density increases but as the density continues to increase there
is a reduction in the radio flux density due to the increasing
synchrotron self-absorption effect. The figure shows that the radio
afterglow is brightest for densities between $n=1-10$ cm$^{-3}$;
suggesting that existing radio samples are biased to a narrow range of
circumburst densities. At lower densities, the afterglow is
intrinsically weak, whereas at higher densities, synchrotron
self-absorption effects are suppressing the radio afterglow strength
for a long time.  This may also explain why some of the bright GRBs
are dim in radio band. They may need late-time observations in order
to be detected.

To determine the dependence of radio afterglow as a function of
$\epsilon_B$, we plot the lightcurves for various values of fractional
magnetic energy density in the top right plot of 
Figure~\ref{fig:synall}.  We derive the light curves for a burst at a $z=3$
and for circumburst density $n=10$cm$^{-3}$. We fix rest of the
parameters as $E_{KE,iso}=10^{53}$ erg, $\theta_j=0.2$ rad,
$\epsilon_e=0.1$, and $p=2.2$ as before. There is a clear trend.  The
higher the magnetic energy density, the brighter the radio afterglow
peak.  This is not surprising since the synchrotron emission is more
efficient in high magnetic field. This trend is seen from $n=0.1$
cm$^{-3}$ to $n=10$ cm$^{-3}$ but it does not continue to the higher
density $n=100$ cm$^{-3}$. At higher densities the peak flux density does not
increase dramatically with increasing $\epsilon_B$, there is only a
shift of the time-to-peak to later times. 
This is due to the effects of the large synchrotron absorption frequency which  
suppresses the emission.

As expected, the radio afterglow strength strongly depends upon the
kinetic energy of the burst (lower left panel of Figure~\ref{fig:synall}). 
The curve with an
isotropic energy of $10^{53}$ erg is identical to that in Figure~\ref{fig:density2} 
and we noted above that it provided a reasonable
description of the average properties of our sample. The
beaming-corrected energy for this burst is $2\times10^{51}$ erg, a
value that is at least consistent with the kinetic energy derived from
radio calorimetry \citep{bkf04,vkr+08}.

\begin{figure}
\centering
\includegraphics[width=0.48\textwidth]{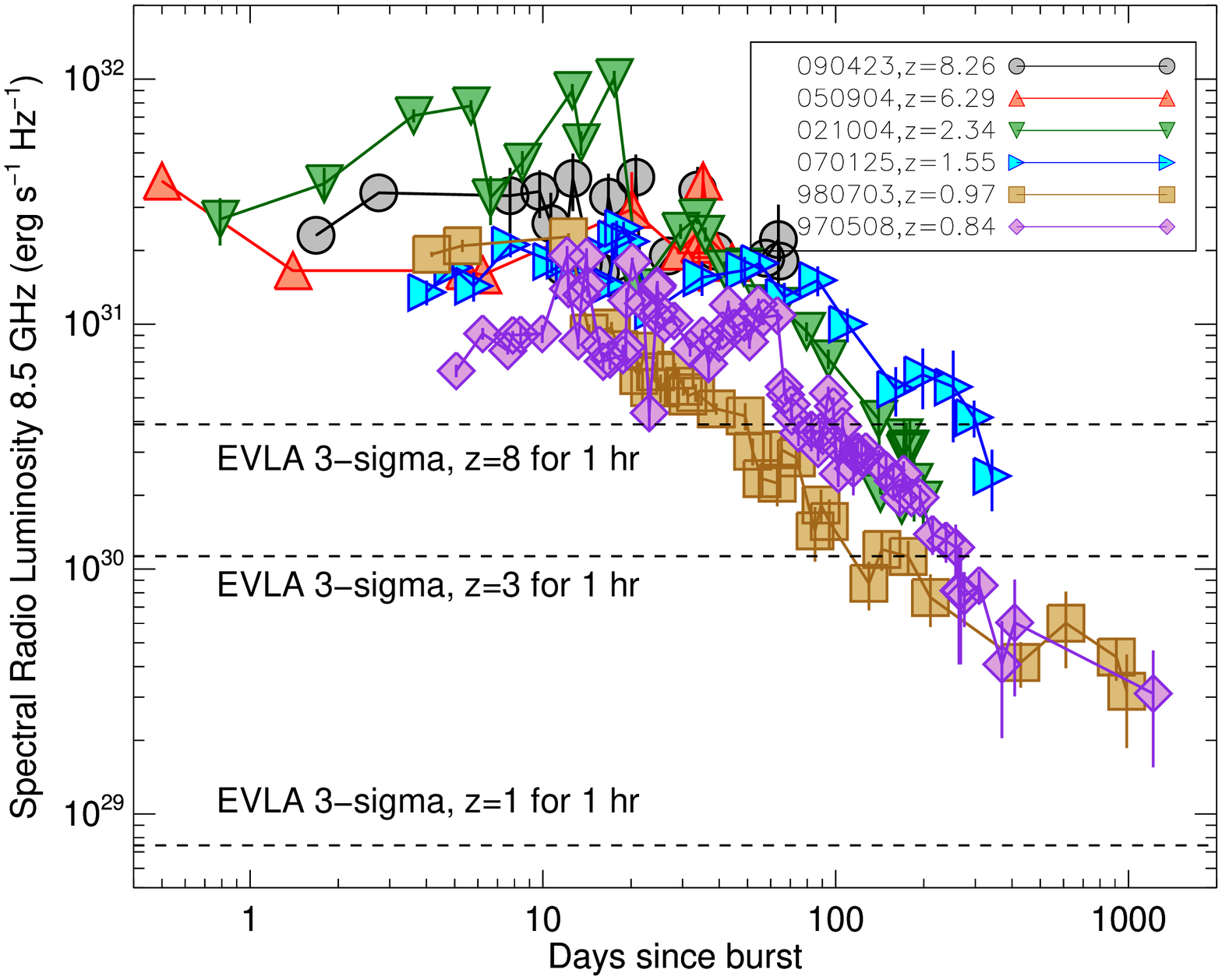}
\caption{8.5 GHz radio light curve luminosity plots for some
radio afterglows at different redshifts. We also plot the
1 hr EVLA 3-$\sigma$ sensitivity limit at a $z=1$, 3 and 8.
}
\label{fig:evla}
\end{figure}

In the lower right panel of the Figure~\ref{fig:synall}, we determine
the redshift dependence of radio afterglows for values of $z=$0.5, 1,
3, 5 and 8. Here we fix other parameters as $E_{KE,iso}=10^{53}$ erg,
$\theta_j=0.2$ rad, $\epsilon_e=0.1$, $\epsilon_B=1$\%, $n=10$
cm$^{-3}$, and $p=2.2$.  Between the redshifts of 0.5 to 3, the
lightcurves get progressively fainter with decreasing redshift, but
beyond about $z\sim 3$, the negative-$k$ correction effects come into
effect and the peak radio flux density falls only slightly with
increasing redshift (see also \S\ref{sec:red}).
Thus, essentially all the bursts with kinetic enegies $E_{KE,iso}\ge10^{53}$ 
erg should be easily detectable with less than 20-30
minutes exposure with the EVLA at any redshift. 
In Figure~\ref{fig:evla}, we also plot some the the LGRBs with known
redshifts between $z=0.8$ to $z=8.3$ and show the capability of the
EVLA to observe these bursts for a much longer duration.

\begin{figure*}
\centering
\includegraphics[width=0.33\textwidth]{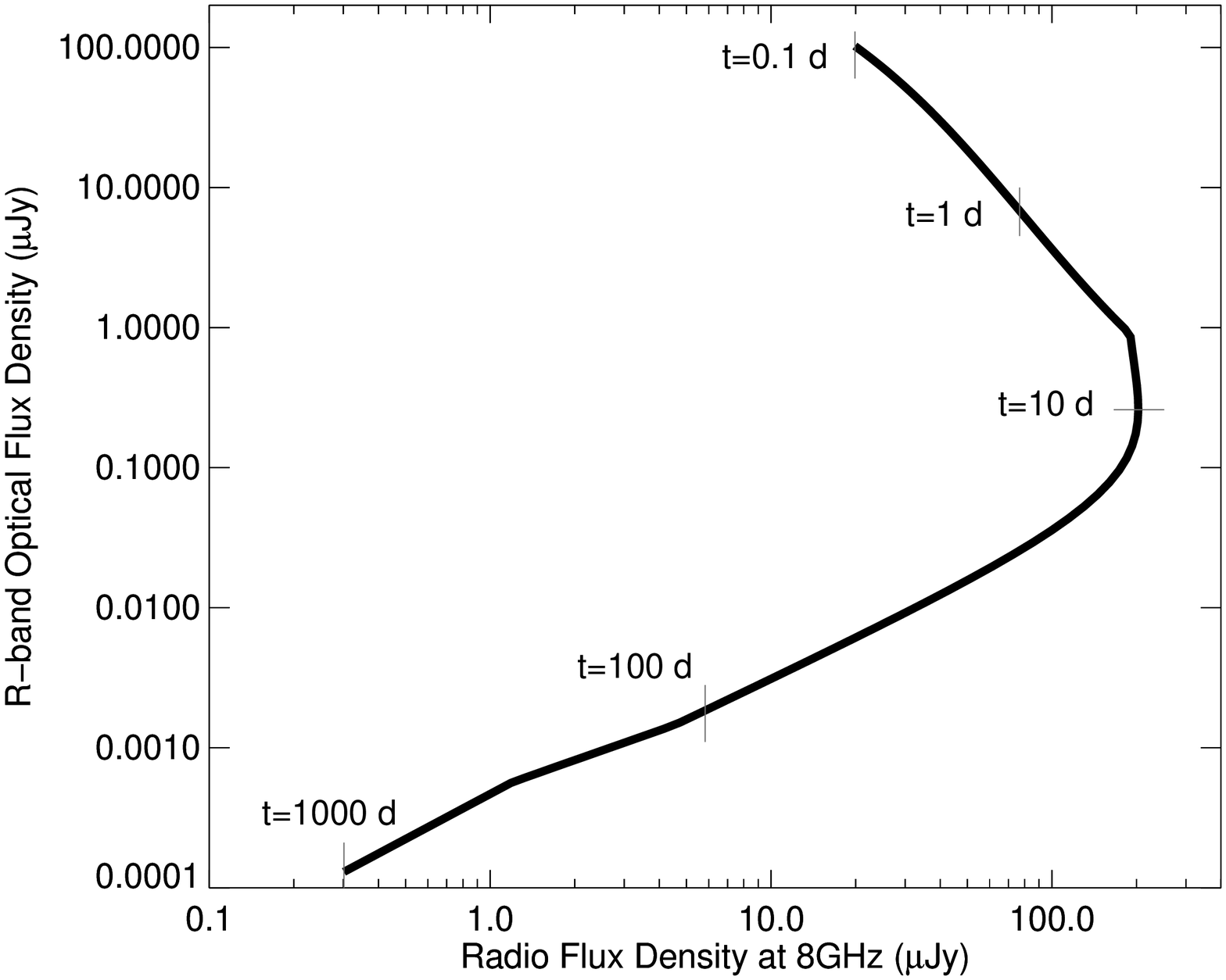}
\includegraphics[width=0.33\textwidth]{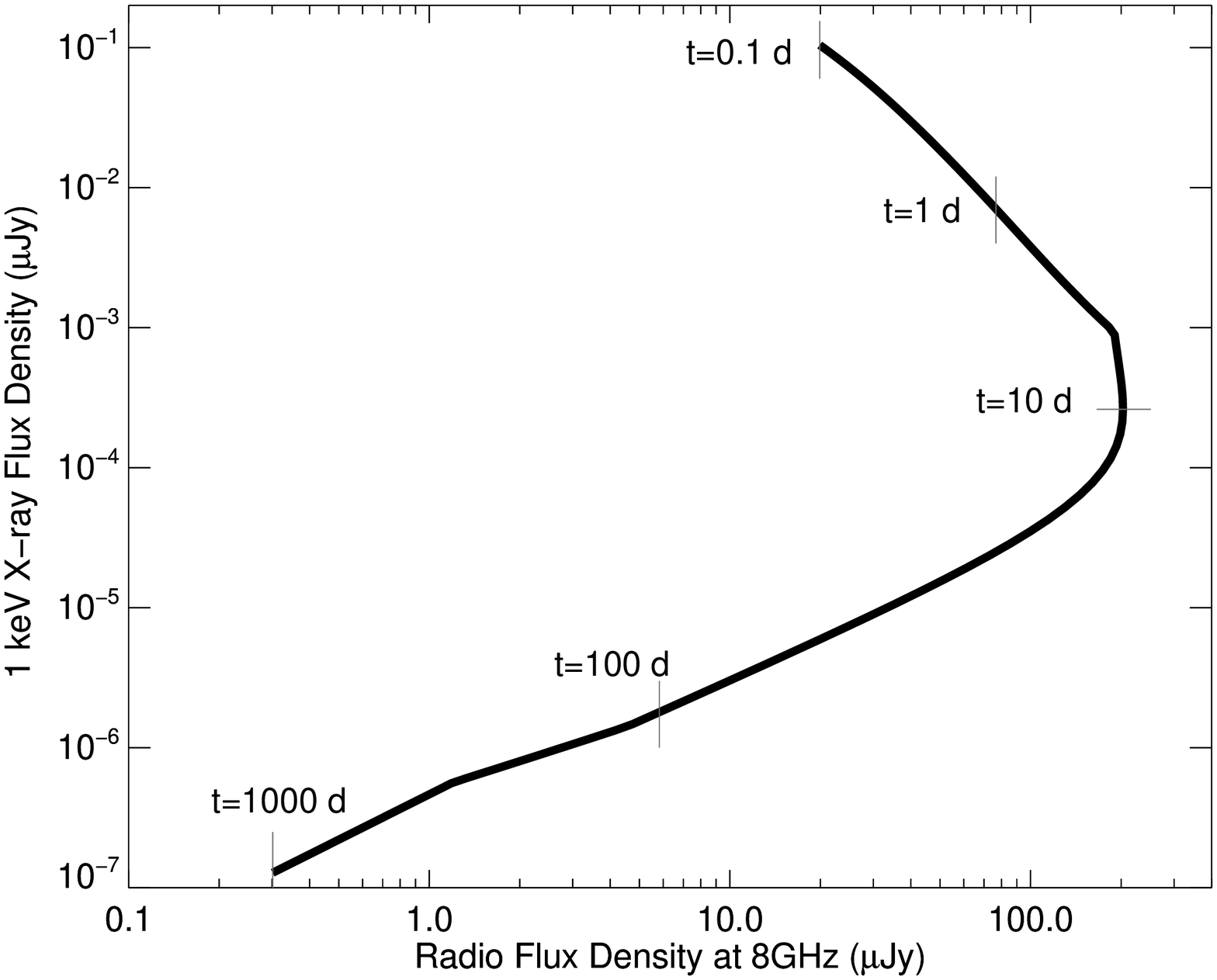}
\includegraphics[width=0.33\textwidth]{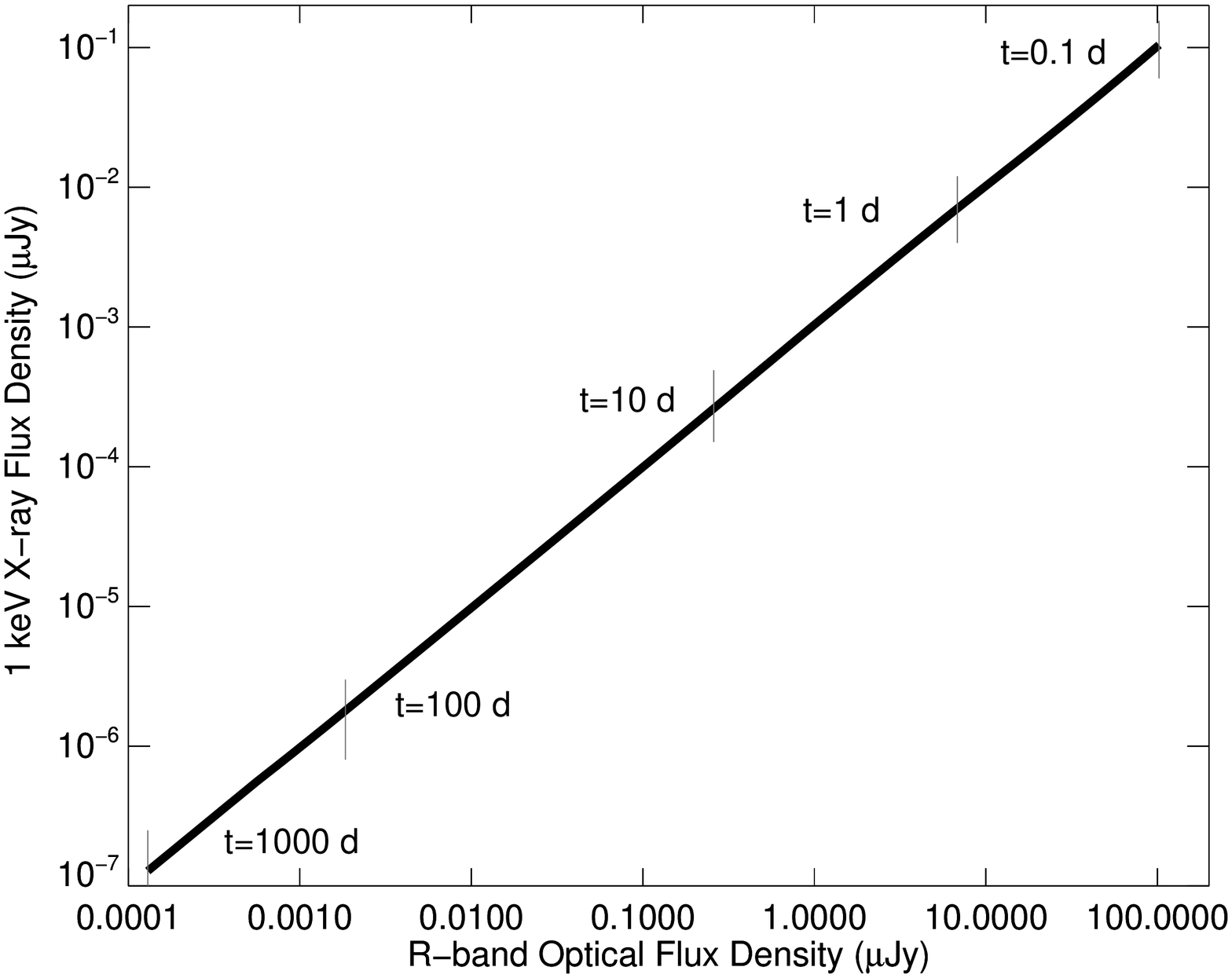}
\caption{Synthetic model radio flux density values at 8.5 GHz 
  with respect to the corresponding R-band optical (left panel) and 1
  keV X-ray (middle panel) flux densities. Time stamps are indicated
  from 0.1 to 1000 days. For comparison, the right panel shows the
  plot of optical versus X-ray flux densities. The model is the same
  as that used in Fig.~\ref{fig:density2}. See text for more details.}
\label{fig:synopt-xray}
\end{figure*}

Finally in Figure~\ref{fig:synopt-xray} we show flux density
relationships for our standard afterglow model at X-ray, optical and
radio wavelengths. These curves are not meant to be a complete
exploration of the phase space but they represent the broad behaviors
to be expected with time. The correlations (or lack thereof) in
\S\ref{corr} can now be understood. As we have already noted, our
radio sample is sensitivity limited. Most 8.5 GHz measurements are
tightly clustered between 30 $\mu$Jy to around 400 $\mu$Jy with a mean
of about 170 $\mu$Jy. In contrast, optical and X-ray measurements are
sensitive to flux densities which vary over many more orders of
magnitude. We can see from the figures that this limitation of the
radio sample restricts the time range over which the correlations can
be investigated.  There is an additional complication in
Figure~\ref{fig:synopt-xray} and that is the temporal evolution of the
radio flux density is complex compared to the nearly monotonic
behavior at X-ray and optical wavelengths.

\section{ALMA predictions}
\label{sec:alma}

Our radio-selected catalog also contains millimeter and submillimeter
radio observations. There are 186 mm/submm observations of 30 GRBs in
our database so an analysis similar to the one we have done for the
centimeter data is possible. However, \citet{alma} have recently done
a comprehensive analysis of all mm and submm data available for GRB
afterglows. It is more useful to simply summarize their findings and
then compare their results with a typical GRB afterglow model
(Figure~\ref{fig:density2}) and explore the dependence of the radio
millimeter afterglow brightness as a function of density, redshift,
energy and $\epsilon_B$ (e.g. \S\ref{sec:synthetic}).

The first mm/submm observations of GRBs were undertaken shortly after
the discovery of the cm afterglow \citep{sgll97,sfkm98,bkg98}. In
subsequent compilations \citep{stv+99,stt+05} it was well-established
that the mm/submm detection rate was strongly limited by sensitivity,
with most detected events with flux densities in excess of one or two
milliJanskys.  The situation has not improved significantly in
subsequent years but the upcoming ALMA represents a jump of two orders
of magnitude in continuum sensitivity.

\citet{alma} report a detection rate of 25\% for a sample of 102  mm/submm
afterglows. 
\citet{alma} use the detections and upper limits
(assuming a Gaussian distribution) to make a rough estimate of the
mean peak mm/submm flux density of 0.3 mJy. The average redshift
of their sample is $z_{av}=1.99$ and the average peak luminosity is
10$^{32.1\pm0.7}$ erg s$^{-1}$ Hz$^{-1}$.

Using typical GRB parameters, we showed in Figure~\ref{fig:density2})
that a radio afterglow will be an order of magnitude brighter in the
mm/submm (e.g. 250 GHz band) than in cm (e.g. 8.5 GHz). Support for
this claim comes from average peak spectral luminosities which also
differ by an order of magnitude (\S\ref{sec:radioafterglows}).

\begin{figure*}
\centering
\includegraphics[width=0.48\textwidth]{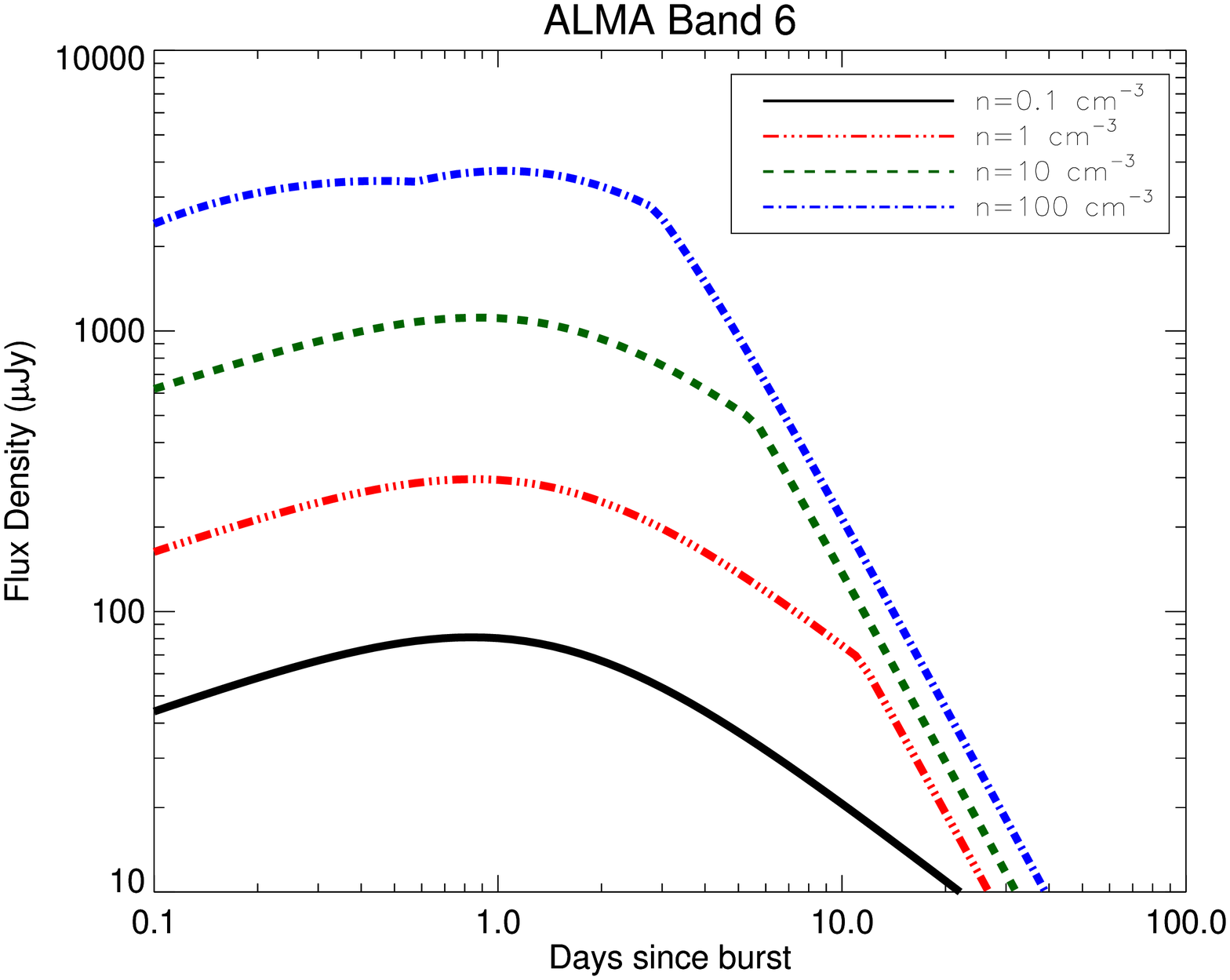}
\includegraphics[width=0.48\textwidth]{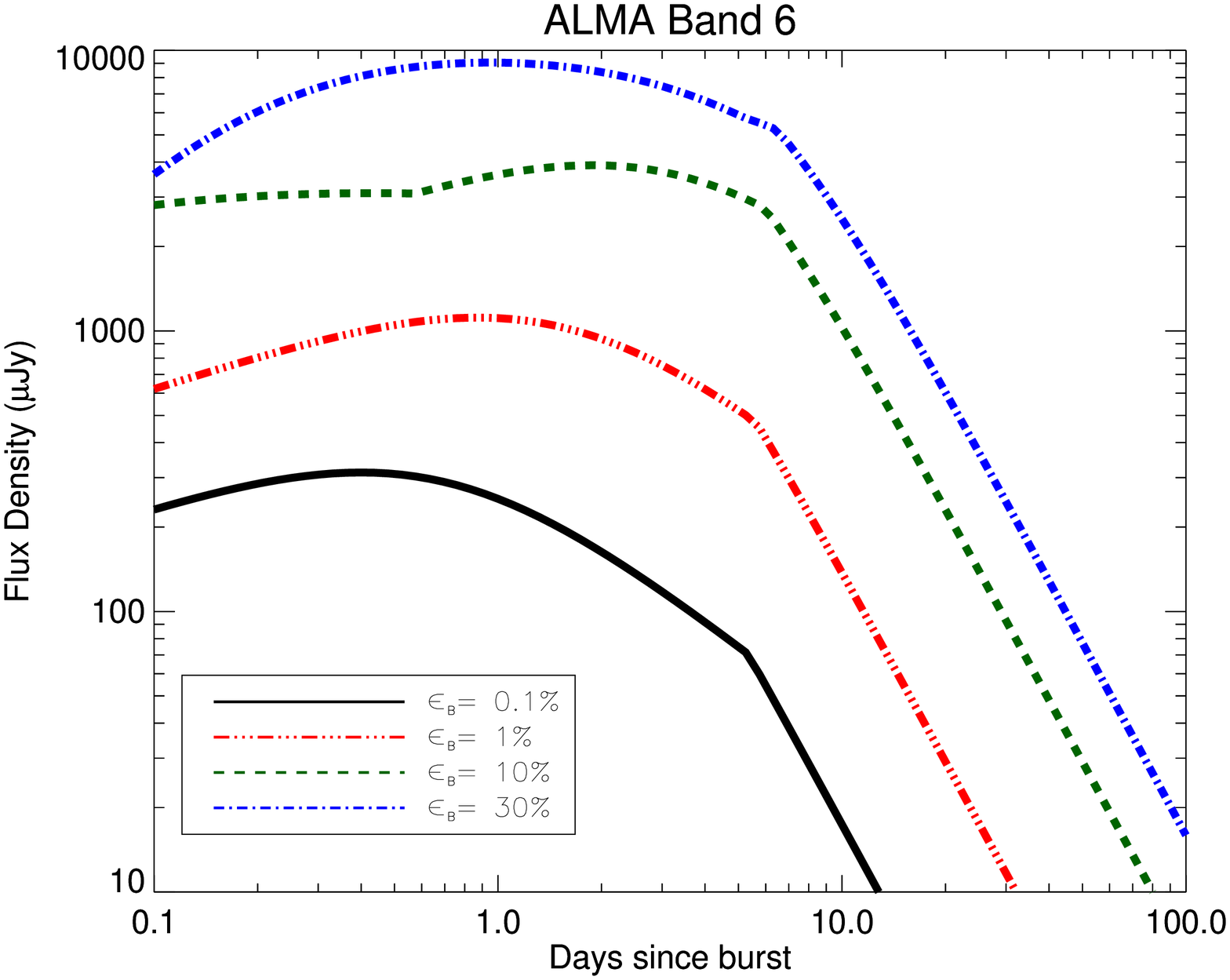}
\includegraphics[width=0.48\textwidth]{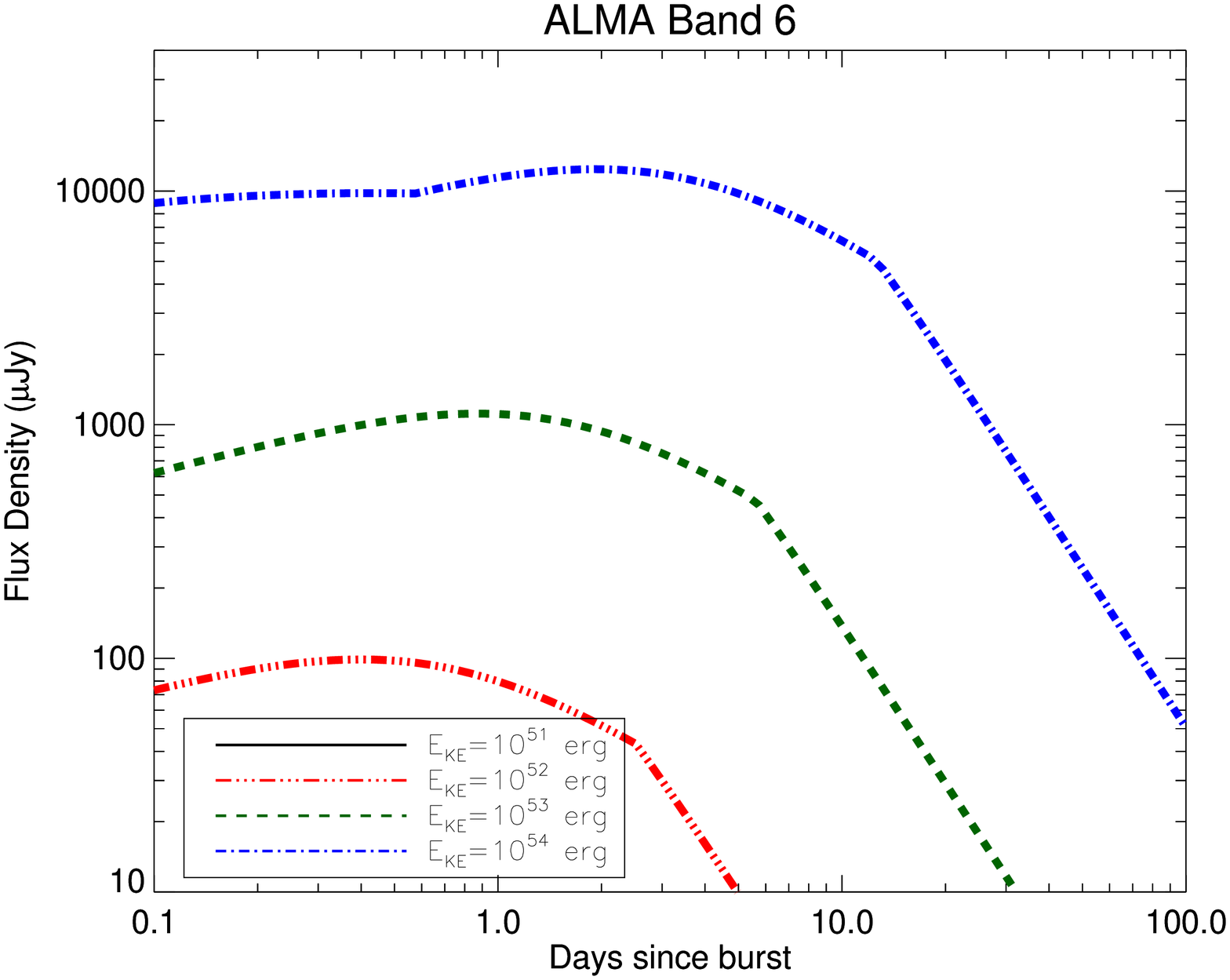}
\includegraphics[width=0.48\textwidth]{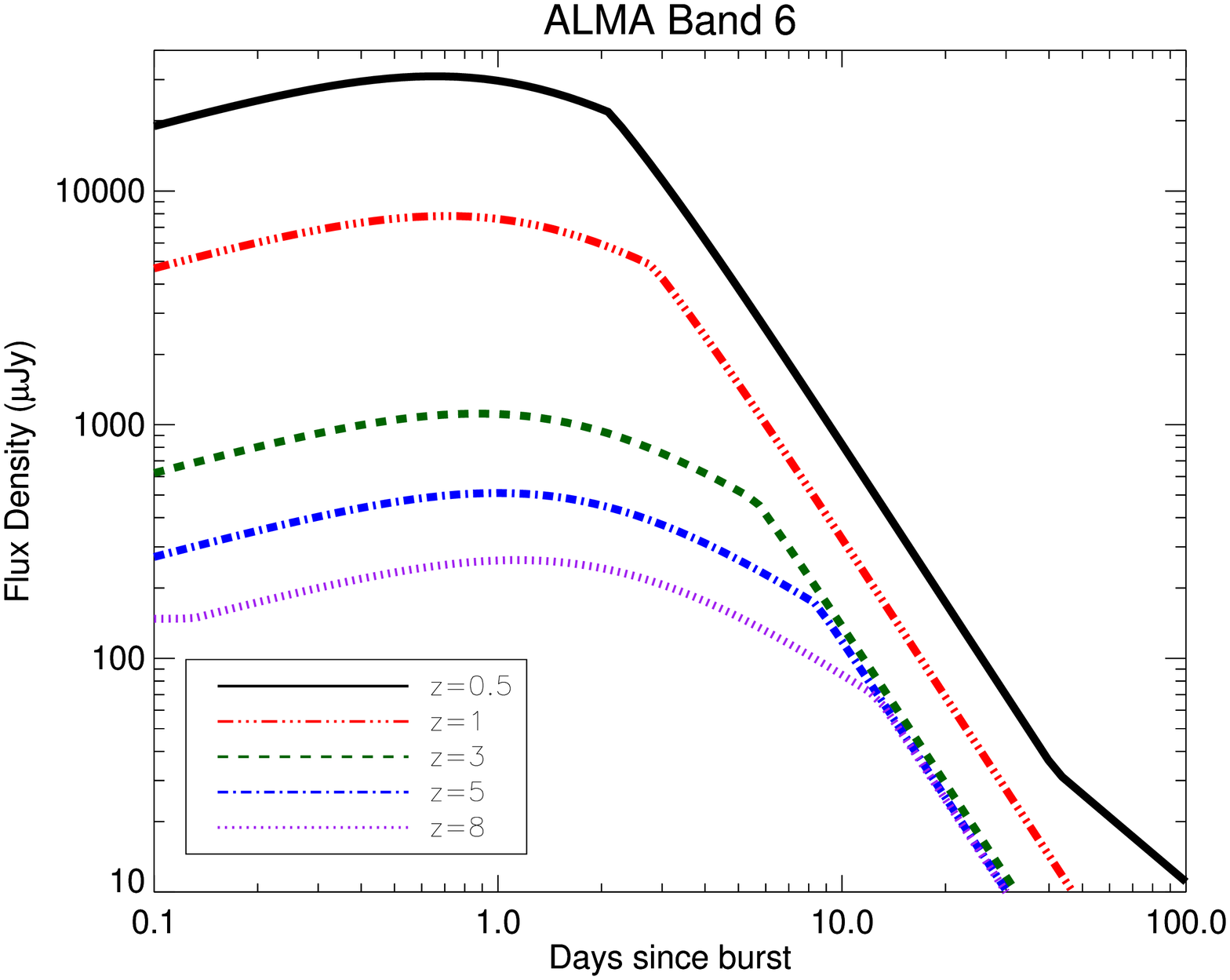}
\caption{ALMA 250 GHz (Band 6) synthetic light curves for various values of density (top left),
magnetic energy density (top right), isotropic kinetic energy (lower left)
and redshift (lower right).
ALMA band 6 sensitivity in its full capability will be 42 $\mu$Jy in
1 hr integration time.}
\label{fig:synall-alma}
\end{figure*}

We also can explore synthetic light curves identical to the cm curves
from \S\ref{sec:synthetic} but applied to the ALMA band 6 (211--275
GHz band) as a function of various parameters.  In the upper left
panel of Figure~\ref{fig:synall-alma}, we plot the synthetic radio
light curves of GRB afterglows for various values of density $n$ at a
redshift $z=3$.  We fix other parameters as given in
\S\ref{sec:synthetic}. The mm/submm radio afterglow is a strong
function of density.  Since the effects of synchrotron self-absorption
are weak in mm/submm bands, there is no competing effect to reduce the
emission at higher densities (unlike at cm bands). 
 With its full collecting area, ALMA's
3-$\sigma$ sensitivity for 1 hr integration time in band 6 (211-275
GHz) will be 42 $\mu$Jy \citep{alma}. 
Thus ALMA should be able to detect
all mm afterglows for $n>0.1$ cm$^{-3}$.  Not only is the mm/submm
flux density brighter than the cm emission, it does not suffer from
diffractive and refractive interstellar scintillation. 

To determine the dependence of radio afterglow brightness as a
function of $\epsilon_B$, we plot the lightcurves for various values
of fractional magnetic energy density in the top right plot of
Figure~\ref{fig:synall-alma}.  We derive the light curves for a burst
at a $z=3$ and for circumburst density $n=10$ cm$^{-3}$. Here again we
fix the rest of the parameters as above.  We note that higher the
magnetic energy density, the brighter the radio afterglow during its
full evolution.  Again this is in contrast to radio light curves at 8
GHz (\S\ref{sec:synthetic}), where higher $\epsilon_B$ was suppressing
the optically thick flux density in high density medium.

As expected, the mm afterglow strength also very strongly depends upon
the kinetic energy of the burst (lower left panel of
Figure~\ref{fig:synall-alma}). ALMA will have some difficulties
detecting a burst with $E_{KE,iso}<10^{52}$.  In the lower right panel
of the Figure~\ref{fig:synall-alma}, we determine the redshift dependence of
radio afterglows for values of $z=$0.5, 1, 3, 5 and 8.  Here we again
fix other parameters as $E_{KE,iso}=10^{53}$ erg, $\theta_j=0.2$ rad,
$\epsilon_e=0.1$, $\epsilon_B=1$\%, $n=10$cm$^{-3}$, and $p=2.2$. The
trend is similar to the cm but the mm/submm lacks the synchrotron
self-absorption contribution to the negative-$k$ correction.

\section{Discussion and Conclusions}\label{sec:discussion}

In this paper we have formed a radio-selected sample of 304 GRBs based
on 14 years of observations. In addition to the radio observations, we
have collected supplementary data including redshift, gamma-ray
fluence, and optical and X-ray flux densities on all bursts. We have
used these data to carry out the first comparative and correlative
study of the prompt and afterglow emission across the full
electromagnetic spectrum. Previous studies
\citep[e.g.][]{gbb+08,nf09,kkz+10} have not included radio data in
the analysis.

We show that the fractional detection rate of radio afterglows is 31\%
(\S\ref{venn}).  These detection statistics do not change
substantially between the pre-\emph{Swift} (42/123 or 34\%) and
post-\emph{Swift} (53/181 or 29\%) samples. This is markedly different
than the factor of two increase in the detection probability of
optical and X-ray afterglows post-\emph{Swift}. The benefits of
\emph{Swift}'s on-board detection, autonomous slews, and the
availability of rapid, well-localized positions are not realized by
the slower evolving radio afterglows.

Next we investigated the flux density distribution of the radio
afterglows for cosmological GRBs (\S\ref{sec:radioafterglows}). The
mean flux density in the 8.5 GHz band between days 2--8 of the 65
detections is 287$\pm$26 $\mu$Jy. But if we include the non-detections
(i.e. total of 157 8.5 GHz flux density points between days 2--8),
using a Kaplan-Meier estimator, the mean is 162$\pm$15 $\mu$Jy. We
also showed that range between the peak flux density of the brightest
afterglow and the faintest afterglow spans only a narrow range (a
factor of 50), in sharp contrast to optical and X-ray afterglows which
range over several orders of magnitude.  Together this suggests
that the radio radio afterglow searches are sensitivity
limited and that this bias will be present in correlative studies (see
below).

Using the measured redshifts for our sample, we calculated the mean
spectral luminosity and the mean time-to-peak to be $\sim10^{31}$ erg
s$^{-1}$ Hz$^{-1}$ and 3-6 days, respectively
(\S\ref{sec:radioafterglows}). We investigated the luminosity
distribution of several sub-classes of GRBs including the
long-duration, cosmological GRBs (LGRB), short-hard bursts (SHB),
X-ray flashes (XRF) and GRBs with firm supernova associations
(SNe/GRB). There are a wide spread of luminosities for these different
subclasses with LGRBs the brightest on average, while SHBs are nearly
two orders of magnitude fainter than LGRBs, and XRFs and SNe/GRBs are
ten times fainter than LGRBs, albeit with large scatter (\S
\ref{sec:different}).  The best sampled radio light curves are those of
the LGRB class with multiple data points that range from nearly 0.01 d
to 1000 d. The canonical LGRB will reach a peak luminosity of
$\sim2\times 10^{31}$ erg s$^{-1}$ Hz$^{-1}$ about 3 to 6 days in the
rest frame, and after about 10 to 20 days will undergo a power
law decline with an index of order unity (\S \ref{sec:cannon}).  There
is some (not strong) evidence of two equally luminous emission
components with a transition occurring at 1 d in the rest frame. The
early component may be the result of the reverse shock as has been
claimed previously for several bursts
\citep{kfs+99,fps+03,bsfk03,cff+10} while the later component is the
standard forward shock which has been studied through extensive
broad-band modeling \citep{pk01}.

Before any correlative analysis was carried out, we investigated the
high energy and non-radio afterglow properties for our sample.  We
find that the distribution of gamma-ray fluence, duration, energy, and
optical/X-ray flux densities for the 304 GRBs in our sample are not
dissimilar those from previous studies. There is a difference in the
redshift distribution (\S\ref{sec:red}).  While the mean redshift for
our pre-\emph{Swift} sample agrees with past work, 
for our post-\emph{Swift} sample we derive a lower
value for the mean redshift (2.0) based on the radio selected sample
compared to published values of
$z=2.2$ \citep{fjp+09} and $z=2.8$ \citep{jlf+06}, with \citet{fjp+09} 
value closer to our mean.
 Owing to
a negative $k$-correction \citep{cl00}, radio afterglows are rather
insensitive to redshift but we show that the mean radio flux density
of high redshift GRBs are close to the sensitivity limit of a typical
observation made with existing instruments. This bias will have an
effect on any distance-dependent correlations.

We examined the observed GRB and afterglow parameters for the radio
sample (\S\ref{sec:multi}) and we found that radio-detected GRBs had
higher fluences, larger energies, and brighter X-ray and optical
fluxes on average, compared to radio non-detections. There is
evidently a statistical relationship between the detectability of a
radio afterglow and these properties. Further correlations were explored
between the peak radio flux density and gamma-ray fluence, optical flux
density and X-ray flux density (\S\ref{corr}).
The only significant correlation that was found was between the peak
radio flux density and the optical flux density at 11 hr.

The synthetic light curves \S\ref{sec:synthetic} show that EVLA Ka
band (26-40 GHz) and ALMA mm/submm bands are the most favorable bands
for radio afterglow studies.  We also show that the centimeter radio
afterglow emission will be brightest for circumburst densities between
$n=1-10$ cm$^{-3}$.  Outside of this range, the flux density will be
weak either due to a low intrinsic emission strength (at lower
densities) or due to increased synchrotron self-absorption (at higher
densities). Within this density range there is a simple relationship
of increasing strength of the centimeter radio emission and the
fraction of the shock energy in the magnetic field ($\epsilon_B$), but
at higher densities synchrotron absorption suppresses the emission.
The synthetic light curves also confirm what is already
well-established and that is that the radio afterglow flux density is
only weakly dependent on redshift at $z \ge 2.5$.

We also make predictions at mm/submm wavelengths (\S\ref{sec:alma}),
showing that ALMA should be able to detect most afterglows assuming
typical parameters. The synchrotron self absorption effect does not play
at a major role in the mm/submm bands, thus ALMA should be able to
detect afterglows very early on. 
ALMA should be able to provide a true distribution of the
mm/submm afterglows, which  is not well known.
With an order of magnitude of
continuum sensitivity of the EVLA and two orders of magnitude sensitivity 
of the ALMA, EVLA and ALMA will dominate GRB afterglow
studies in the future. These telescopes will increase the temporal
phase space and the redshift space over which GRBs can be observed. We
illustrate this in Figure~\ref{fig:evla} with the EVLA 3-$\sigma$
spectral luminosity sensitivity at redshifts of 1, 3 and 8.  

\acknowledgments

We thank the anonymous referee for helpful comments.  We thank all our
numerous colleagues over the last 15 years who took the original data
upon which this paper is based. Special mention is due to Shri
Kulkarni, Edo Berger, Alicia Soderberg, Brian Cameron, Josh Bloom,
Brian Metzger, and Brad Cenko for their roles in the VLA program, Mark
Wieringa for the ATCA data and Guy Pooley for the Ryle telescope data.
We thank David Alexander Kann, Bing Zhang, Neil Gehrels and Alexander
J.  van der Horst for useful suggestions.  We acknowledge the frequent
use of GRBlog (\url{http://grblog.org/grblog.php}), GCN circulars
(\url{http://gcn.gsfc.nasa.gov/gcn3\_archive.html}), Nat Butler's
Swift BAT+XRT(+optical) lightcurve repository
(\url{http://astro.berkeley.edu/~nat/swift/}) and Jochen Greiner's GRB
afterglow page (\url{http://www.mpe.mpg.de/~jcg/grb.html}).  We also
thank online Swift/XRT GRB lightcurve repository
\url{http://www.swift.ac.uk/xrt\_curves/}.  The National Radio
Astronomy Observatory is a facility of the National Science Foundation
operated under cooperative agreement by Associated Universities, Inc.
This research has made use of the GHostS database
(\url{www.grbhosts.org}), which is partly funded by Spitzer/NASA grant
RSA Agreement No.  1287913.  P.C. is supported by her NSERC Discovery
grants as well as the DND-ARP grants held by Kristine Spekkens and
Gregg Wade at the Royal Military College of Canada.


\newpage\clearpage

\LongTables

\tabletypesize{\tiny}
\begin{landscape}
\begin{deluxetable}{clllccccllcclllllll}
\tablecolumns{19}
\tablewidth{0pt}
\tablecaption{Master table}
\tablehead{
\colhead{1} & \colhead{2} & \colhead{3} & \colhead{4} & \colhead{5} & \colhead{6} & \colhead{7} &
\colhead{8} & \colhead{9} & \colhead{10} & \colhead{11} & \colhead{12} & \colhead{13} &
\colhead{14} & \colhead{15} & \colhead{16} & \colhead{17} & \colhead{18} & \colhead{19} \\
\colhead{GRB}&  
\colhead{Inst.\tablenotemark{a}}&       
\colhead{RA}&   
\colhead{Decl.}&
\colhead{X\tablenotemark{b}}&
\colhead{O\tablenotemark{c}}&
\colhead{R\tablenotemark{d}}&
\colhead{Radio}&        
\colhead{$T_{90}$}&
\colhead{$z$}&  
\colhead{$S_{15-150}$\tablenotemark{e}} &
\colhead{$E_{iso}^{bol}$\tablenotemark{f}}      &       
\colhead{$F_{X}^{11h}$\tablenotemark{g}}        &       
\colhead{$F_{R}^{11h}$\tablenotemark{h}}&       
\colhead{$t_j$} &       
\colhead{$n$\tablenotemark{i}}  &
\colhead{$\theta_j$}    &       
\colhead{$E_{true}^{bol}$\tablenotemark{j}}     &       
\colhead{References\tablenotemark{k}} 
\\
&       &       \colhead{J2000} &       
\colhead{J2000} &               &               &               &
\colhead{Tel}   &       \colhead{s}     &               &       \colhead{erg cm$^{-2}$}    &       \colhead{erg}   &
\colhead{CGS}   &       \colhead{$\mu$Jy}       &       \colhead{d}     &       \colhead{cm$^{-3}$}     &
\colhead{deg.}  &       \colhead{erg}   &               
}
\startdata
970111  &       B, T    &       15:28:15.0      &       +19:36:18       &       N       &       N       &       N       &
VLA\tablenotemark{r}    &       31      &       \ldots  &       3.30E-05        &       \ldots  &       \ldots  &       \ldots  &
\ldots  &       \ldots  &       \ldots  &       \ldots  &       19,[],[],[],[],[],[],[] \\
970228\tablenotemark{n} &       B       &       05:01:46.7      &       +11:46:53       &       Y       &       Y       &       N
&       VLA     &       56      &       0.695   &       5.55E-06        &       1.60E+52        &       20.8    &       22.0    &
\ldots  &       \ldots  &       \ldots  &       \ldots  &       19,7,[],22,37,37,[],[]  \\
970402  &       B       &       14:50:06.0      &       -69:20:00       &       Y       &       N       &       N       &       ATCA
&       105     &       \ldots  &       4.84E-06        &       \ldots  &       1.4     &       \ldots  &       \ldots  &
\ldots  &       \ldots  &       \ldots  &       19,[],[],[],37,[],[],[] \\
970508  &       B       &       06:53:49.2      &       +79:16:19       &       Y       &       Y       &       Y       &
VLA\tablenotemark{r}    &       14      &       0.835   &       1.11E-06        &       7.10E+51        &       5.7     &       6.5
&       25      &       1       &       20.93   &       4.74E+50        &       19,7,[],22,37,37,17,61  \\
970616  &       B, T    &       01:18:48.0      &       -05:26:18       &       Y       &       N       &       N       &
VLA     &       200     &       \ldots  &       1.78E-05        &       \ldots  &       38.6    &       \ldots  &       \ldots  &
\ldots  &       \ldots  &       \ldots  &       28,[],[],[],37,[],[],[] \\
\enddata
\end{deluxetable}
\clearpage
\end{landscape}

\end{document}